\documentclass[aps,pre,twocolumn,superscriptaddress,a4paper,english,longbibliography]{revtex4-1}
\usepackage{silence}
\usepackage{times}
\usepackage{color}
\usepackage[colorlinks=true,urlcolor=blue,citecolor=blue]{hyperref}
\usepackage{url}
\usepackage{breakurl}
\usepackage{soul}
\usepackage{graphicx}
\usepackage{amsmath}
\usepackage{subfigure}
\usepackage{xcolor}
\usepackage{amssymb}
\usepackage{latexsym}
\usepackage{hyperref}
\DeclareGraphicsExtensions{.png,.eps}
\usepackage{float}
\usepackage{appendix}
\usepackage{etoolbox}
\usepackage{physics}

\usepackage{xcolor}
\usepackage[urlcolor=blue]{hyperref}      
\hypersetup{
    colorlinks = true,                    
    citecolor = {blue},
    linkcolor = {purple},
           }

\usepackage[normalem]{ulem}

\begin{document}

\title{Eigenstate Thermalization and its Breakdown in Quantum Spin Chains with Inhomogeneous Interactions}

\author{Ding-Zu Wang} \thanks{These two authors contributed equally to this work}
\affiliation{School of Physics, Beihang University,100191,Beijing, China}

\author{Hao Zhu} \thanks{These two authors contributed equally to this work}
\affiliation{School of Physics, Beihang University,100191,Beijing, China}

\author{Jian Cui}
\affiliation{School of Physics, Beihang University,100191,Beijing, China}

\author{Javier Argüello-Luengo}
\affiliation{ICFO - Institut de Ciencies Fotoniques, The Barcelona Institute of Science and Technology, Av. Carl Friedrich Gauss 3, 08860 Castelldefels (Barcelona), Spain}

\author{Maciej Lewenstein}
\affiliation{ICFO - Institut de Ciencies Fotoniques, The Barcelona Institute of Science and Technology, Av. Carl Friedrich Gauss 3, 08860 Castelldefels (Barcelona), Spain}
\affiliation{ICREA, Pg. Llu\'is Companys 23, 08010 Barcelona, Spain}

\author{Guo-Feng Zhang}
\email[Corresponding author: ]{gf1978zhang@buaa.edu.cn}
\affiliation{School of Physics, Beihang University,100191,Beijing, China}

\author{Piotr Sierant}
\email[Corresponding author: ]{Piotr.Sierant@icfo.eu}
\affiliation{ICFO - Institut de Ciencies Fotoniques, The Barcelona Institute of Science and Technology, Av. Carl Friedrich Gauss 3, 08860 Castelldefels (Barcelona), Spain}

\author{Shi-Ju Ran}
\email[Corresponding author: ]{sjran@cnu.edu.cn}
\affiliation{Center for Quantum Physics and Intelligent Sciences, Department of Physics, Capital Normal University, Beijing 10048, China}

\date{\today}

\begin{abstract} 
	The eigenstate thermalization hypothesis (ETH) is a successful theory that establishes the criteria for ergodicity and thermalization in isolated quantum many-body systems. In this work, we investigate the thermalization properties of spin-$ 1/2 $ \textit{XXZ} chain with linearly-inhomogeneous interactions. We demonstrate that introduction of the inhomogeneous interactions leads to an onset of quantum chaos and thermalization, which, however, becomes inhibited for sufficiently strong inhomogeneity. To exhibit ETH, and to display its breakdown upon varying the strength of interactions, we probe statistics of energy levels and properties of matrix elements of local observables in eigenstates of the inhomogeneous \textit{XXZ} spin chain. Moreover, we investigate the dynamics of the entanglement entropy and the survival probability which further evidence the thermalization and its breakdown in the considered model. We outline a way to experimentally realize the \textit{XXZ} chain with linearly-inhomogeneous interactions in systems of ultracold atoms. Our results highlight a mechanism of emergence of ETH due to insertion of inhomogeneities in an otherwise integrable system and illustrate the arrest of quantum dynamics in presence of strong interactions.
\end{abstract}

\maketitle

\section{Introduction}
Understanding the thermalization and equilibration in isolated quantum many-body systems has been a central topic since the birth of quantum mechanics~\cite{1929.ZPhy.57.30N, 2010.EPJH.35.173G}. Its growing interest is closely linked to the remarkable progress in the ultracold atomic experiments~\cite{2008.RevModPhys.80.885, 2015.ARCMP.6.201L, 2015.NatPh.11.124E}, where advancements in control and isolation have enabled the coherence in many-body systems over unprecedented time scales~\cite{2007.AdPhy.56.243L, 2012.NatPh.8.267B}. The experiments on non-equilibrium dynamics have revealed thermalization in the chaotic quantum systems~\cite{2012.NatPh.8.325T, 2016.Sci.353.794K, 2016.NatPh.12.1037N, 2018.PhysRevX.8.021030, 2016.PhysRevLett.117.170401}, which is inhibited in the integrable systems~\cite{2006.Natur.440.900K, 2012.Sci.337.1318G, 2015.Sci.348.207L, Vidmar16}.

Thermalization in generic (quantum-chaotic, non-integrable) isolated quantum many-body systems can be explained by the eigenstate thermalization hypothesis (ETH)~\cite{1991.PhysRevA.43.2046, 1994.PhysRevE.50.888}. The ETH is usually formulated as an ansatz for matrix elements of physical observables in the eigenbasis of the Hamiltonian~\cite{1991.PhysRevA.43.2046, 1994.PhysRevE.50.888, 2008.Natur.452.854R, pappalardi2023microcanonical}. This ansatz guarantees that the local observables after relaxation can be described by appropriate ensembles of statistical mechanics, while the fluctuations in a steady state satisfy the fluctuation dissipation theorem~\cite{2016.AdvancesinPhysics}.
Recently, the connections between the notions of: $k$-designs \cite{Dankert09unitary}, the theory of free probability \cite{Voiculescu1991} and ETH were made explicit~\cite{Pappalardi22,Pappalardi23,fava2023designs}.

The validity of the ETH ansatz has been confirmed in a wide range of quantum many-body systems, including spin chains, bosonic and fermionic models, or systems with electron-photon coupling~\cite{Rigol09, Steinigeweg13, 2013.PhysRevLett.111.050403, Beugeling14,Sorg14, Steinigeweg14,2015.PhysRevE.91.012144,2016.PhysRevE.93.032104,Yoshizawa18, 2019.PhysRevB.99.155130, 2019.PhysRevLett.122.070601, 2021.PhysRevB.103.235137, 2023.PhysRevE.107.014130, 2023.SciPostPhys.15.2.048}. One-dimensional spin-$1/2$ \textit{XXZ} chain with various types of integrability breaking terms has become a paradigmatic system for studies of ETH \cite{2009.PhysRevB.80.125118, 2018.PhysRevB.98.235128, 2019.PhysRevE.100.062134,  2020.PhysRevE.102.062113,  2020.PhysRevE.102.042127, 2020.PhysRevB.102.075127, 2020.PhysRevLett.125.070605}. Introduction of spatial inhomogeneities is an intriguing way of integrability breaking, especially considering its relevance to nonequilibrium physics~\cite{2014.PhysRevX.4.041007, 2016.EL.11540011V, 2016.JSMTE.05.3108A, 2017.ScPP.2.2D, 2019.PhysRevB.99.214514, 2019.NatCo.10.4820B, 2020.CMaPh.377.1227G, 2022.ScPP.12.60D, 2022.PhysRevB.106.054314, 2024.QST.9a5008W} and generalized hydrodynamics~\cite{2019.PhysRevLett.123.130602, 2020.PhysRevB.102.180409, 2021.PhysRevB.103.165121, 2021.JPhA.54W4001D, 2021.JSMTE2021k4004A, 2021.JSMTE2021k4003B}.

\begin{figure}[tbp]
	\centering
	\includegraphics[angle=0,width=0.95\linewidth]{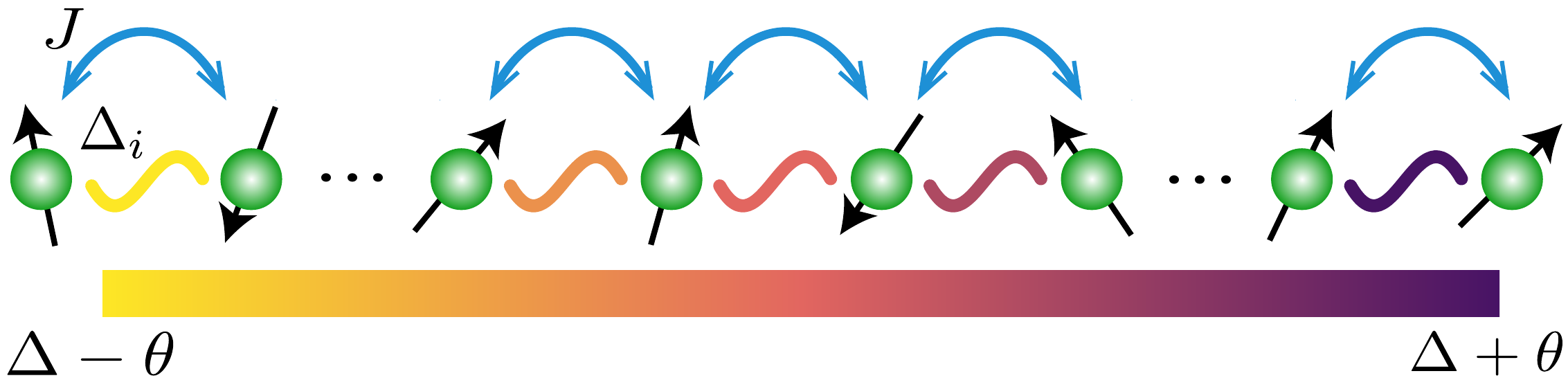}
	\caption{(Color online) Schematic representation of the \textit{XXZ} chain with linearly varying $z$-$z$ interactions. The uniform hopping term is denoted by $J$, while $\Delta_{i}$ represents an inhomogeneous $z$-$z$ coupling whose strength varies linearly with the spatial position, as depicted by the color scheme in the figure. Interaction strength is $\Delta - \theta$ ($\Delta + \theta$) at the leftmost (rightmost) site of the chain.
	}
	\label{Fig1}
\end{figure}

In this work, we investigate the eigenstate thermalization properties of the spin-$ 1/2 $ \textit{XXZ} chain with spatially inhomogeneous interaction strength, see Fig. \ref{Fig1}. The considered Hamiltonian consist of the spatially uniform hopping terms and inhomogeneous $z$-$z$ spin coupling whose strength varies linearly across the chain. By means of exact diagonalization (ED), we show that the system is driven from an integrable point to a quantum chaotic region upon introduction of the linear variation of the $z$-$z$ spin coupling. However, when the inhomogeneity becomes sufficiently strong, it inhibits the thermalization of the system.

This manuscript is structured as follows. In Sec.~\ref{sec:Model}, we detail the \textit{XXZ} model with spatially inhomogeneous interactions. In Sec.~\ref{sec:int-cha}, we formulate our predictions about integrability and thermalization in the system by investigating its level statistics. Sec.~\ref{sec:eth} contains a detailed study of ETH and its breakdown in the considered model, with particular attention devoted to properties of the matrix elements of local operators. In Sec.~\ref{sec:dyn}, we demonstrate qualitative changes in the dynamics of the system in the identified ETH and non-ergodic regimes. Finally, Sec.\ref{sec:exp}, details a blueprint proposal for realization of the considered model in cold atomic systems. We summarize our findings and provide an outlook in Sec.~\ref{sec:sum}.

\begin{figure*}[tbp]
	\centering
	\includegraphics[angle=0,width=0.85\linewidth]{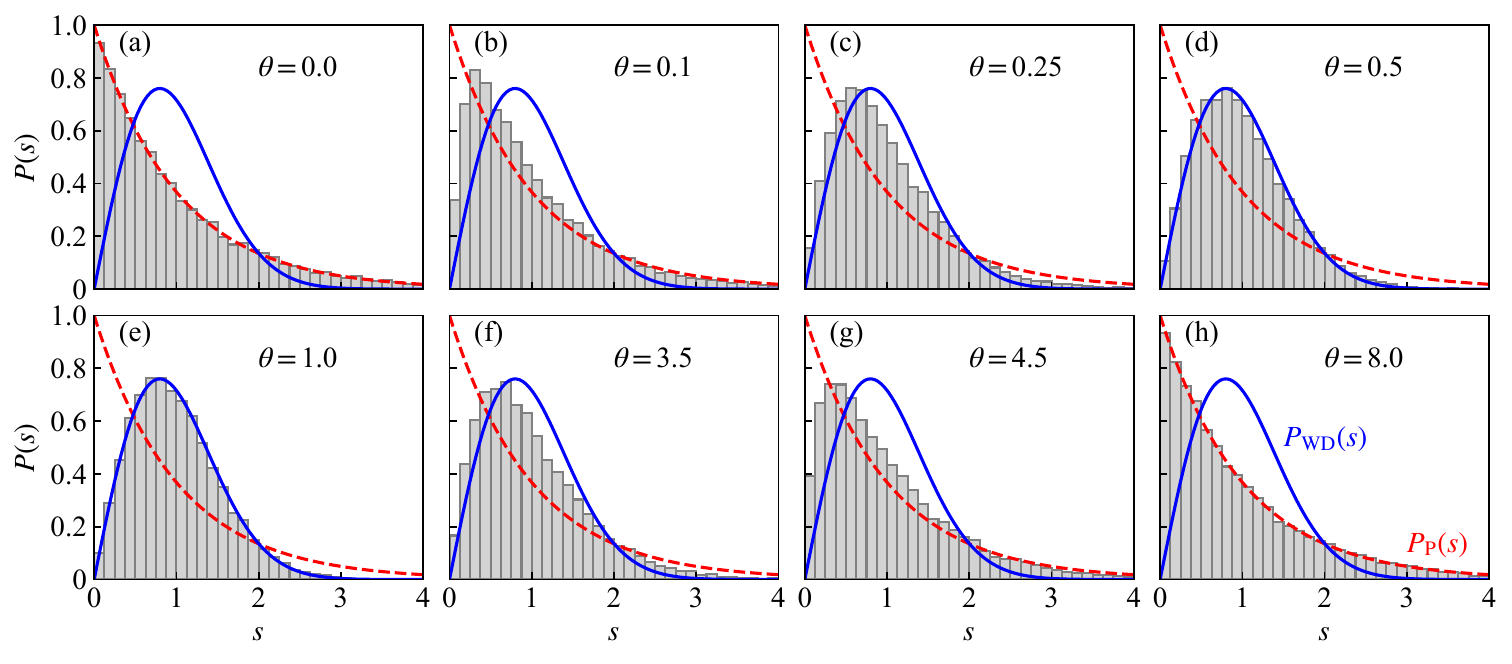}
	\caption{(Color online) Level spacing distribution $ P(s) $ of the unfolded energy spectra for the inhomogeneous \textit{XXZ} model [see Eq.~(\ref{Hami})]. The red dash and blue solid lines correspond to the Poisson and the Wigner-Dyson distribution, respectively. The results shown are for chains with open boundary conditions and $ N = 18, \Delta = 1, \text{even-$Z_{2}$ within} \sum_{i = 1}^{N } \langle \hat{\sigma}_{i}^{z} \rangle = 0$ sector. The 19448 eigenvalues in the middle of the spectrum are used for the calculation.} 
	\label{Fig2}
\end{figure*}

\section{\textit{XXZ} chain with spatially-inhomogeneous interaction}\label{sec:Model}

We consider the spin-$1/2$ \textit{XXZ} chain with inhomogeneous interaction and open boundary conditions (Fig. \ref{Fig1}), whose Hamiltonian can be written as (putting $\hbar = 1$):
\begin{eqnarray}\label{Hami}
	\hat{H}_{\text{inho}}
	=
	\sum_{i = 1}^{N - 1} \left[ J (\sigma_{i}^{x}\sigma_{i + 1}^{x} + \sigma_{i}^{y}\sigma_{i + 1}^{y}) + \Delta_{i}\sigma_{i}^{z}\sigma_{i + 1}^{z} \right],
\end{eqnarray}
where $ \sigma_{i}^{\alpha} $ represents the Pauli operator of the $ i $-th spin in the $ \alpha \in \{x, y, z\} $ direction, and $N$ is the length of chain that is taken to be even. 
The XXZ spin chain  can be mapped via the Jordan-Wigner transformation to a 
system of interacting spinless fermions, with the hopping strength equal to $J$ and  
nearest-neighbor density-density interaction strength $\Delta_i$.
We set the hopping strength $ J = 1 $ as the energy unit. The strength of the $z$-$z$ coupling terms is assumed to vary linearly with the spatial position as
\begin{eqnarray}\label{inhomogeneity}
	\Delta_{i}
		=
		\Delta + \theta \frac{2i - N}{N - 2},
\end{eqnarray}
where $ \theta $ is the slope of the linear dependence characterizing the strength of inhomogeneity and $ \Delta $ represents the average interaction strength. A homogeneous \textit{XXZ} chain will be obtained by taking  $\theta = 0$, which is a quintessential interacting integrable model~\cite{1990.PhysRevLett.65.243, 2011.RevModPhys.83.1405}.

The Hamiltonian $\hat{H}_{\text{inho}}$ in Eq. (\ref{Hami}) has the $U(1)$ symmetry as it conserves the total magnetization along the spin $z$-direction, $ [\hat{H}_{\text{inho}}, \sum_{i}^{} \sigma_{i}^{z}] = 0$. The zero magnetization sector ($\sum_{i}^{} \langle \sigma_{i}^{z} \rangle = 0$) is the largest sector that maintains the $Z_{2}$ spin inversion symmetry (with operator $\prod_{j=1}^N \sigma^x_j$ being conserved). In our investigation, we focus on the even-$Z_{2}$ sector within $\sum_{i}^{} \expval{ \sigma_{i}^{z} } = 0 $ and resolve all the symmetries of the Hamiltonian~\cite{2016.AdvancesinPhysics}. The length of chain we consider here is up to $ N = 20 $, where the dimension of the Hilbert space of the considered sector is $ D = N!/[(N/2)!]^{2}/2 = 92378$.

\begin{figure}[tpb]
	\centering
	\includegraphics[width=0.99\linewidth]{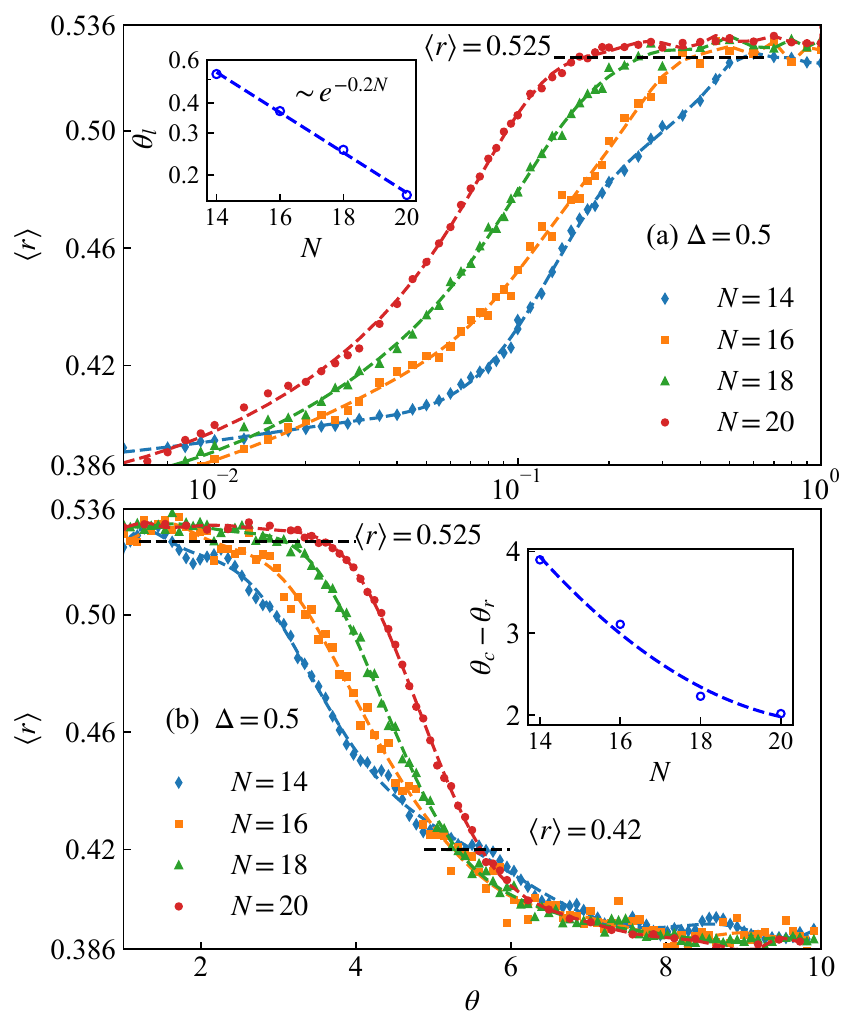}
	\caption{(Color online) The average ratio of consecutive level spacings, $ \langle r \rangle $, as the function of  inhomogeneity $ \theta $ (a) from $0.01$ to $1$ and (b) from $ 1 $ to $ 10 $. We take $ \Delta= 0.5$ and $ N = 14, 16, 18, 20  $. Considering the finite-size effects, the chaotic region (the yellow region in Fig.~\ref{Fig3}) is identified by $ \langle r \rangle > 0.525$. The inset of (a) shows the fitting of the left boundary $\theta_l$ of the chaotic region, where we have $ \langle r \rangle \simeq 0.525$. The $\theta_c \simeq 0.55$ in (b) is the collapsing point with $ \langle r \rangle \simeq 0.42$. The inset of (b) shows the fitting of the right boundary $\theta_r$ of the chaotic region. To ensure smoothness in the curve, each blue point for $ N = 14 $ is given by the average of $ 10 $ simulations within its closest range of two neighboring $ \theta $ values.}
	\label{fig:crossover_vs_size}
\end{figure}

\section{Integrable-chaotic-integrable crossovers driven by spatial inhomogeneity of interactions}
\label{sec:int-cha}

\subsection{Distribution of level spacing}

We first investigate the distribution of energy level spacing $P(s)$ with $s$ the spacing between neighboring unfolded levels. The spectral unfolding is performed so that the mean level spacing is unity~\cite{1981.RevModPhys.53.385, 2010.PhysRevE.81.036206}, allowing to extract short-range spectral correlations of many-body systems, such as $P(s)$, in a robust manner~\cite{PhysRevB.99.104205}. The distribution $P(s)$ exhibits distinct characteristics depending on whether the system is chaotic or integrable. 

For an integrable system, the eigenvalues are uncorrelated and crossings between energy levels are not prohibited, which leads to the Poissonian distribution of level spacings, i.e., $P_{\text{P}}(s) = \exp(-s)$~\cite{1991.PhysRevA.43.4237, 2013.PhysRevE.88.032913, 2018.NJPh.20k3039G}. In contrast, the energy levels in a quantum chaotic system are correlated, and the crossings are avoided as level repulsion emerges. Consequently, the level statistics follow the Wigner-Dyson distribution according to the random matrix theory~\cite{1998.PhR.299.189G}. In our model, which preserves the time-reversal invariance, the appropriate symmetry class is the Gaussian orthogonal ensemble (GOE), and the corresponding Wigner surmise reads $P_{\text{WD}}(s) = (\pi s/2)\exp(-\pi s^2/4)$.

The unfolding procedure mentioned above is performed by introducing a cumulative spectral function $N(E) = \sum_{n}^{} \Theta(E - E_{n})$, where $\Theta$ represents the unit step function. We fit $N(E)$ with the polynomials up to $12$-th order. For robustness, we consider $80 \%$ of the energy levels in the regions with high density of states.

The level spacing distributions $P(s)$, for several values of the slope $\theta$, are shown in Fig.~\ref{Fig2}, where we take $N=18$ and  $\Delta=1$. For the homogeneous case with $\theta=0$, $P(s)$ matches accurately the Poissonian distribution (depicted by the red dash line). As the inhomogeneity strength increases, $P(s)$ gradually changes and shows an excellent agreement with the Wigner-Dyson distribution (depicted by the blue solid line) at $\theta \sim 0.5$ [Figs. \ref{Fig2}(a-d)]. This indicates the presence of level repulsion and onset of applicability of random matrix statistics.
As $\theta$ continues to increase, $P(s)$ gradually changes back to  the Poissonian distribution [Fig. \ref{Fig2}(e-h)]. These results suggest that sufficiently strong inhomogeneity ($\theta \simeq 8$) drives the system from quantum chaos back to integrability.

The observed behavior can be readily understood. At $\theta=0$, the system \eqref{Hami} is integrable. The multiple conserved quantities break the considered $D$ dimensional sector of Hilbert space into smaller subspaces, giving rise to Poissonian level statistics. Introduction of inhomogeneous interactions, $\theta > 0$, breaks the integrability of the Hamiltonian, giving rise to GOE level statistics for $\theta$ of the order of unity. In contrast, for $\theta \gg 1$, the $z$-$z$ coupling term dominates in the Hamiltonian $\hat{H}_{\text{inho}}$ and the hopping term is not sufficiently strong to delocalize the eigenstates in the eigenbasis of $ \sigma_{i}^{z}$ operators, giving rise to Poissonian level statistics.

\subsection{Ratio of consecutive level spacing}

To understand better the system size dependence of the observed crossovers as well as to pin-point the roles of $\Delta $ and $ \theta $, we investigate the level spacing ratio~\cite{Oganesyan07, 2013.PhysRevLett.110.084101} that is defined as
\begin{eqnarray}\label{lsratio}
	r = \text{min} \{r_{n}, \frac{1}{r_{n}}\},\ \ \ \ \ r_{n}=\frac{E_{n+1}-E_{n}}{E_{n}-E_{n-1}},
\end{eqnarray}
where $\{ E_{n} \}$ are sorted eigenvalues of $\hat{H}_{\text{inho}}$.
This ratio serves as another important signature of quantum chaos, and it does not require the unfolding procedure. 

Its average value $ \langle r \rangle $, computed from all eigenenergies, is known to be approximately $ \langle r \rangle = r_{\text{GOE}} \simeq 0.5307 $ for GOE level statistics and $ \langle r \rangle = r_\text{PS} \simeq 0.3863 $ for Poisson level statistics, respectively.

We begin by studying the integrable to chaotic crossover observed at small values of $\theta$. To that end, we fix $\Delta=0.5$ and plot $ \langle r \rangle$ as function of $\theta$ for system sizes $14 \leq N \leq 20$, as shown in the top panel of Fig.~\ref{fig:crossover_vs_size}. At each $N$, we observe that the average level spacing ratio grows from $r_{\text{PS}}$ to $r_{\text{GOE}}$ with the increase of $\theta$. Notably, this crossover shifts towards smaller values of $\theta$ with increasing $N$, so that the $\theta_l$, i.e. the slope at which the level spacing ratio becomes close to the GOE value $\langle r \rangle = 0.525$, shifts exponentially with $N$ towards smaller values of $\theta$, see the inset in the top panel of Fig.~\ref{fig:crossover_vs_size}. This behavior indicates that in the large system size limit, $N\gg 1$, the system \eqref{Hami} possesses an integrable point at $\theta = 0$, and is quantum chaotic for $0<\theta<\theta_r$, where $\theta_r$ is the inhomogeneity at which the $z$-$z$ coupling term starts to dominate and the systems becomes integrable.

To probe the latter behavior, we plot $\langle r \rangle$ for larger values of $\theta$ and several system sizes $N$, see the bottom panel of Fig.~\ref{fig:crossover_vs_size}. The interval of $\theta$ in which $\langle r \rangle$ is close to GOE (say, bigger than $0.525$), extends towards larger and larger values of $\theta$ with increasing $N$. Notably, for $\theta > 5.5$, the $\langle r \rangle(\theta)$ curves for different $N$ approximately collapse on top of each other. This system size dependence of $\langle r \rangle$ is immediately reminiscent of phenomenology observed for \textit{XXZ} spin chain with disordered on-site magnetic field~\cite{Oganesyan07, Luitz15, Sierant20poly}. Understanding of implications of the numerical results for the fate of the disordered system in thermodynamic limit remains an outstanding challenge in the field of many-body localization (MBL)~\cite{Suntajs20, Sierant20thou, Abanin21,Kiefer20, Panda19}. Indeed, a crossover between  $\langle r \rangle = r_\text{GOE} $ and $\langle r \rangle = r_\text{PS}$ does not necessarily signify ergodicity breaking in the large $N$ limit~\cite{Sierant21constr}. One possible scenario for the \textit{XXZ} chain with inhomogeneous interactions consistent with our results is that there exists a finite $\theta_c$ (for instance $\theta_c \approx 5.5$) such that for $\theta < \theta_c$ the system becomes quantum chaotic, $\langle r \rangle \stackrel{ N \to \infty}{ \longrightarrow} r_\text{GOE} $, while for $\theta > \theta_c$ the system is integrable, $\langle r \rangle \stackrel{ N \to \infty}{ \longrightarrow} r_\text{PS}$. Based on the ED data we cannot, however, exclude other scenarios for the $N \to \infty$ limit. Nevertheless, the observed behavior $\langle r \rangle$ has significant implications for the dynamics of the  \textit{XXZ} chain with inhomogeneous interactions at finite times and system sizes, akin to Stark-MBL \cite{2019PNAS.116.9269V, 2019.PhysRevLett.122.040606,Chanda20, Yao21, 2022.PhysRevB.106.075107}, as we demonstrate in Sec.~\ref{sec:dyn}.

\begin{figure}[tbp]
	\centering
	\includegraphics[angle=0,width=0.99\linewidth]{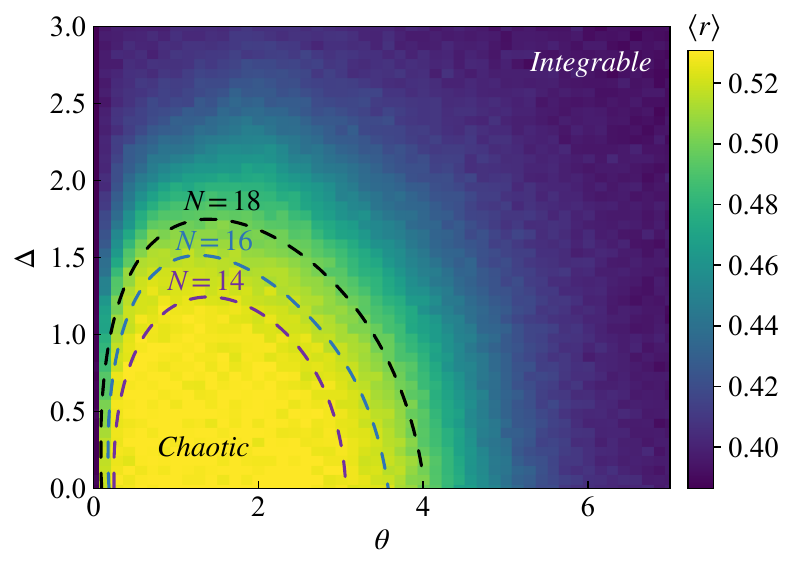}
	\caption{(Color online) The average ratio of consecutive level spacings $ \langle r \rangle $ for the inhomogeneous \textit{XXZ} chain [Eq. (\ref{Hami})], the system size $ N = 18 $ with open boundary condition, as a function of the inhomogeneity $ \theta $ and the average strength $ \Delta $. The black (blue, purple) dashed line is the contour of $ \langle r \rangle_{c} \simeq 0.52 $ for the system with $ N = 18(16, 14) $.}
	\label{Fig3}
\end{figure}

To understand the interplay of  $ \Delta $ and $ \theta $ we plot $ \langle r \rangle $ for $ N = 18 $ in Fig.~\ref{Fig3} on the $ \Delta$, $ \theta $ plane. Note that we only focus on the positive $ \theta $ and $ \Delta $ since the plot would be nearly centrosymmetric about the origin ($ \theta=0 $ and $ \Delta=0 $). For a negative $ \theta $ for instance, one can introduce a unitary reflection operation along the chain, which effectively transforms the system back to the positive $ \theta $ without altering the eigenenergies. For a negative $ \Delta $, one can apply $ \exp \{ \sum_{\nu}^{} i \frac{\pi}{2} \sigma^{z}_{2\nu} \} $, a $ \pi /2 $ rotation along the spin-$ z $ direction on the even sites. This operation changes the sign of $ J $ while preserving the interactions along $ z $ direction. Although this adjustment could introduce a minus sign to the eigenenergies, it does not impact the averaged level spacing ratio according to Eq.~(\ref{lsratio}). 

The system is integrable when $ \theta $ is small or large enough, as confirmed for various values of $\Delta$ in Fig.~\ref{Fig3}. In the former limit, the system reduces to the typical homogeneous \textit{XXZ} chain. In the latter limit, the system is dominated by the $z$-$z$ interactions.

When the inhomogeneity $ \theta $ is comparable to the hopping term, the system is chaotic, which is consistent with the suggestion that, in finite systems, quantum-chaotic properties usually emerge when there are no simplifying descriptions of the model that would emerge if one of the terms in the Hamiltonian is dominating~\cite{2019.PhysRevB.99.155130}. Notably, as $ \Delta $ increases, the range of $ \theta $ in which the system is chaotic shrinks, indicating the suppression of chaos and thermalization by the $z$-$z$ interactions. For a sufficiently large $ \Delta $, the system is integrable for all $ \theta $, giving another integrable limit. This behavior is analogous to Hilbert space fragmentation in clean systems \cite{Khemani20, Sala20} associated with strong presence of quasi-conservation laws due to strong interactions~\cite{Lan18, Li21, Korbmacher23} and observed experimentally in Hubbard model \cite{Scherg21, Kohlert23}.

\section{Eigenstate Thermalization}\label{sec:eth}

In the previous section, we investigated the signatures of quantum chaos, which is one of the consequences of thermalization in the inhomogeneous \textit{XXZ} chain. We will now investigate directly the eigenstate thermalization hypothesis (ETH) by exploring the statistics of the matrix elements of local operators in the eigenstates of \eqref{Hami}. The properties of matrix elements of the inhomogeneous interacting \textit{XXZ} chain in the integrable regime will be presented for comparison and to enhance our understanding of the latter regime~\cite{2019.PhysRevLett.123.240603, 2019.PhysRevE.100.062134}.

The ETH ansatz for the matrix element of an observable, denoted as $ O_{nm} = \mel*{n}{O}{m} $ in energy eigenstates (with $ \hat{H}\ket*{m} = E_{m}\ket*{m} $) can be written as 
\begin{eqnarray}\label{ETH}
	O_{nm}
		=
		O(\bar{E})\delta_{nm} + e^{-S( \bar{E})/2} f_{O}( \bar{E}, \omega) R_{nm},
\end{eqnarray}
with $\bar{E} = (E_{n} + E_{m})/2 $ and $ \omega = E_{m} - E_{n} $. Here, $ S( \bar{E}) $ denotes the thermodynamic entropy at the energy $ \bar{E} $ equal to the logarithm of the density of states \cite{Burke23}; $ O( \bar{E}) $ and $ f_{O}( \bar{E}, \omega) $ are smooth functions; $ R_{nm} $ is a Gaussian-distributed variable with zero mean and unit variance.

The first term in Eq. (\ref{ETH}) ensures that when the energy fluctuations in the initial state are sub-extensive, the equilibrated result can be described using statistical mechanical ensembles. The factor $ e^{-S(\bar{E})/2} $ in the second term suggests that the off-diagonal matrix elements decrease exponentially with system size. Up to random fluctuations, these elements are characterized by a smooth function $ f_{O}(\bar{E}, \omega) $~\cite{2013.PhysRevLett.111.050403, 2015.PhysRevE.91.012144, 2016.AdvancesinPhysics, 2017.PhysRevE.96.012157, 2019.PhysRevE.100.062134, 2019.PhysRevB.99.155130} that carry crucial information on the quantum thermalization and fluctuation dissipation relations~\cite{2012.PhysRevLett.109.247206, 2016.AdvancesinPhysics, 1999.JPhA.32.1163S, 2013.PhysRevLett.111.050403, 2019.PhysRevE.99.052139, 2020.PhysRevLett.125.050603, 2021.PhysRevB.103.235137, Serbyn17}. 
It is worth noting that $ R_{nm} $ is similar to the random matrices in the GOE. However, the higher-order correlations are not described by GOE or the random matrix theory~\cite{2019.PhysRevE.99.052139, 2019.PhysRevE.99.042139, 2019.PhysRevLett.122.220601, 2020.PhysRevE.102.042127, 2021.PhysRevE.104.034120, 2022.PhysRevLett.128.180601}. Here, our primary target is to probe the nature of the  distribution of matrix elements, while the higher-order statistical correlations remain beyond the scope of this work.

We consider the matrix elements of two operators, $ \hat{T}$ and $ \hat{Z} $. $ \hat{T} $ is the next-nearest-neighboring ``kinetic'' energy per site
\begin{align}\label{T_obs}
	\hat{T}
		=
		\frac{1}{N} \sum_{i = 1}^{N - 2} \left( \sigma_{i}^{x}\sigma_{i + 2}^{x} + \sigma_{i}^{y}\sigma_{i + 2}^{y} \right).
\end{align}
$\hat{Z}$ contains the nearest-neighboring  $ z $-$ z $ interactions with spatially-inhomogeneous coefficient, which is defined as
\begin{align}\label{Z_obs}
	\hat{Z}
		=
		\frac{1}{N} \sum_{i = 1}^{N - 1}\tilde{\Delta}_{i} \hat{\sigma}_{i}^{z} \hat{\sigma}_{i + 1}^{z}.
\end{align}
The coefficient $\tilde{\Delta}_{i}$ takes the same expression as Eq.~\eqref{inhomogeneity} with $ \Delta = 1 $ and $ \theta = 1 $. We expect that the specific choices of operators do not affect our main conclusions.

\subsection{Diagonal Matrix Elements}

\begin{figure}[tbp]
	\centering
	\includegraphics[angle=0,width=0.99\linewidth]{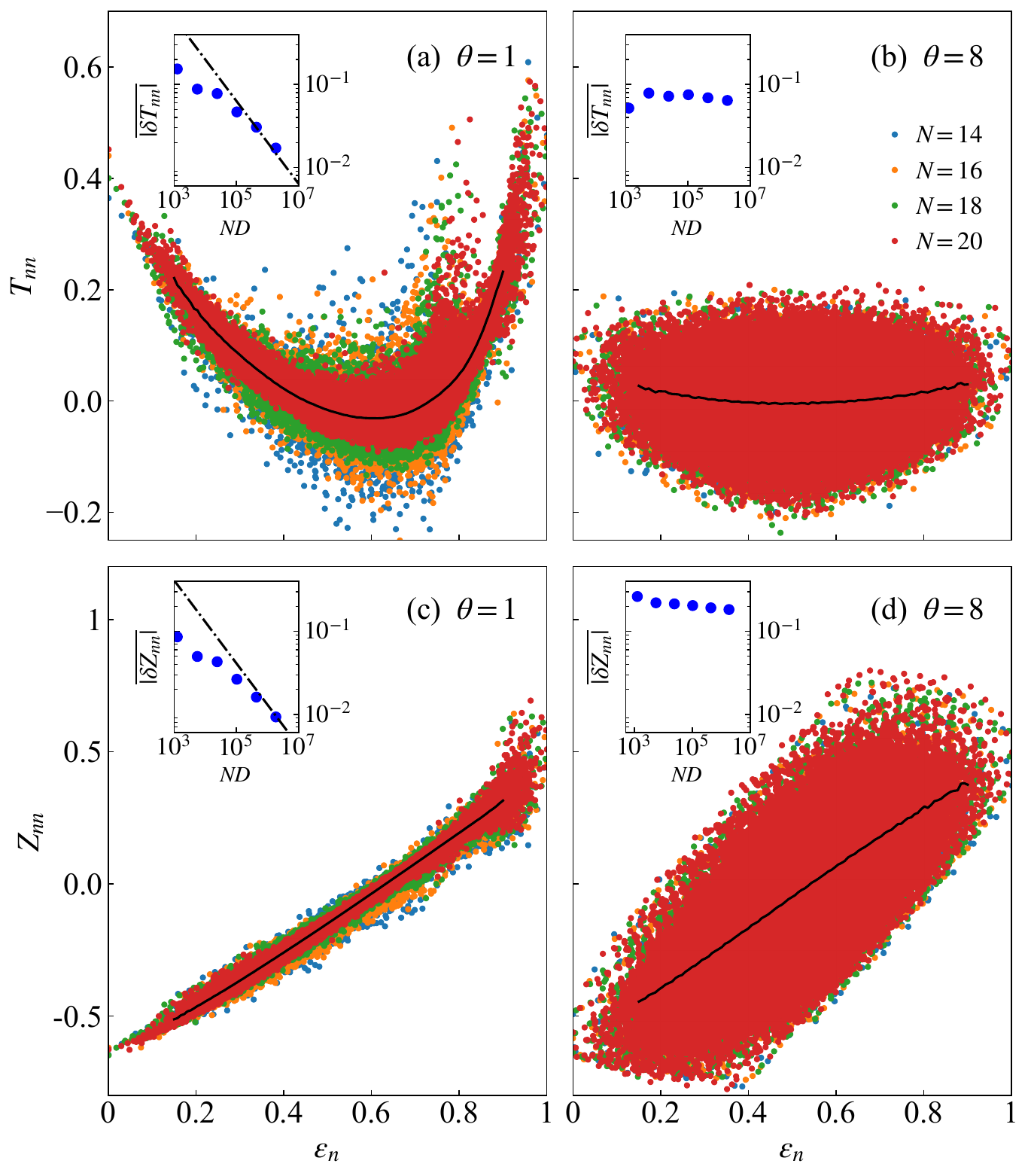}
	\caption{(Color online) Diagonal matrix elements of $ \hat{T} $ [(a), (b)] and $ \hat{Z} $ [(c), (d)] in the eigenstates of $ \hat{H}_{\text{Inho}} $ in the quantum chaotic regime ($ \Delta = 1, \theta = 1 $) [(a), (c)] and in the integrable regime ($ \Delta = 1, \theta = 8 $) [(b), (d)]. The black lines represent the microcanonical averages (within windows of $ \delta \varepsilon_{n} = 0.01 $) for the largest chain size ($ N = 20 $). The insets exhibit the scaling behavior of $ \overline{| \delta O_{nn} |} = \overline{| O_{nn} - O_{n + 1n + 1} | }$ (for $O=T,Z$) with respect to $ ND $ (the dashed lines indicate the line $ \propto (ND)^{-1/2} $). The averaging is performed over the central $ 20\% $ of the eigenstates in Hamiltonian described by Eq.~\ref{Hami} with $ N $ from $ 10 $ to $ 20 $.}
	\label{Fig4} 
\end{figure}

Figs.~\ref{Fig4}(a-b) and (c-d) show the diagonal matrix elements $ T_{nn} = \mel*{n}{\hat{T}}{n} $ and $ Z_{nn} = \mel*{n}{\hat{Z}}{n} $, respectively. These results are obtained in the even-$ Z_{2} $ sector within the $\sum_{i}^{} \langle \sigma_{i}^{z} \rangle = 0$ sector (see Sec. \ref{sec:Model}). The matrix elements are plotted as functions of the energy density, defined as $ \varepsilon_{n} = (E_{n}-E_{\text{min}})/(E_{\text{max}} - E_{\text{min}}) $, where $ E_{n} $ represents the $ n $-th  eigenvalue of \eqref{Hami}, while $E_{\text{max}}$ and $E_{\text{min}}$ are the highest and lowest eigenvalues.

In the quantum chaotic regime, at $ \Delta =  \theta = 1 $, we observe a decrease of the support of both $ T_{nn} $ and $ Z_{nn} $ around the $ \varepsilon_{n} $ away from the edges of spectrum as the system size $ N $ increases [see Figs.~\ref{Fig4}(a) and (c)]. Meanwhile, an exponential decay of the average strength of the eigenstate-to-eigenstate fluctuations, $ \overline{ | \delta T_{nn} | } $~\cite{2014.PhysRevE.90.052105, Luitz16long, 2016.PhysRevE.93.032104, 2019.PhysRevB.99.155130,2019.PhysRevE.100.062134, 2020.PhysRevLett.125.070605}, is shown in the insets of Fig.~\ref{Fig4} (a). Similar observations are shown in the inset of Fig.~\ref{Fig4} (c) for $ \delta Z_{nn} $. Thus, the diagonal matrix elements, up to the exponentially-decaying fluctuations, follow a smooth function of energy which agrees with the microcanonical predictions of these observables as shown by the black solid lines in the main panels of (a) and (c). These results are also consistent with the ETH, regardless of whether the observable is homogeneous ($ \hat{T} $) or not ($ \hat{Z} $). 

In the integrable regime, at [$ \Delta = 1, \theta = 8 $; see Figs.~\ref{Fig4}(b) and (d)], we observe that the support of distributions of $ T_{nn}$ and $Z_{nn}$ remains wide and does not shrink with the system size $N$. The insets also show that the eigenstate-to-eigenstate fluctuations exhibit a very slow or even no decay as $N$ increases. This wide and non-shrinking support indicates the absence of diagonal eigenstate thermalization for these observables in the integrable inhomogeneous \textit{XXZ} chain. This observation is consistent with Poissonian level statistics of the system and shows that the model violates the ETH.

\subsection{Off-diagonal Matrix Elements}
\begin{figure}[tbp]
	\centering
	\includegraphics[angle=0,width=0.98\linewidth]{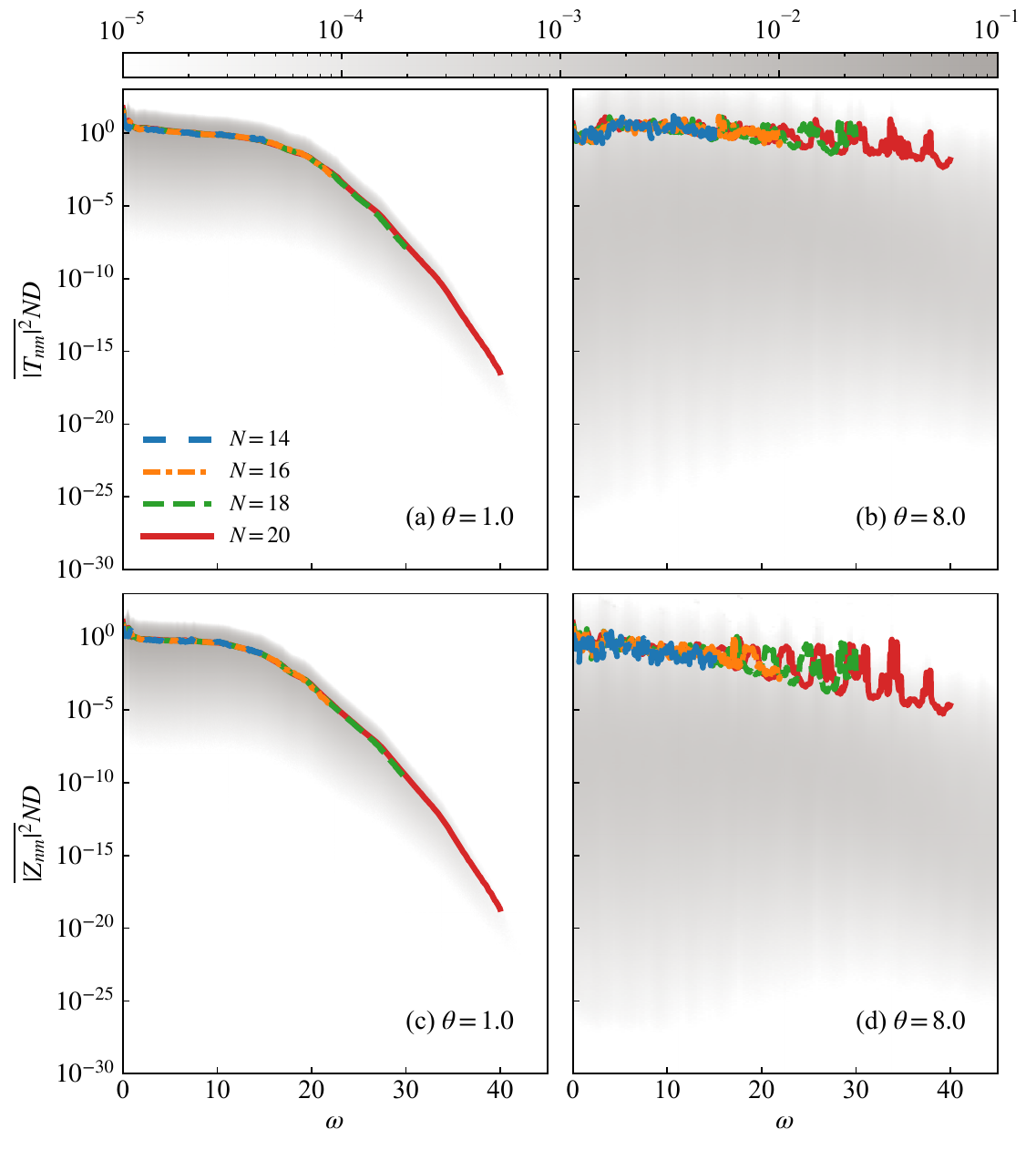}
	\caption{(Color online) Normalized 2D histograms of the off-diagonal matrix elements and the corresponding coarse-grained average of $ \hat{T} $ [(a), (b)] and $ \hat{Z} $ [(c), (d)] for different chain sizes as a function of $ \omega $. Figs. [(a), (c)] correspond to the nonintegrable point of the inhomogeneous \textit{XXZ} model with parameters $ \Delta = 1, \theta = 1 $, while Figs. [(b), (d)] present results for the integrable point with $ \Delta = 1, \theta = 8 $. The matrix elements were computed within a small energy window around $ \bar{E} \simeq 0 $, the center of the spectrum, with a width of $0.075 \varepsilon $, where $ \varepsilon = E_{\text{max}} - E_{\text{min}} $. The coarse-grained averages in $ \omega $ were calculated using windows of width $ \delta \omega = 0.1 $. }
	\label{Fig5} 
\end{figure}

We now turn to the off-diagonal matrix elements of observables $ T_{nm} \equiv \mel*{n}{\hat{T}}{m} $ and $ Z_{nm} \equiv \mel*{n}{\hat{Z}}{m}$ in the energy eigenbasis and focus on the second term of the ETH ansatz [Eq.~(\ref{ETH})]. Since the Hilbert-Schmidt norms of $ \hat{T} $ and $ \hat{Z} $ scale as $ 1/ \sqrt{N} $, the ETH ansatz for the off-diagonal matrix elements should be modified to~\cite{2019.PhysRevE.100.062134, 2020.PhysRevLett.124.040603, 2020.PhysRevLett.125.070605}
\begin{align}\label{eq:ETHanz}
	O_{nm}
		=
		\frac{e^{-S( \bar{E})/2}}{\sqrt{N}}f_{O}( \bar{E}, \omega)R_{nm}.
\end{align}
We focus on the region with $ \bar{E}\simeq 0 $, which corresponds to the ``infinite-temperature'' region as $ S( \bar{E}) \simeq \ln D$. 

Fig.~\ref{Fig5} illustrates the distribution of the off-diagonal matrix elements $|T_{nm}|^{2}$ and $|Z_{nm}|^{2}$ using normalized 2D histograms, along with the coarse-grained averages $\overline{|T_{nm}|^{2}}$ and $\overline{|Z_{nm}|^{2}}$ as a function of $\omega$. These averages correspond to the variances of the off-diagonal matrix elements as $ \overline{T_{nm}} = \overline{Z_{nm}} = 0 $. In the chaotic regime ($ \Delta = \theta = 1$), the variances change smoothly with $\omega$~\cite{2019.PhysRevE.100.062134,2020.PhysRevLett.125.070605} [see Figs.~\ref{Fig5}(a) and (c)]. Both the homogeneous observable $\hat{T}$ and inhomogeneous $\hat{Z}$ exhibit a slow decay at the intermediate values of $ \omega $ and a relative rapid decay at larger $ \omega $. Nearly perfect collapses of variances for different system sizes are demonstrated, indicating that the variances of the off-diagonal matrix elements satisfy $\overline{|O_{nm}|^{2}} \propto (N\mathcal{D})^{-1}$ for $O=T,Z$. These results are in full agreement with the ETH~\cite{2016.AdvancesinPhysics, 2019.PhysRevE.100.062134, 2020.PhysRevLett.125.070605, 2020.PhysRevE.102.042127}. 

For our inhomogeneous \textit{XXZ} chain in the integrable regime, Figs.~\ref{Fig5}(b) and (d) show the distributions of the off-diagonal matrix elements and their coarse-grained averages. Remarkable differences are observed in comparison to the chaotic region. First, the overall dispersion is much larger than that at the quantum-chaotic points, consistent with the previous results~\cite{2019.PhysRevE.100.062134, 2020.PhysRevLett.125.070605}. Second, the coarse-grained averages ($\overline{|O_{nm}|^{2}}$) at an integrable point do not evidently show the trend to drop as $ \omega $ increases, and change non-smoothly with $ \omega $. Thirdly, we find no data collapsing for different system sizes $N$. These differences between integrability and chaos in our inhomogeneous system are inconsistent with the ETH. Note that for the conventional homogeneous integrable \textit{XXZ} model, the off-diagonal matrix element variances behave smoothly and the data collapse for different system sizes, which are similar to its chaotic counterpart~\cite{2019.PhysRevE.100.062134, 2020.PhysRevLett.125.070605, 2020.PhysRevE.102.062113, 2020.PhysRevB.102.075127}.

\begin{figure}[tbp]
	\centering
	\includegraphics[angle=0,width=\linewidth]{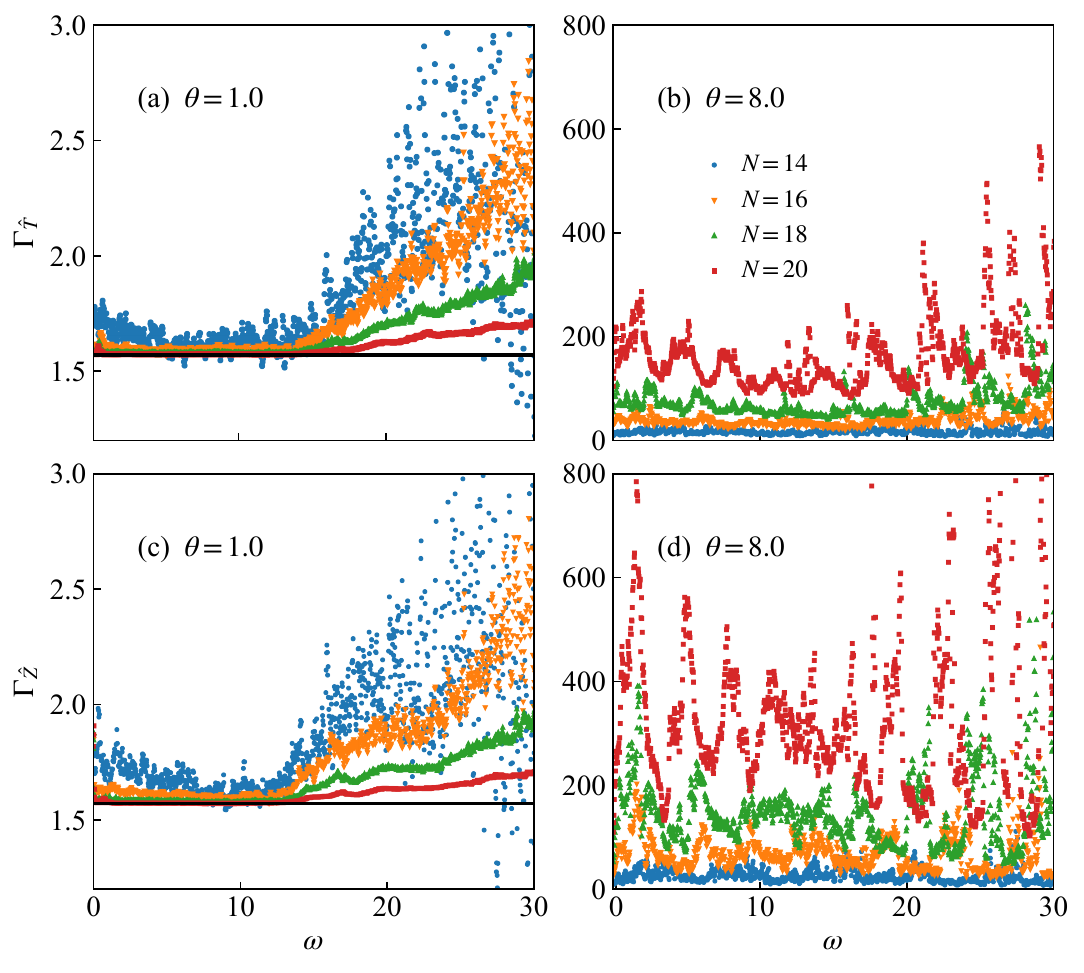}
	\caption{(Color online) $ \Gamma_{\hat{O}}(\omega) $ [see Eq.~(\ref{ratio})] for $ \hat{T} $ [(a), (b)] and for $ \hat{Z} $ [(c), (d)] in the nonintegrable inhomogeneous \textit{XXZ} chain with $ \Delta = 1, \theta = 1 $ [(a), (c)], and the integrable one with $ \Delta = 1, \theta = 8 $ [(b), (d)]. In (a) and (c), the horizontal line denotes $ \pi/2 $. The matrix elements were computed within a narrow energy window of width $0.075 \varepsilon$, where $ \varepsilon = E_{\text{max}} - E_{\text{min}} $. The coarse-grained averages were calculated using a window size of $ \delta \omega = 0.1 $.}
	\label{Fig6}
\end{figure}

To test the normality of the distribution of the off-diagonal matrix elements, we evaluate the frequency-dependent ratio
\begin{align}
\label{ratio}
	\Gamma_{ \hat{O}}(\omega)
		= \overline{| O_{nm} |^{2}} / \overline{| O_{nm} | }^{2}.
\end{align}
This ratio equals to $ \pi / 2 $ when $ O_{nm} $ obeys the Gaussian distribution with zero mean~\cite{2019.PhysRevE.100.062134}. Fig.~\ref{Fig6} illustrates $ \Gamma_{\hat{T}}(\omega) $ and $ \Gamma_{\hat{Z}}(\omega) $ for the inhomogeneous \textit{XXZ} chain. In the chaotic regime [$\theta=1$, see Figs.~\ref{Fig6} (a) and (c)], the ratios are close to $ \pi/2 $ at the intermediate frequencies, and deviate from $ \pi/2 $ for lower and higher frequencies. The deviations are caused by the finite-size effects, and are suppressed by increasing the system size $ N $. The presented results indicate that the distribution of the off-diagonal matrix elements $ T_{nm} $ and $ Z_{nm} $ is the Gaussian distribution in a wide frequency range for sufficiently large $N$, consistently with the prediction of ETH. 

In contrast, the behaviors of the ratios for the integrable inhomogeneous chain ($\theta=8$) are strongly affected by the system size $ N $ [see Figs. \ref{Fig6}(b) and (d)]. The value deviates further from $ \pi/2 $ as $ N $ increases, indicating that the distribution of off-diagonal matrix elements $ T_{nm} $ and $ Z_{nm} $ is not Gaussian.

\begin{figure}[tpb]
	\centering
	\includegraphics[width=0.99\linewidth]{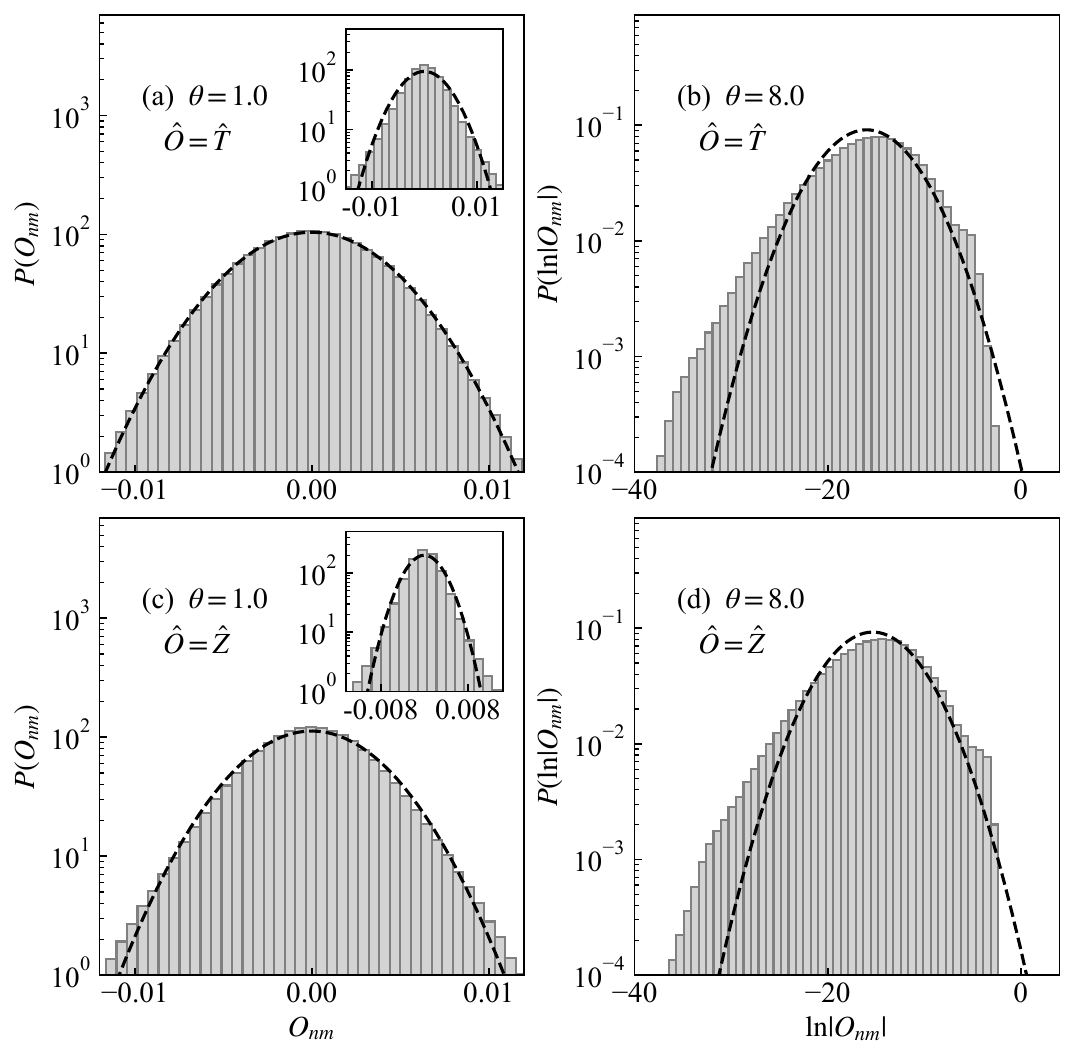}
	\caption{
		Probability distributions $ P(O_{nm}) $ for observables $ \hat{T} $ and $ \hat{Z} $ [(a), (c)] for quantum chaotic Hamiltonian with $ \theta = 1, \Delta = 0 $ ($\theta = 1, \Delta = 1 $) in the main panel (inset), along with Gaussian distributions (dash lines) 
		with the same mean and variance. We consider the eigen pairs around $ \bar{E} \approx 0 $ within a narrow energy window of width $0.075 \varepsilon$, where $ \varepsilon = E_{\text{max}} - E_{\text{min}} $, and $ \omega < 0.1 $. 
		The probability distributions $ P(\ln | O_{nm} | ) $, of the matrix elements of $ \hat{T} $ and $ \hat{Z} $ respectively, along with the log-normal distributions (dash line), are shown in figure (b) and (d) for the integrable Hamiltonian with $ \theta = 8 $ and $ \Delta = 0 $.}
	\label{Fig7}
\end{figure}

The distributions of off-diagonal matrix elements near zero frequency are depicted in Fig.~\ref{Fig7}. 
In the chaotic regime ($ \theta = 1 $, panels (a) and (c)), the reliability of ETH is confirmed by the remarkable agreement with the Gaussian distribution. 
Additionally, comparing the case of  $ \Delta = 0 $ with $\Delta=1$(see the insets), we find that the agreement with Gaussian distribution is better for $ \Delta = 0 $. Nevertheless, both distributions tend towards a Gaussian distribution as the system size increases, as supported by Fig.~\ref{Fig6}. In the integrable regime ($ \theta = 8 $, panels (b) and (d)), the $\ln|O_{nm}|$ distribution has a skewed normal-like shapes as typically observed in integrable \textit{XXZ} model~\cite{2019.PhysRevE.100.062134}, even though the overall frequency behavior depicted in Fig.~\ref{Fig5} is different.

\section{Dynamics}
\label{sec:dyn}

The results shown above concern the matrix elements of local operators, which demonstrate the ETH behavior of the inhomogeneous \textit{XXZ} spin chain. In this section, we explore the thermalization and ergodicity breaking in the system by investigating time evolution of the entanglement entropy and the survival probability, of which both are relevant to the quench experiments with quantum simulators.


\subsection{Entanglement Entropy}

\begin{figure}[tpb]
	\centering
	\includegraphics[width=0.95\linewidth]{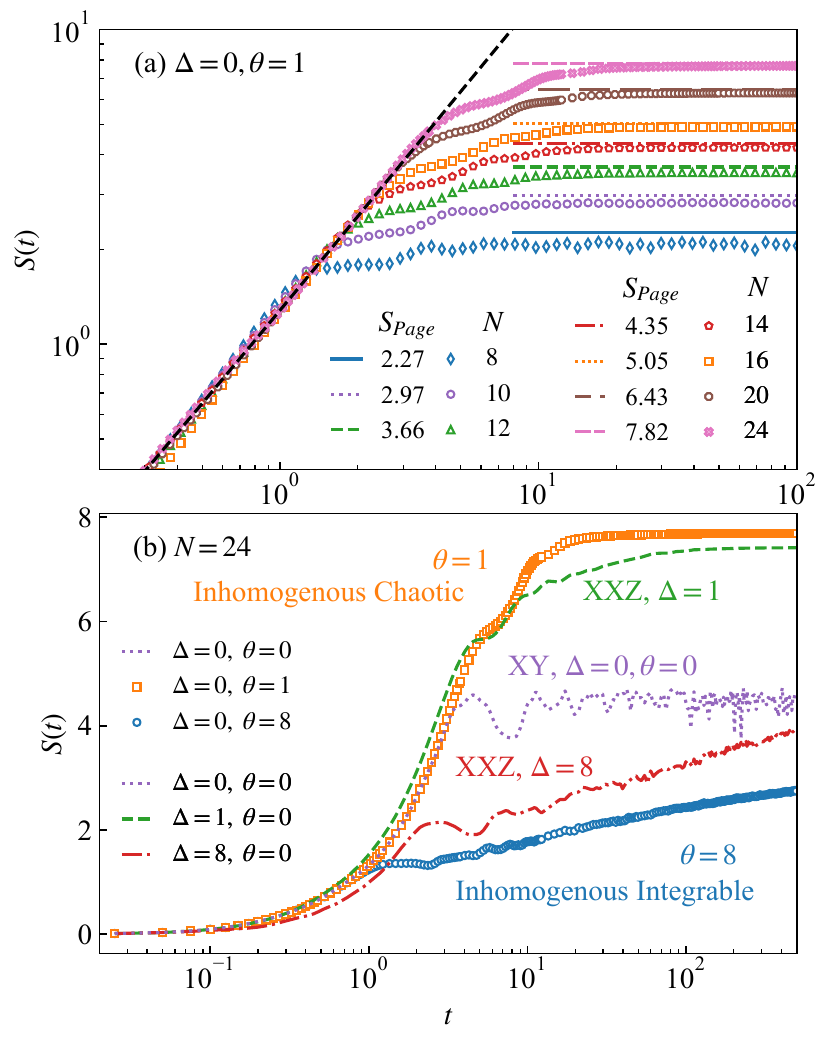}
	\caption{(Color online) (a) The evolution of entanglement entropy $ S(t) $ in the chaotic region ($\Delta=0$ and $\theta=1$) with different system sizes $N$. The black dashed line represents the fitting at short times ($t~O(10^0)$). The asymptotic saturation values of $S(t)$ for large $t$ are indicated by the Page values (see the horizontal dashed lines). (b) The $ S(t) $ in the integrable system with various $\Delta$'s and $\theta$'s. See the analyses in the main text. Note each data point in this figure is the average over the simulations from about $10^3$ distinct initial states.}
	\label{entropy}
\end{figure}

We first investigate the time evolution of the entanglement entropy for a bipartition of the chain into two halves. The entanglement entropy at the time $t$ is defined as
\begin{align}
	S(t) 
		=
		- \operatorname{Tr}\left[ \hat{\rho}_{\text{R}}(t) \ln \hat{\rho}_{\text{R}}(t) \right],
\end{align}
where $ \rho_{\text{R}}(t) $ represents the reduced density matrix at the time $ t $ after tracing over the degrees of freedom residing in one of the halves of the chain. Here, as the initial states, we consider the product states, where the spin on each site is drawn randomly to be oriented either up or down in the \textit{z} direction.

The behavior of $ S(t) $ for different choices of $ \Delta$ and $\theta$ is illustrated in Fig.~\ref{entropy}. The results are averaged over $ 10^3 $ time evolutions with different initial product states. In (a), we show the $ S(t) $ in the chaotic region ($ \Delta = 0$ and $\theta = 1 $) for various system sizes ranging from $ N = 8 $ to $ 16 $. There is an early-time ballistic linear growth of $ S(t) $ that persist to longer times with increasing $ N $~\cite{2013.PhysRevLett.111.127205}. Subsequently, the growth of $ S(t) $ eventually saturates at the Page value $ S = N/2 \ln2 - 1/2 $~\cite{1993.PhysRevLett.71.1291} up to an $O(1)$ correction associated with the symmetry of $\hat{H}_{\text{inho}}$ \cite{Vidmar17,2019.NuPhB.938.594H}. These results indicate the ergodic dynamics, implying the occurrence of thermalization.

In Fig.~\ref{entropy}(b), we compare the ergodic dynamics observed for $\Delta=0$, $\theta=1$ with non-ergodic dynamics that arise at large $\theta$. In the integrable regime, at $ \Delta = 0, \theta = 8 $, the entanglement entropy exhibits logarithmic growth at long-times, $S(t) \sim \ln(t)$. This logarithmic behavior of $S(t)$ in our inhomogeneous integrable \textit{XXZ} spin chain resembles the observations in the MBL systems, where $S(t)$ grows logarithmically with time~\cite{Chiara06, Znidaric08, Bardarson12, Serbyn13a, iemini2016signatures}. 
Interestingly, increasing either $ \theta $ in the inhomogeneous \textit{XXZ} chain or $ \Delta $ in the homogeneous chain would have a similar effect which, in the case of MBL systems, stems from the presence of localized integrals of motion~\cite{Serbyn13b, Huse14, Ros15, Mierzejewski18}. This observation suggests potential connections between the phenomenology of MBL and the dynamics of systems with shattered Hilbert space due to significant \textit{z-z} interactions.

\subsection{Survival Probability}
\begin{figure}[tpb]
	\centering
	\includegraphics[width=\linewidth]{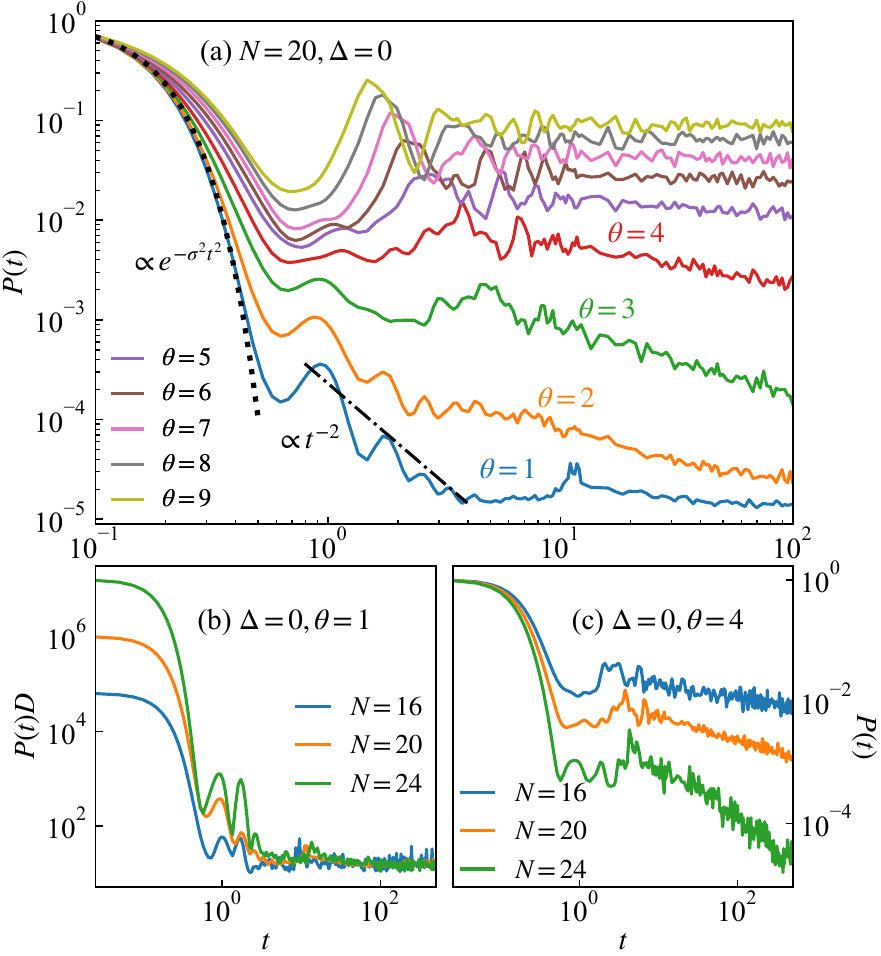}
	\caption{(Color online) (a) The decay of the survival probability $P(t)$ for various $ \theta $ with $ \Delta = 0 $ and $ N = 20$. The dotted line illustrates a Gaussian decay with $ \sigma $ (the variance of the LDOS with $ \theta = 1 $). The dash-dotted line is proportional to $ t^{-2} $. Each curve is given by the average over $10^3$ different initial states. (b) Survival probability multiplied by Hilbert space dimension, $P(t) D$,  as function of time in the ETH regime, $\Delta=0$, $\theta =1$. (c) Persistent decay of the survival probability $P(t)$ at the crossover between ETH and integrable regions. }
	\label{survival}
\end{figure}

To further probe the dynamics of the \textit{XXZ} spin chain with inhomogeneous interactions, and to highlight the memory effects occurring in the integrable regime, we consider the survival probability of the initial state is defined as
\begin{align}
	P(t)
		=
		| \langle \psi(0) |e^{-i \hat{H}_{\text{inho}} t}| \psi(0) \rangle|^{2},
\end{align}
where $\psi(0)$ is the initial state and $e^{-i\hat{H}_{\text{inho}} t}$ is the time evolution operator for our system.
Alternatively, the survival probability can also be expressed as the norm square of the Fourier transformation of the local density of state (LDOS), obeying
\begin{align}
	P(t)
		=
		\left| \int \mathrm{d}E \, \rho(E) e^{-iEt} \right|^{2},
\end{align}
where the LDOS is given by $ \rho(E) = \sum_{n} | C_{n} |^{2} \delta(E - E_{n}) $ (with $ C_{n} = \langle E_{n} | \psi(0) \rangle $ and $ \ket*{ E_{n} } $ the $n$-th eigenstate of $\hat{H}_{\text{inho}}$). Fig.~\ref{survival} shows the survival probability $P(t)$ averaged over  $ 10^3 $ initial random product states polarized in $z$ direction for different values of $ \theta $ ranging from $ 1 $ to $ 9 $, with fixed \(\Delta = 0\).

In the chaotic region (see, as illustrated by the results for $ \theta = 1\sim 3 $ in Fig.~\ref{survival}(a)), the decay is Gaussian for the early time~\cite{2001.PhysRevE.64.026124,2006.PhLA.350.355I,2014.PhysRevA.89.043620,2016.PhysRevA.94.041603}. This suggests that the LDOS is Gaussian-shaped, i.e., the coefficients $C_n$ do not add any specific structure on top of the Gaussian density of states at large energy scales. Subsequently, the decay of $P(t)$ changes to a power-law, and eventually  saturates at a value that is exponentially small in the systems size $N$~\cite{Schiulaz19, Lezama21}, as shown in Fig.~\ref{survival}(b). This behavior signals occurrence of thermalization in the system.

With increasing inhomogeneity strength $\theta$, we observe a significant slow down of the decay of the survival probability. Upon entering into the integrable, non-ergodic region, for $ \theta > 4 $, we observe only a residual power-law decay of $P(t)$ with the exponent that decreases rapidly with $\theta$. We have checked that the survival probability at large times is decreasing slower than exponentially quickly with system size in this regime (data not shown). While our results in the integrable region of the inhomogeneous \textit{XXZ} spin chain demonstrate the memory of the initial state at the considered system sizes $N$ and time scales $t$, our numerical data are insufficient to decide whether this behavior persists in the asymptotic limit of large $t$ and $N$. The decay of survival probability is slow, but the decreasing of $P(t)$ tends to be faster with increasing $N$ even at the largest values of $\theta$ considered here. This behavior is especially well pronounced in the region intermediate between the ETH and integrable regimes, as illustrated in Fig.~\ref{survival}(c).

The uncovered features of time evolution in \textit{XXZ} spin chain with inhomogeneous interactions are similar to the behavior of strongly disordered many-body systems~\cite{Luitz16slow, Bera17} ultimately preventing us from distinguishing a regime of very slow thermalization from ergodicity breaking phenomenon in the asymptotic limit $t \to \infty$, $N \to \infty$~\cite{Sierant22obs}.

\section{Experimental implementation}
\label{sec:exp}

The experimental realization of the \textit{XXZ} model presented in Eq.~\eqref{Hami} can benefit from recent advances in atomic systems and waveguide QED. In particular, the spin degree of freedom is mapped to two metastable states of the atoms, whose position can be optically controlled with the use of optical lattices or atomic tweezers. Spin-exchange naturally appears when atomic dipolar interactions are strong enough to compete with the finite lifetime ($\tau$) of their internal levels, $(J\tau\gg1)$. This is the case of magnetic atoms in short optical lattices, where single atom addressing becomes challenging due to the diffraction limit, and new strategies are being put into place~\cite{schollMicrowave2022}. Another emergent platform consists of atoms coupled to waveguides, where spin exchange can be mediated by exponentially localized photons emitted into the fibre, and additional control fields can be used to engineer the desired \textit{XXZ} interactions~\cite{hungQuantum2016}. In this open-quantum system, other additional terms not conserving the number of excitations would enter as well, though~\cite{tabaresVariational2023}. 

An alternative consists of using Rydberg atoms, where the strength of dipolar interactions scales with the quantum number $n$ of the valence electron as $n^{4}$, leading to strong forces even for typical atomic separations of $r \sim 1 \mu$m. While the resulting spin-exchange terms are of the form \textit{XX}, the \textit{XXZ} Hamiltonian can be engineered in a Floquet manner by appropriately rotating the spin axis at regular time intervals, as it has been experimentally realized in~\cite{PRXQuantum.3.020303}. Single-atom addressing can then be used to modulate the desired linear tilt $\Delta_i$ in the $\sigma_i^z \sigma_{i+1}^z$ term of each atom in the array in the regime $0\leq \Delta \pm \theta \leq 2J$ where both the chaotic and integrable regimes can be accessed. Using randomized measurements one can extract arbitrary observables~\cite{notarnicolaRandomized2023}, including the entanglement entropy of the chain~\cite{bluvsteinQuantum2022}. Following this approach, dipolar interactions decay polynomially as $r^{-3}$, and the role of next-nearest neighbor interactions (which are one order of magnitude weaker than the nearest-neighbor ones) will be the subject of future work.

\section{Summary and perspective}
\label{sec:sum}



Our work extends the explorations of the interplay of ETH and quantum chaos with ergodicity breaking to the systems with linearly inhomogeneous interactions. We demonstrate that insertion of a suitable inhomogeneity of the \textit{z}-\textit{z} interactions leads to the onset of quantum chaos in spin-$ 1/2 $ \textit{XXZ} chain, while a sufficiently large inhomogeneity restores the integrability of the system. While our results hold for a clean system, similar phenomenology can be found also in systems with disordered interactions, see~\cite{Sierant17ran, Bar16, Li17, Sierant18bos}.

To support our conclusions, we probe level statistics of the inhomogeneous \textit{XXZ} spin chain and study statistical properties of matrix elements of local observables in eigenstates of the system. In the quantum chaotic regime, the support and average eigenstate-to-eigenstate fluctuations of the diagonal elements vanish exponentially with the system size. Furthermore, the off-diagonal elements follow a Gaussian distribution and their variances exhibit a well-defined smooth function $ | f_{O}( \bar{E}\simeq 0, \omega) |^{2} $ with respect to the frequency $ \omega $. These results are fully consistent with the ETH. In contrast, the system exhibits essentially different behavior in the integrable regions with strong inhomogeneity. The variances of off-diagonal matrix elements are not anymore the smooth function of energy. Notably, the observed behavior is also different from other integrable models such as the homogeneous \textit{XXZ} chain

We also investigate the dynamics of the entanglement entropy and survival probability, both of which can be probed in quench experiments with quantum simulators. In the chaotic region, we find a ballistic spreading of entanglement entropy and an abrupt decay of the survival probability. These results indicate the ergodic dynamics, implying the occurrence of thermalization. Conversely, in the integrable region with the large inhomogeneity, entropy exhibits logarithmic growth, and the survival probability remains significant even at longer times, indicating the presence of the memory in the system. This closely resembles the observations for strongly disordered MBL phase and stems from Hilbert space shattering due to strong interactions in the system. 

\begin{acknowledgments}
	This work was supported in part by the National Key Research and Development Program of China (Grant No. 2021YFA1402001), NSFC (Grant No. 12004266 and No. 11834014, and No. 12074027 and No.12375007), Beijing Natural Science Foundation (Grant No. 1232025). Academy for Multidisciplinary Studies, Capital Normal University. J.A, M.L. and P.S. acknowledge support from: ERC AdG NOQIA; MICIN/AEI (PGC2018-0910.13039/501100011033,  CEX2019-000910-S/10.13039/501100011033, Plan National FIDEUA PID2019-106901GB-I00, Plan National STAMEENA PID2022-139099NB-I00 project funded by the MICIN/AEI/10.13039/501100011033/FEDER, EU, FPI); MICIIN with funding from European Union NextGenerationEU (PRTR-C17.I1): QUANTERA MAQS PCI2019-111828-2); MCIN/AEI/10.13039/501100011033 and by the “European Union NextGeneration EU/PRTR"  QUANTERA DYNAMITE PCI2022-132919 (QuantERA II Programme co-funded by European Union’s Horizon 2020 programme under Grant Agreement No 101017733), Ministry of Economic Affairs and Digital Transformation of the Spanish Government through the QUANTUM ENIA project call – Quantum Spain project, and by the European Union through the Recovery, Transformation and Resilience Plan – NextGenerationEU within the framework of the Digital Spain 2026 Agenda.Fundació Cellex; Fundació Mir-Puig; Generalitat de Catalunya (European Social Fund FEDER and CERCA program, AGAUR Grant No. 2021 SGR 01452, QuantumCAT \ U16-011424, co-funded by ERDF Operational Program of Catalonia 2014-2020); Barcelona Supercomputing Center MareNostrum (FI-2023-1-0013); EU Quantum Flagship (PASQuanS2.1, 101113690); EU Horizon 2020 FET-OPEN OPTOlogic (Grant No 899794); EU Horizon Europe Program (Grant Agreement 101080086 — NeQST), ICFO Internal “QuantumGaudi” project; European Union’s Horizon 2020 program under the Marie Sklodowska-Curie grant agreement No 847648;  “La Caixa” Junior Leaders fellowships, La Caixa” Foundation (ID 100010434): LCF/BQ/PR23/11980043. Views and opinions expressed are, however, those of the author(s) only and do not necessarily reflect those of the European Union, European Commission, European Climate, Infrastructure and Environment Executive Agency (CINEA), or any other granting authority.  Neither the European Union nor any granting authority can be held responsible for them.
\end{acknowledgments}

\bibliography{Primary_manuscript}

\begin{thebibliography}{141}%
\makeatletter
\providecommand \@ifxundefined [1]{%
 \@ifx{#1\undefined}
}%
\providecommand \@ifnum [1]{%
 \ifnum #1\expandafter \@firstoftwo
 \else \expandafter \@secondoftwo
 \fi
}%
\providecommand \@ifx [1]{%
 \ifx #1\expandafter \@firstoftwo
 \else \expandafter \@secondoftwo
 \fi
}%
\providecommand \natexlab [1]{#1}%
\providecommand \enquote  [1]{``#1''}%
\providecommand \bibnamefont  [1]{#1}%
\providecommand \bibfnamefont [1]{#1}%
\providecommand \citenamefont [1]{#1}%
\providecommand \href@noop [0]{\@secondoftwo}%
\providecommand \href [0]{\begingroup \@sanitize@url \@href}%
\providecommand \@href[1]{\@@startlink{#1}\@@href}%
\providecommand \@@href[1]{\endgroup#1\@@endlink}%
\providecommand \@sanitize@url [0]{\catcode `\\12\catcode `\$12\catcode `\&12\catcode `\#12\catcode `\^12\catcode `\_12\catcode `\%12\relax}%
\providecommand \@@startlink[1]{}%
\providecommand \@@endlink[0]{}%
\providecommand \url  [0]{\begingroup\@sanitize@url \@url }%
\providecommand \@url [1]{\endgroup\@href {#1}{\urlprefix }}%
\providecommand \urlprefix  [0]{URL }%
\providecommand \Eprint [0]{\href }%
\providecommand \doibase [0]{http://dx.doi.org/}%
\providecommand \selectlanguage [0]{\@gobble}%
\providecommand \bibinfo  [0]{\@secondoftwo}%
\providecommand \bibfield  [0]{\@secondoftwo}%
\providecommand \translation [1]{[#1]}%
\providecommand \BibitemOpen [0]{}%
\providecommand \bibitemStop [0]{}%
\providecommand \bibitemNoStop [0]{.\EOS\space}%
\providecommand \EOS [0]{\spacefactor3000\relax}%
\providecommand \BibitemShut  [1]{\csname bibitem#1\endcsname}%
\let\auto@bib@innerbib\@empty
\bibitem [{\citenamefont {{Neumann}}(1929)}]{1929.ZPhy.57.30N}%
  \BibitemOpen
  \bibfield  {author} {\bibinfo {author} {\bibfnamefont {J.~V.}\ \bibnamefont {{Neumann}}},\ }\bibfield  {title} {\enquote {\bibinfo {title} {{Beweis des Ergodensatzes und des H-Theorems in der neuen Mechanik}},}\ }\href {\doibase 10.1007/BF01339852} {\bibfield  {journal} {\bibinfo  {journal} {Zeitschrift fur Physik}\ }\textbf {\bibinfo {volume} {57}},\ \bibinfo {pages} {30--70} (\bibinfo {year} {1929})}\BibitemShut {NoStop}%
\bibitem [{\citenamefont {{Goldstein}}\ \emph {et~al.}(2010)\citenamefont {{Goldstein}}, \citenamefont {{Lebowitz}}, \citenamefont {{Tumulka}},\ and\ \citenamefont {{Zangh{\`\i}}}}]{2010.EPJH.35.173G}%
  \BibitemOpen
  \bibfield  {author} {\bibinfo {author} {\bibfnamefont {S.}~\bibnamefont {{Goldstein}}}, \bibinfo {author} {\bibfnamefont {J.~L.}\ \bibnamefont {{Lebowitz}}}, \bibinfo {author} {\bibfnamefont {R.}~\bibnamefont {{Tumulka}}}, \ and\ \bibinfo {author} {\bibfnamefont {N.}~\bibnamefont {{Zangh{\`\i}}}},\ }\bibfield  {title} {\enquote {\bibinfo {title} {{Long-time behavior of macroscopic quantum systems. Commentary accompanying the English translation of John von Neumann's 1929 article on the quantum ergodic theorem}},}\ }\href {\doibase 10.1140/epjh/e2010-00007-7} {\bibfield  {journal} {\bibinfo  {journal} {European Physical Journal H}\ }\textbf {\bibinfo {volume} {35}},\ \bibinfo {pages} {173--200} (\bibinfo {year} {2010})}\BibitemShut {NoStop}%
\bibitem [{\citenamefont {Bloch}\ \emph {et~al.}(2008)\citenamefont {Bloch}, \citenamefont {Dalibard},\ and\ \citenamefont {Zwerger}}]{2008.RevModPhys.80.885}%
  \BibitemOpen
  \bibfield  {author} {\bibinfo {author} {\bibfnamefont {Immanuel}\ \bibnamefont {Bloch}}, \bibinfo {author} {\bibfnamefont {Jean}\ \bibnamefont {Dalibard}}, \ and\ \bibinfo {author} {\bibfnamefont {Wilhelm}\ \bibnamefont {Zwerger}},\ }\bibfield  {title} {\enquote {\bibinfo {title} {Many-body physics with ultracold gases},}\ }\href {\doibase 10.1103/RevModPhys.80.885} {\bibfield  {journal} {\bibinfo  {journal} {Rev. Mod. Phys.}\ }\textbf {\bibinfo {volume} {80}},\ \bibinfo {pages} {885--964} (\bibinfo {year} {2008})}\BibitemShut {NoStop}%
\bibitem [{\citenamefont {{Langen}}\ \emph {et~al.}(2015{\natexlab{a}})\citenamefont {{Langen}}, \citenamefont {{Geiger}},\ and\ \citenamefont {{Schmiedmayer}}}]{2015.ARCMP.6.201L}%
  \BibitemOpen
  \bibfield  {author} {\bibinfo {author} {\bibfnamefont {Tim}\ \bibnamefont {{Langen}}}, \bibinfo {author} {\bibfnamefont {Remi}\ \bibnamefont {{Geiger}}}, \ and\ \bibinfo {author} {\bibfnamefont {J{\"o}rg}\ \bibnamefont {{Schmiedmayer}}},\ }\bibfield  {title} {\enquote {\bibinfo {title} {{Ultracold Atoms Out of Equilibrium}},}\ }\href {\doibase 10.1146/annurev-conmatphys-031214-014548} {\bibfield  {journal} {\bibinfo  {journal} {Annual Review of Condensed Matter Physics}\ }\textbf {\bibinfo {volume} {6}},\ \bibinfo {pages} {201--217} (\bibinfo {year} {2015}{\natexlab{a}})}\BibitemShut {NoStop}%
\bibitem [{\citenamefont {{Eisert}}\ \emph {et~al.}(2015)\citenamefont {{Eisert}}, \citenamefont {{Friesdorf}},\ and\ \citenamefont {{Gogolin}}}]{2015.NatPh.11.124E}%
  \BibitemOpen
  \bibfield  {author} {\bibinfo {author} {\bibfnamefont {J.}~\bibnamefont {{Eisert}}}, \bibinfo {author} {\bibfnamefont {M.}~\bibnamefont {{Friesdorf}}}, \ and\ \bibinfo {author} {\bibfnamefont {C.}~\bibnamefont {{Gogolin}}},\ }\bibfield  {title} {\enquote {\bibinfo {title} {{Quantum many-body systems out of equilibrium}},}\ }\href {\doibase 10.1038/nphys3215} {\bibfield  {journal} {\bibinfo  {journal} {Nature Physics}\ }\textbf {\bibinfo {volume} {11}},\ \bibinfo {pages} {124--130} (\bibinfo {year} {2015})}\BibitemShut {NoStop}%
\bibitem [{\citenamefont {{Lewenstein}}\ \emph {et~al.}(2007)\citenamefont {{Lewenstein}}, \citenamefont {{Sanpera}}, \citenamefont {{Ahufinger}}, \citenamefont {{Damski}}, \citenamefont {{Sen}},\ and\ \citenamefont {{Sen}}}]{2007.AdPhy.56.243L}%
  \BibitemOpen
  \bibfield  {author} {\bibinfo {author} {\bibfnamefont {Maciej}\ \bibnamefont {{Lewenstein}}}, \bibinfo {author} {\bibfnamefont {Anna}\ \bibnamefont {{Sanpera}}}, \bibinfo {author} {\bibfnamefont {Veronica}\ \bibnamefont {{Ahufinger}}}, \bibinfo {author} {\bibfnamefont {Bogdan}\ \bibnamefont {{Damski}}}, \bibinfo {author} {\bibfnamefont {Aditi}\ \bibnamefont {{Sen}}}, \ and\ \bibinfo {author} {\bibfnamefont {Ujjwal}\ \bibnamefont {{Sen}}},\ }\bibfield  {title} {\enquote {\bibinfo {title} {{Ultracold atomic gases in optical lattices: mimicking condensed matter physics and beyond}},}\ }\href {\doibase 10.1080/00018730701223200} {\bibfield  {journal} {\bibinfo  {journal} {Advances in Physics}\ }\textbf {\bibinfo {volume} {56}},\ \bibinfo {pages} {243--379} (\bibinfo {year} {2007})}\BibitemShut {NoStop}%
\bibitem [{\citenamefont {{Bloch}}\ \emph {et~al.}(2012)\citenamefont {{Bloch}}, \citenamefont {{Dalibard}},\ and\ \citenamefont {{Nascimb{\`e}ne}}}]{2012.NatPh.8.267B}%
  \BibitemOpen
  \bibfield  {author} {\bibinfo {author} {\bibfnamefont {Immanuel}\ \bibnamefont {{Bloch}}}, \bibinfo {author} {\bibfnamefont {Jean}\ \bibnamefont {{Dalibard}}}, \ and\ \bibinfo {author} {\bibfnamefont {Sylvain}\ \bibnamefont {{Nascimb{\`e}ne}}},\ }\bibfield  {title} {\enquote {\bibinfo {title} {{Quantum simulations with ultracold quantum gases}},}\ }\href {\doibase 10.1038/nphys2259} {\bibfield  {journal} {\bibinfo  {journal} {Nature Physics}\ }\textbf {\bibinfo {volume} {8}},\ \bibinfo {pages} {267--276} (\bibinfo {year} {2012})}\BibitemShut {NoStop}%
\bibitem [{\citenamefont {{Trotzky}}\ \emph {et~al.}(2012)\citenamefont {{Trotzky}}, \citenamefont {{Chen}}, \citenamefont {{Flesch}}, \citenamefont {{McCulloch}}, \citenamefont {{Schollw{\"o}ck}}, \citenamefont {{Eisert}},\ and\ \citenamefont {{Bloch}}}]{2012.NatPh.8.325T}%
  \BibitemOpen
  \bibfield  {author} {\bibinfo {author} {\bibfnamefont {S.}~\bibnamefont {{Trotzky}}}, \bibinfo {author} {\bibfnamefont {Y.~A.}\ \bibnamefont {{Chen}}}, \bibinfo {author} {\bibfnamefont {A.}~\bibnamefont {{Flesch}}}, \bibinfo {author} {\bibfnamefont {I.~P.}\ \bibnamefont {{McCulloch}}}, \bibinfo {author} {\bibfnamefont {U.}~\bibnamefont {{Schollw{\"o}ck}}}, \bibinfo {author} {\bibfnamefont {J.}~\bibnamefont {{Eisert}}}, \ and\ \bibinfo {author} {\bibfnamefont {I.}~\bibnamefont {{Bloch}}},\ }\bibfield  {title} {\enquote {\bibinfo {title} {{Probing the relaxation towards equilibrium in an isolated strongly correlated one-dimensional Bose gas}},}\ }\href {\doibase 10.1038/nphys2232} {\bibfield  {journal} {\bibinfo  {journal} {Nature Physics}\ }\textbf {\bibinfo {volume} {8}},\ \bibinfo {pages} {325--330} (\bibinfo {year} {2012})}\BibitemShut {NoStop}%
\bibitem [{\citenamefont {{Kaufman}}\ \emph {et~al.}(2016)\citenamefont {{Kaufman}}, \citenamefont {{Tai}}, \citenamefont {{Lukin}}, \citenamefont {{Rispoli}}, \citenamefont {{Schittko}}, \citenamefont {{Preiss}},\ and\ \citenamefont {{Greiner}}}]{2016.Sci.353.794K}%
  \BibitemOpen
  \bibfield  {author} {\bibinfo {author} {\bibfnamefont {Adam~M.}\ \bibnamefont {{Kaufman}}}, \bibinfo {author} {\bibfnamefont {M.~Eric}\ \bibnamefont {{Tai}}}, \bibinfo {author} {\bibfnamefont {Alexander}\ \bibnamefont {{Lukin}}}, \bibinfo {author} {\bibfnamefont {Matthew}\ \bibnamefont {{Rispoli}}}, \bibinfo {author} {\bibfnamefont {Robert}\ \bibnamefont {{Schittko}}}, \bibinfo {author} {\bibfnamefont {Philipp~M.}\ \bibnamefont {{Preiss}}}, \ and\ \bibinfo {author} {\bibfnamefont {Markus}\ \bibnamefont {{Greiner}}},\ }\bibfield  {title} {\enquote {\bibinfo {title} {{Quantum thermalization through entanglement in an isolated many-body system}},}\ }\href {\doibase 10.1126/science.aaf6725} {\bibfield  {journal} {\bibinfo  {journal} {Science}\ }\textbf {\bibinfo {volume} {353}},\ \bibinfo {pages} {794--800} (\bibinfo {year} {2016})}\BibitemShut {NoStop}%
\bibitem [{\citenamefont {{Neill}}\ \emph {et~al.}(2016)\citenamefont {{Neill}}, \citenamefont {{Roushan}}, \citenamefont {{Fang}}, \citenamefont {{Chen}}, \citenamefont {{Kolodrubetz}}, \citenamefont {{Chen}}, \citenamefont {{Megrant}}, \citenamefont {{Barends}}, \citenamefont {{Campbell}}, \citenamefont {{Chiaro}}, \citenamefont {{Dunsworth}}, \citenamefont {{Jeffrey}}, \citenamefont {{Kelly}}, \citenamefont {{Mutus}}, \citenamefont {{O'Malley}}, \citenamefont {{Quintana}}, \citenamefont {{Sank}}, \citenamefont {{Vainsencher}}, \citenamefont {{Wenner}}, \citenamefont {{White}}, \citenamefont {{Polkovnikov}},\ and\ \citenamefont {{Martinis}}}]{2016.NatPh.12.1037N}%
  \BibitemOpen
  \bibfield  {author} {\bibinfo {author} {\bibfnamefont {C.}~\bibnamefont {{Neill}}}, \bibinfo {author} {\bibfnamefont {P.}~\bibnamefont {{Roushan}}}, \bibinfo {author} {\bibfnamefont {M.}~\bibnamefont {{Fang}}}, \bibinfo {author} {\bibfnamefont {Y.}~\bibnamefont {{Chen}}}, \bibinfo {author} {\bibfnamefont {M.}~\bibnamefont {{Kolodrubetz}}}, \bibinfo {author} {\bibfnamefont {Z.}~\bibnamefont {{Chen}}}, \bibinfo {author} {\bibfnamefont {A.}~\bibnamefont {{Megrant}}}, \bibinfo {author} {\bibfnamefont {R.}~\bibnamefont {{Barends}}}, \bibinfo {author} {\bibfnamefont {B.}~\bibnamefont {{Campbell}}}, \bibinfo {author} {\bibfnamefont {B.}~\bibnamefont {{Chiaro}}}, \bibinfo {author} {\bibfnamefont {A.}~\bibnamefont {{Dunsworth}}}, \bibinfo {author} {\bibfnamefont {E.}~\bibnamefont {{Jeffrey}}}, \bibinfo {author} {\bibfnamefont {J.}~\bibnamefont {{Kelly}}}, \bibinfo {author} {\bibfnamefont {J.}~\bibnamefont {{Mutus}}}, \bibinfo {author} {\bibfnamefont {P.~J.~J.}\ \bibnamefont {{O'Malley}}}, \bibinfo {author} {\bibfnamefont {C.}~\bibnamefont {{Quintana}}}, \bibinfo {author} {\bibfnamefont {D.}~\bibnamefont {{Sank}}}, \bibinfo {author} {\bibfnamefont {A.}~\bibnamefont {{Vainsencher}}}, \bibinfo {author} {\bibfnamefont {J.}~\bibnamefont {{Wenner}}}, \bibinfo {author} {\bibfnamefont {T.~C.}\ \bibnamefont {{White}}}, \bibinfo {author} {\bibfnamefont {A.}~\bibnamefont {{Polkovnikov}}}, \ and\ \bibinfo {author} {\bibfnamefont {J.~M.}\ \bibnamefont {{Martinis}}},\ }\bibfield  {title} {\enquote {\bibinfo {title} {{Ergodic dynamics and thermalization in an isolated quantum system}},}\ }\href {\doibase 10.1038/nphys3830} {\bibfield  {journal} {\bibinfo  {journal} {Nature Physics}\ }\textbf {\bibinfo {volume} {12}},\ \bibinfo {pages} {1037--1041} (\bibinfo {year} {2016})}\BibitemShut {NoStop}%
\bibitem [{\citenamefont {Tang}\ \emph {et~al.}(2018)\citenamefont {Tang}, \citenamefont {Kao}, \citenamefont {Li}, \citenamefont {Seo}, \citenamefont {Mallayya}, \citenamefont {Rigol}, \citenamefont {Gopalakrishnan},\ and\ \citenamefont {Lev}}]{2018.PhysRevX.8.021030}%
  \BibitemOpen
  \bibfield  {author} {\bibinfo {author} {\bibfnamefont {Yijun}\ \bibnamefont {Tang}}, \bibinfo {author} {\bibfnamefont {Wil}\ \bibnamefont {Kao}}, \bibinfo {author} {\bibfnamefont {Kuan-Yu}\ \bibnamefont {Li}}, \bibinfo {author} {\bibfnamefont {Sangwon}\ \bibnamefont {Seo}}, \bibinfo {author} {\bibfnamefont {Krishnanand}\ \bibnamefont {Mallayya}}, \bibinfo {author} {\bibfnamefont {Marcos}\ \bibnamefont {Rigol}}, \bibinfo {author} {\bibfnamefont {Sarang}\ \bibnamefont {Gopalakrishnan}}, \ and\ \bibinfo {author} {\bibfnamefont {Benjamin~L.}\ \bibnamefont {Lev}},\ }\bibfield  {title} {\enquote {\bibinfo {title} {Thermalization near integrability in a dipolar quantum newton's cradle},}\ }\href {\doibase 10.1103/PhysRevX.8.021030} {\bibfield  {journal} {\bibinfo  {journal} {Phys. Rev. X}\ }\textbf {\bibinfo {volume} {8}},\ \bibinfo {pages} {021030} (\bibinfo {year} {2018})}\BibitemShut {NoStop}%
\bibitem [{\citenamefont {Clos}\ \emph {et~al.}(2016)\citenamefont {Clos}, \citenamefont {Porras}, \citenamefont {Warring},\ and\ \citenamefont {Schaetz}}]{2016.PhysRevLett.117.170401}%
  \BibitemOpen
  \bibfield  {author} {\bibinfo {author} {\bibfnamefont {Govinda}\ \bibnamefont {Clos}}, \bibinfo {author} {\bibfnamefont {Diego}\ \bibnamefont {Porras}}, \bibinfo {author} {\bibfnamefont {Ulrich}\ \bibnamefont {Warring}}, \ and\ \bibinfo {author} {\bibfnamefont {Tobias}\ \bibnamefont {Schaetz}},\ }\bibfield  {title} {\enquote {\bibinfo {title} {Time-resolved observation of thermalization in an isolated quantum system},}\ }\href {\doibase 10.1103/PhysRevLett.117.170401} {\bibfield  {journal} {\bibinfo  {journal} {Phys. Rev. Lett.}\ }\textbf {\bibinfo {volume} {117}},\ \bibinfo {pages} {170401} (\bibinfo {year} {2016})}\BibitemShut {NoStop}%
\bibitem [{\citenamefont {{Kinoshita}}\ \emph {et~al.}(2006)\citenamefont {{Kinoshita}}, \citenamefont {{Wenger}},\ and\ \citenamefont {{Weiss}}}]{2006.Natur.440.900K}%
  \BibitemOpen
  \bibfield  {author} {\bibinfo {author} {\bibfnamefont {Toshiya}\ \bibnamefont {{Kinoshita}}}, \bibinfo {author} {\bibfnamefont {Trevor}\ \bibnamefont {{Wenger}}}, \ and\ \bibinfo {author} {\bibfnamefont {David~S.}\ \bibnamefont {{Weiss}}},\ }\bibfield  {title} {\enquote {\bibinfo {title} {{A quantum Newton's cradle}},}\ }\href {\doibase 10.1038/nature04693} {\bibfield  {journal} {\bibinfo  {journal} {\nat}\ }\textbf {\bibinfo {volume} {440}},\ \bibinfo {pages} {900--903} (\bibinfo {year} {2006})}\BibitemShut {NoStop}%
\bibitem [{\citenamefont {{Gring}}\ \emph {et~al.}(2012)\citenamefont {{Gring}}, \citenamefont {{Kuhnert}}, \citenamefont {{Langen}}, \citenamefont {{Kitagawa}}, \citenamefont {{Rauer}}, \citenamefont {{Schreitl}}, \citenamefont {{Mazets}}, \citenamefont {{Smith}}, \citenamefont {{Demler}},\ and\ \citenamefont {{Schmiedmayer}}}]{2012.Sci.337.1318G}%
  \BibitemOpen
  \bibfield  {author} {\bibinfo {author} {\bibfnamefont {M.}~\bibnamefont {{Gring}}}, \bibinfo {author} {\bibfnamefont {M.}~\bibnamefont {{Kuhnert}}}, \bibinfo {author} {\bibfnamefont {T.}~\bibnamefont {{Langen}}}, \bibinfo {author} {\bibfnamefont {T.}~\bibnamefont {{Kitagawa}}}, \bibinfo {author} {\bibfnamefont {B.}~\bibnamefont {{Rauer}}}, \bibinfo {author} {\bibfnamefont {M.}~\bibnamefont {{Schreitl}}}, \bibinfo {author} {\bibfnamefont {I.}~\bibnamefont {{Mazets}}}, \bibinfo {author} {\bibfnamefont {D.~Adu}\ \bibnamefont {{Smith}}}, \bibinfo {author} {\bibfnamefont {E.}~\bibnamefont {{Demler}}}, \ and\ \bibinfo {author} {\bibfnamefont {J.}~\bibnamefont {{Schmiedmayer}}},\ }\bibfield  {title} {\enquote {\bibinfo {title} {{Relaxation and Prethermalization in an Isolated Quantum System}},}\ }\href {\doibase 10.1126/science.1224953} {\bibfield  {journal} {\bibinfo  {journal} {Science}\ }\textbf {\bibinfo {volume} {337}},\ \bibinfo {pages} {1318} (\bibinfo {year} {2012})}\BibitemShut {NoStop}%
\bibitem [{\citenamefont {{Langen}}\ \emph {et~al.}(2015{\natexlab{b}})\citenamefont {{Langen}}, \citenamefont {{Erne}}, \citenamefont {{Geiger}}, \citenamefont {{Rauer}}, \citenamefont {{Schweigler}}, \citenamefont {{Kuhnert}}, \citenamefont {{Rohringer}}, \citenamefont {{Mazets}}, \citenamefont {{Gasenzer}},\ and\ \citenamefont {{Schmiedmayer}}}]{2015.Sci.348.207L}%
  \BibitemOpen
  \bibfield  {author} {\bibinfo {author} {\bibfnamefont {Tim}\ \bibnamefont {{Langen}}}, \bibinfo {author} {\bibfnamefont {Sebastian}\ \bibnamefont {{Erne}}}, \bibinfo {author} {\bibfnamefont {Remi}\ \bibnamefont {{Geiger}}}, \bibinfo {author} {\bibfnamefont {Bernhard}\ \bibnamefont {{Rauer}}}, \bibinfo {author} {\bibfnamefont {Thomas}\ \bibnamefont {{Schweigler}}}, \bibinfo {author} {\bibfnamefont {Maximilian}\ \bibnamefont {{Kuhnert}}}, \bibinfo {author} {\bibfnamefont {Wolfgang}\ \bibnamefont {{Rohringer}}}, \bibinfo {author} {\bibfnamefont {Igor~E.}\ \bibnamefont {{Mazets}}}, \bibinfo {author} {\bibfnamefont {Thomas}\ \bibnamefont {{Gasenzer}}}, \ and\ \bibinfo {author} {\bibfnamefont {J{\"o}rg}\ \bibnamefont {{Schmiedmayer}}},\ }\bibfield  {title} {\enquote {\bibinfo {title} {{Experimental observation of a generalized Gibbs ensemble}},}\ }\href {\doibase 10.1126/science.1257026} {\bibfield  {journal} {\bibinfo  {journal} {Science}\ }\textbf {\bibinfo {volume} {348}},\ \bibinfo {pages} {207--211} (\bibinfo {year} {2015}{\natexlab{b}})}\BibitemShut {NoStop}%
\bibitem [{\citenamefont {Vidmar}\ and\ \citenamefont {Rigol}(2016)}]{Vidmar16}%
  \BibitemOpen
  \bibfield  {author} {\bibinfo {author} {\bibfnamefont {Lev}\ \bibnamefont {Vidmar}}\ and\ \bibinfo {author} {\bibfnamefont {Marcos}\ \bibnamefont {Rigol}},\ }\bibfield  {title} {\enquote {\bibinfo {title} {Generalized gibbs ensemble in integrable lattice models},}\ }\href {\doibase 10.1088/1742-5468/2016/06/064007} {\bibfield  {journal} {\bibinfo  {journal} {Journal of Statistical Mechanics: Theory and Experiment}\ }\textbf {\bibinfo {volume} {2016}},\ \bibinfo {pages} {064007} (\bibinfo {year} {2016})}\BibitemShut {NoStop}%
\bibitem [{\citenamefont {Deutsch}(1991)}]{1991.PhysRevA.43.2046}%
  \BibitemOpen
  \bibfield  {author} {\bibinfo {author} {\bibfnamefont {J.~M.}\ \bibnamefont {Deutsch}},\ }\bibfield  {title} {\enquote {\bibinfo {title} {Quantum statistical mechanics in a closed system},}\ }\href {\doibase 10.1103/PhysRevA.43.2046} {\bibfield  {journal} {\bibinfo  {journal} {Phys. Rev. A}\ }\textbf {\bibinfo {volume} {43}},\ \bibinfo {pages} {2046--2049} (\bibinfo {year} {1991})}\BibitemShut {NoStop}%
\bibitem [{\citenamefont {Srednicki}(1994)}]{1994.PhysRevE.50.888}%
  \BibitemOpen
  \bibfield  {author} {\bibinfo {author} {\bibfnamefont {Mark}\ \bibnamefont {Srednicki}},\ }\bibfield  {title} {\enquote {\bibinfo {title} {Chaos and quantum thermalization},}\ }\href {\doibase 10.1103/PhysRevE.50.888} {\bibfield  {journal} {\bibinfo  {journal} {Phys. Rev. E}\ }\textbf {\bibinfo {volume} {50}},\ \bibinfo {pages} {888--901} (\bibinfo {year} {1994})}\BibitemShut {NoStop}%
\bibitem [{\citenamefont {{Rigol}}\ \emph {et~al.}(2008)\citenamefont {{Rigol}}, \citenamefont {{Dunjko}},\ and\ \citenamefont {{Olshanii}}}]{2008.Natur.452.854R}%
  \BibitemOpen
  \bibfield  {author} {\bibinfo {author} {\bibfnamefont {Marcos}\ \bibnamefont {{Rigol}}}, \bibinfo {author} {\bibfnamefont {Vanja}\ \bibnamefont {{Dunjko}}}, \ and\ \bibinfo {author} {\bibfnamefont {Maxim}\ \bibnamefont {{Olshanii}}},\ }\bibfield  {title} {\enquote {\bibinfo {title} {{Thermalization and its mechanism for generic isolated quantum systems}},}\ }\href {\doibase 10.1038/nature06838} {\bibfield  {journal} {\bibinfo  {journal} {\nat}\ }\textbf {\bibinfo {volume} {452}},\ \bibinfo {pages} {854--858} (\bibinfo {year} {2008})}\BibitemShut {NoStop}%
\bibitem [{\citenamefont {Pappalardi}\ \emph {et~al.}(2023{\natexlab{a}})\citenamefont {Pappalardi}, \citenamefont {Foini},\ and\ \citenamefont {Kurchan}}]{pappalardi2023microcanonical}%
  \BibitemOpen
  \bibfield  {author} {\bibinfo {author} {\bibfnamefont {Silvia}\ \bibnamefont {Pappalardi}}, \bibinfo {author} {\bibfnamefont {Laura}\ \bibnamefont {Foini}}, \ and\ \bibinfo {author} {\bibfnamefont {Jorge}\ \bibnamefont {Kurchan}},\ }\href@noop {} {\enquote {\bibinfo {title} {Microcanonical windows on quantum operators},}\ } (\bibinfo {year} {2023}{\natexlab{a}}),\ \Eprint {http://arxiv.org/abs/2304.10948} {arXiv:2304.10948 [cond-mat.stat-mech]} \BibitemShut {NoStop}%
\bibitem [{\citenamefont {D'Alessio}\ \emph {et~al.}(2016)\citenamefont {D'Alessio}, \citenamefont {Kafri}, \citenamefont {Polkovnikov},\ and\ \citenamefont {Rigol}}]{2016.AdvancesinPhysics}%
  \BibitemOpen
  \bibfield  {author} {\bibinfo {author} {\bibfnamefont {Luca}\ \bibnamefont {D'Alessio}}, \bibinfo {author} {\bibfnamefont {Yariv}\ \bibnamefont {Kafri}}, \bibinfo {author} {\bibfnamefont {Anatoli}\ \bibnamefont {Polkovnikov}}, \ and\ \bibinfo {author} {\bibfnamefont {Marcos}\ \bibnamefont {Rigol}},\ }\bibfield  {title} {\enquote {\bibinfo {title} {From quantum chaos and eigenstate thermalization to statistical mechanics and thermodynamics},}\ }\href {\doibase 10.1080/00018732.2016.1198134} {\bibfield  {journal} {\bibinfo  {journal} {Advances in Physics}\ }\textbf {\bibinfo {volume} {65}},\ \bibinfo {pages} {239--362} (\bibinfo {year} {2016})}\BibitemShut {NoStop}%
\bibitem [{\citenamefont {Dankert}\ \emph {et~al.}(2009)\citenamefont {Dankert}, \citenamefont {Cleve}, \citenamefont {Emerson},\ and\ \citenamefont {Livine}}]{Dankert09unitary}%
  \BibitemOpen
  \bibfield  {author} {\bibinfo {author} {\bibfnamefont {Christoph}\ \bibnamefont {Dankert}}, \bibinfo {author} {\bibfnamefont {Richard}\ \bibnamefont {Cleve}}, \bibinfo {author} {\bibfnamefont {Joseph}\ \bibnamefont {Emerson}}, \ and\ \bibinfo {author} {\bibfnamefont {Etera}\ \bibnamefont {Livine}},\ }\bibfield  {title} {\enquote {\bibinfo {title} {Exact and approximate unitary 2-designs and their application to fidelity estimation},}\ }\href {\doibase 10.1103/PhysRevA.80.012304} {\bibfield  {journal} {\bibinfo  {journal} {Phys. Rev. A}\ }\textbf {\bibinfo {volume} {80}},\ \bibinfo {pages} {012304} (\bibinfo {year} {2009})}\BibitemShut {NoStop}%
\bibitem [{\citenamefont {Voiculescu}(1991)}]{Voiculescu1991}%
  \BibitemOpen
  \bibfield  {author} {\bibinfo {author} {\bibfnamefont {Dan}\ \bibnamefont {Voiculescu}},\ }\bibfield  {title} {\enquote {\bibinfo {title} {Limit laws for random matrices and free products},}\ }\href {\doibase 10.1007/BF01245072} {\bibfield  {journal} {\bibinfo  {journal} {Inventiones mathematicae}\ }\textbf {\bibinfo {volume} {104}},\ \bibinfo {pages} {201--220} (\bibinfo {year} {1991})}\BibitemShut {NoStop}%
\bibitem [{\citenamefont {Pappalardi}\ \emph {et~al.}(2022)\citenamefont {Pappalardi}, \citenamefont {Foini},\ and\ \citenamefont {Kurchan}}]{Pappalardi22}%
  \BibitemOpen
  \bibfield  {author} {\bibinfo {author} {\bibfnamefont {Silvia}\ \bibnamefont {Pappalardi}}, \bibinfo {author} {\bibfnamefont {Laura}\ \bibnamefont {Foini}}, \ and\ \bibinfo {author} {\bibfnamefont {Jorge}\ \bibnamefont {Kurchan}},\ }\bibfield  {title} {\enquote {\bibinfo {title} {Eigenstate thermalization hypothesis and free probability},}\ }\href {\doibase 10.1103/PhysRevLett.129.170603} {\bibfield  {journal} {\bibinfo  {journal} {Phys. Rev. Lett.}\ }\textbf {\bibinfo {volume} {129}},\ \bibinfo {pages} {170603} (\bibinfo {year} {2022})}\BibitemShut {NoStop}%
\bibitem [{\citenamefont {Pappalardi}\ \emph {et~al.}(2023{\natexlab{b}})\citenamefont {Pappalardi}, \citenamefont {Fritzsch},\ and\ \citenamefont {Prosen}}]{Pappalardi23}%
  \BibitemOpen
  \bibfield  {author} {\bibinfo {author} {\bibfnamefont {Silvia}\ \bibnamefont {Pappalardi}}, \bibinfo {author} {\bibfnamefont {Felix}\ \bibnamefont {Fritzsch}}, \ and\ \bibinfo {author} {\bibfnamefont {Tomaž}\ \bibnamefont {Prosen}},\ }\href@noop {} {\enquote {\bibinfo {title} {General eigenstate thermalization via free cumulants in quantum lattice systems},}\ } (\bibinfo {year} {2023}{\natexlab{b}}),\ \Eprint {http://arxiv.org/abs/2303.00713} {arXiv:2303.00713 [cond-mat.stat-mech]} \BibitemShut {NoStop}%
\bibitem [{\citenamefont {Fava}\ \emph {et~al.}(2023)\citenamefont {Fava}, \citenamefont {Kurchan},\ and\ \citenamefont {Pappalardi}}]{fava2023designs}%
  \BibitemOpen
  \bibfield  {author} {\bibinfo {author} {\bibfnamefont {Michele}\ \bibnamefont {Fava}}, \bibinfo {author} {\bibfnamefont {Jorge}\ \bibnamefont {Kurchan}}, \ and\ \bibinfo {author} {\bibfnamefont {Silvia}\ \bibnamefont {Pappalardi}},\ }\href@noop {} {\enquote {\bibinfo {title} {Designs via free probability},}\ } (\bibinfo {year} {2023}),\ \Eprint {http://arxiv.org/abs/2308.06200} {arXiv:2308.06200 [quant-ph]} \BibitemShut {NoStop}%
\bibitem [{\citenamefont {Rigol}(2009)}]{Rigol09}%
  \BibitemOpen
  \bibfield  {author} {\bibinfo {author} {\bibfnamefont {Marcos}\ \bibnamefont {Rigol}},\ }\bibfield  {title} {\enquote {\bibinfo {title} {Breakdown of thermalization in finite one-dimensional systems},}\ }\href {\doibase 10.1103/PhysRevLett.103.100403} {\bibfield  {journal} {\bibinfo  {journal} {Phys. Rev. Lett.}\ }\textbf {\bibinfo {volume} {103}},\ \bibinfo {pages} {100403} (\bibinfo {year} {2009})}\BibitemShut {NoStop}%
\bibitem [{\citenamefont {Steinigeweg}\ \emph {et~al.}(2013)\citenamefont {Steinigeweg}, \citenamefont {Herbrych},\ and\ \citenamefont {Prelov\ifmmode~\check{s}\else \v{s}\fi{}ek}}]{Steinigeweg13}%
  \BibitemOpen
  \bibfield  {author} {\bibinfo {author} {\bibfnamefont {R.}~\bibnamefont {Steinigeweg}}, \bibinfo {author} {\bibfnamefont {J.}~\bibnamefont {Herbrych}}, \ and\ \bibinfo {author} {\bibfnamefont {P.}~\bibnamefont {Prelov\ifmmode~\check{s}\else \v{s}\fi{}ek}},\ }\bibfield  {title} {\enquote {\bibinfo {title} {Eigenstate thermalization within isolated spin-chain systems},}\ }\href {\doibase 10.1103/PhysRevE.87.012118} {\bibfield  {journal} {\bibinfo  {journal} {Phys. Rev. E}\ }\textbf {\bibinfo {volume} {87}},\ \bibinfo {pages} {012118} (\bibinfo {year} {2013})}\BibitemShut {NoStop}%
\bibitem [{\citenamefont {Khatami}\ \emph {et~al.}(2013)\citenamefont {Khatami}, \citenamefont {Pupillo}, \citenamefont {Srednicki},\ and\ \citenamefont {Rigol}}]{2013.PhysRevLett.111.050403}%
  \BibitemOpen
  \bibfield  {author} {\bibinfo {author} {\bibfnamefont {Ehsan}\ \bibnamefont {Khatami}}, \bibinfo {author} {\bibfnamefont {Guido}\ \bibnamefont {Pupillo}}, \bibinfo {author} {\bibfnamefont {Mark}\ \bibnamefont {Srednicki}}, \ and\ \bibinfo {author} {\bibfnamefont {Marcos}\ \bibnamefont {Rigol}},\ }\bibfield  {title} {\enquote {\bibinfo {title} {Fluctuation-dissipation theorem in an isolated system of quantum dipolar bosons after a quench},}\ }\href {\doibase 10.1103/PhysRevLett.111.050403} {\bibfield  {journal} {\bibinfo  {journal} {Phys. Rev. Lett.}\ }\textbf {\bibinfo {volume} {111}},\ \bibinfo {pages} {050403} (\bibinfo {year} {2013})}\BibitemShut {NoStop}%
\bibitem [{\citenamefont {Beugeling}\ \emph {et~al.}(2014)\citenamefont {Beugeling}, \citenamefont {Moessner},\ and\ \citenamefont {Haque}}]{Beugeling14}%
  \BibitemOpen
  \bibfield  {author} {\bibinfo {author} {\bibfnamefont {W.}~\bibnamefont {Beugeling}}, \bibinfo {author} {\bibfnamefont {R.}~\bibnamefont {Moessner}}, \ and\ \bibinfo {author} {\bibfnamefont {Masudul}\ \bibnamefont {Haque}},\ }\bibfield  {title} {\enquote {\bibinfo {title} {Finite-size scaling of eigenstate thermalization},}\ }\href {\doibase 10.1103/PhysRevE.89.042112} {\bibfield  {journal} {\bibinfo  {journal} {Phys. Rev. E}\ }\textbf {\bibinfo {volume} {89}},\ \bibinfo {pages} {042112} (\bibinfo {year} {2014})}\BibitemShut {NoStop}%
\bibitem [{\citenamefont {Sorg}\ \emph {et~al.}(2014)\citenamefont {Sorg}, \citenamefont {Vidmar}, \citenamefont {Pollet},\ and\ \citenamefont {Heidrich-Meisner}}]{Sorg14}%
  \BibitemOpen
  \bibfield  {author} {\bibinfo {author} {\bibfnamefont {S.}~\bibnamefont {Sorg}}, \bibinfo {author} {\bibfnamefont {L.}~\bibnamefont {Vidmar}}, \bibinfo {author} {\bibfnamefont {L.}~\bibnamefont {Pollet}}, \ and\ \bibinfo {author} {\bibfnamefont {F.}~\bibnamefont {Heidrich-Meisner}},\ }\bibfield  {title} {\enquote {\bibinfo {title} {Relaxation and thermalization in the one-dimensional bose-hubbard model: A case study for the interaction quantum quench from the atomic limit},}\ }\href {\doibase 10.1103/PhysRevA.90.033606} {\bibfield  {journal} {\bibinfo  {journal} {Phys. Rev. A}\ }\textbf {\bibinfo {volume} {90}},\ \bibinfo {pages} {033606} (\bibinfo {year} {2014})}\BibitemShut {NoStop}%
\bibitem [{\citenamefont {Steinigeweg}\ \emph {et~al.}(2014)\citenamefont {Steinigeweg}, \citenamefont {Khodja}, \citenamefont {Niemeyer}, \citenamefont {Gogolin},\ and\ \citenamefont {Gemmer}}]{Steinigeweg14}%
  \BibitemOpen
  \bibfield  {author} {\bibinfo {author} {\bibfnamefont {R.}~\bibnamefont {Steinigeweg}}, \bibinfo {author} {\bibfnamefont {A.}~\bibnamefont {Khodja}}, \bibinfo {author} {\bibfnamefont {H.}~\bibnamefont {Niemeyer}}, \bibinfo {author} {\bibfnamefont {C.}~\bibnamefont {Gogolin}}, \ and\ \bibinfo {author} {\bibfnamefont {J.}~\bibnamefont {Gemmer}},\ }\bibfield  {title} {\enquote {\bibinfo {title} {Pushing the limits of the eigenstate thermalization hypothesis towards mesoscopic quantum systems},}\ }\href {\doibase 10.1103/PhysRevLett.112.130403} {\bibfield  {journal} {\bibinfo  {journal} {Phys. Rev. Lett.}\ }\textbf {\bibinfo {volume} {112}},\ \bibinfo {pages} {130403} (\bibinfo {year} {2014})}\BibitemShut {NoStop}%
\bibitem [{\citenamefont {Beugeling}\ \emph {et~al.}(2015)\citenamefont {Beugeling}, \citenamefont {Moessner},\ and\ \citenamefont {Haque}}]{2015.PhysRevE.91.012144}%
  \BibitemOpen
  \bibfield  {author} {\bibinfo {author} {\bibfnamefont {Wouter}\ \bibnamefont {Beugeling}}, \bibinfo {author} {\bibfnamefont {Roderich}\ \bibnamefont {Moessner}}, \ and\ \bibinfo {author} {\bibfnamefont {Masudul}\ \bibnamefont {Haque}},\ }\bibfield  {title} {\enquote {\bibinfo {title} {Off-diagonal matrix elements of local operators in many-body quantum systems},}\ }\href {\doibase 10.1103/PhysRevE.91.012144} {\bibfield  {journal} {\bibinfo  {journal} {Phys. Rev. E}\ }\textbf {\bibinfo {volume} {91}},\ \bibinfo {pages} {012144} (\bibinfo {year} {2015})}\BibitemShut {NoStop}%
\bibitem [{\citenamefont {Mondaini}\ \emph {et~al.}(2016)\citenamefont {Mondaini}, \citenamefont {Fratus}, \citenamefont {Srednicki},\ and\ \citenamefont {Rigol}}]{2016.PhysRevE.93.032104}%
  \BibitemOpen
  \bibfield  {author} {\bibinfo {author} {\bibfnamefont {Rubem}\ \bibnamefont {Mondaini}}, \bibinfo {author} {\bibfnamefont {Keith~R.}\ \bibnamefont {Fratus}}, \bibinfo {author} {\bibfnamefont {Mark}\ \bibnamefont {Srednicki}}, \ and\ \bibinfo {author} {\bibfnamefont {Marcos}\ \bibnamefont {Rigol}},\ }\bibfield  {title} {\enquote {\bibinfo {title} {Eigenstate thermalization in the two-dimensional transverse field ising model},}\ }\href {\doibase 10.1103/PhysRevE.93.032104} {\bibfield  {journal} {\bibinfo  {journal} {Phys. Rev. E}\ }\textbf {\bibinfo {volume} {93}},\ \bibinfo {pages} {032104} (\bibinfo {year} {2016})}\BibitemShut {NoStop}%
\bibitem [{\citenamefont {Yoshizawa}\ \emph {et~al.}(2018)\citenamefont {Yoshizawa}, \citenamefont {Iyoda},\ and\ \citenamefont {Sagawa}}]{Yoshizawa18}%
  \BibitemOpen
  \bibfield  {author} {\bibinfo {author} {\bibfnamefont {Toru}\ \bibnamefont {Yoshizawa}}, \bibinfo {author} {\bibfnamefont {Eiki}\ \bibnamefont {Iyoda}}, \ and\ \bibinfo {author} {\bibfnamefont {Takahiro}\ \bibnamefont {Sagawa}},\ }\bibfield  {title} {\enquote {\bibinfo {title} {Numerical large deviation analysis of the eigenstate thermalization hypothesis},}\ }\href {\doibase 10.1103/PhysRevLett.120.200604} {\bibfield  {journal} {\bibinfo  {journal} {Phys. Rev. Lett.}\ }\textbf {\bibinfo {volume} {120}},\ \bibinfo {pages} {200604} (\bibinfo {year} {2018})}\BibitemShut {NoStop}%
\bibitem [{\citenamefont {Jansen}\ \emph {et~al.}(2019)\citenamefont {Jansen}, \citenamefont {Stolpp}, \citenamefont {Vidmar},\ and\ \citenamefont {{Heidrich-Meisner}}}]{2019.PhysRevB.99.155130}%
  \BibitemOpen
  \bibfield  {author} {\bibinfo {author} {\bibfnamefont {David}\ \bibnamefont {Jansen}}, \bibinfo {author} {\bibfnamefont {Jan}\ \bibnamefont {Stolpp}}, \bibinfo {author} {\bibfnamefont {Lev}\ \bibnamefont {Vidmar}}, \ and\ \bibinfo {author} {\bibfnamefont {Fabian}\ \bibnamefont {{Heidrich-Meisner}}},\ }\bibfield  {title} {\enquote {\bibinfo {title} {Eigenstate thermalization and quantum chaos in the holstein polaron model},}\ }\href {\doibase 10.1103/PhysRevB.99.155130} {\bibfield  {journal} {\bibinfo  {journal} {Phys. Rev. B}\ }\textbf {\bibinfo {volume} {99}},\ \bibinfo {pages} {155130} (\bibinfo {year} {2019})}\BibitemShut {NoStop}%
\bibitem [{\citenamefont {Khaymovich}\ \emph {et~al.}(2019)\citenamefont {Khaymovich}, \citenamefont {Haque},\ and\ \citenamefont {McClarty}}]{2019.PhysRevLett.122.070601}%
  \BibitemOpen
  \bibfield  {author} {\bibinfo {author} {\bibfnamefont {Ivan~M.}\ \bibnamefont {Khaymovich}}, \bibinfo {author} {\bibfnamefont {Masudul}\ \bibnamefont {Haque}}, \ and\ \bibinfo {author} {\bibfnamefont {Paul~A.}\ \bibnamefont {McClarty}},\ }\bibfield  {title} {\enquote {\bibinfo {title} {Eigenstate thermalization, random matrix theory, and behemoths},}\ }\href {\doibase 10.1103/PhysRevLett.122.070601} {\bibfield  {journal} {\bibinfo  {journal} {Phys. Rev. Lett.}\ }\textbf {\bibinfo {volume} {122}},\ \bibinfo {pages} {070601} (\bibinfo {year} {2019})}\BibitemShut {NoStop}%
\bibitem [{\citenamefont {Sch\"onle}\ \emph {et~al.}(2021)\citenamefont {Sch\"onle}, \citenamefont {Jansen}, \citenamefont {{Heidrich-Meisner}},\ and\ \citenamefont {Vidmar}}]{2021.PhysRevB.103.235137}%
  \BibitemOpen
  \bibfield  {author} {\bibinfo {author} {\bibfnamefont {Christoph}\ \bibnamefont {Sch\"onle}}, \bibinfo {author} {\bibfnamefont {David}\ \bibnamefont {Jansen}}, \bibinfo {author} {\bibfnamefont {Fabian}\ \bibnamefont {{Heidrich-Meisner}}}, \ and\ \bibinfo {author} {\bibfnamefont {Lev}\ \bibnamefont {Vidmar}},\ }\bibfield  {title} {\enquote {\bibinfo {title} {Eigenstate thermalization hypothesis through the lens of autocorrelation functions},}\ }\href {\doibase 10.1103/PhysRevB.103.235137} {\bibfield  {journal} {\bibinfo  {journal} {Phys. Rev. B}\ }\textbf {\bibinfo {volume} {103}},\ \bibinfo {pages} {235137} (\bibinfo {year} {2021})}\BibitemShut {NoStop}%
\bibitem [{\citenamefont {Noh}(2023)}]{2023.PhysRevE.107.014130}%
  \BibitemOpen
  \bibfield  {author} {\bibinfo {author} {\bibfnamefont {Jae~Dong}\ \bibnamefont {Noh}},\ }\bibfield  {title} {\enquote {\bibinfo {title} {Eigenstate thermalization hypothesis in two-dimensional $xxz$ model with or without su(2) symmetry},}\ }\href {\doibase 10.1103/PhysRevE.107.014130} {\bibfield  {journal} {\bibinfo  {journal} {Phys. Rev. E}\ }\textbf {\bibinfo {volume} {107}},\ \bibinfo {pages} {014130} (\bibinfo {year} {2023})}\BibitemShut {NoStop}%
\bibitem [{\citenamefont {Anh-Tai}\ \emph {et~al.}(2023)\citenamefont {Anh-Tai}, \citenamefont {Mikkelsen}, \citenamefont {Busch},\ and\ \citenamefont {Fogarty}}]{2023.SciPostPhys.15.2.048}%
  \BibitemOpen
  \bibfield  {author} {\bibinfo {author} {\bibfnamefont {Tran~Duong}\ \bibnamefont {Anh-Tai}}, \bibinfo {author} {\bibfnamefont {Mathias}\ \bibnamefont {Mikkelsen}}, \bibinfo {author} {\bibfnamefont {Thomas}\ \bibnamefont {Busch}}, \ and\ \bibinfo {author} {\bibfnamefont {Thomás}\ \bibnamefont {Fogarty}},\ }\bibfield  {title} {\enquote {\bibinfo {title} {{Quantum chaos in interacting Bose-Bose mixtures}},}\ }\href {\doibase 10.21468/SciPostPhys.15.2.048} {\bibfield  {journal} {\bibinfo  {journal} {SciPost Phys.}\ }\textbf {\bibinfo {volume} {15}},\ \bibinfo {pages} {048} (\bibinfo {year} {2023})}\BibitemShut {NoStop}%
\bibitem [{\citenamefont {Bari\ifmmode \check{s}\else \v{s}\fi{}i\ifmmode~\acute{c}\else \'{c}\fi{}}\ \emph {et~al.}(2009)\citenamefont {Bari\ifmmode \check{s}\else \v{s}\fi{}i\ifmmode~\acute{c}\else \'{c}\fi{}}, \citenamefont {Prelov\ifmmode~\check{s}\else \v{s}\fi{}ek}, \citenamefont {Metavitsiadis},\ and\ \citenamefont {Zotos}}]{2009.PhysRevB.80.125118}%
  \BibitemOpen
  \bibfield  {author} {\bibinfo {author} {\bibfnamefont {O.~S.}\ \bibnamefont {Bari\ifmmode \check{s}\else \v{s}\fi{}i\ifmmode~\acute{c}\else \'{c}\fi{}}}, \bibinfo {author} {\bibfnamefont {P.}~\bibnamefont {Prelov\ifmmode~\check{s}\else \v{s}\fi{}ek}}, \bibinfo {author} {\bibfnamefont {A.}~\bibnamefont {Metavitsiadis}}, \ and\ \bibinfo {author} {\bibfnamefont {X.}~\bibnamefont {Zotos}},\ }\bibfield  {title} {\enquote {\bibinfo {title} {Incoherent transport induced by a single static impurity in a heisenberg chain},}\ }\href {\doibase 10.1103/PhysRevB.80.125118} {\bibfield  {journal} {\bibinfo  {journal} {Phys. Rev. B}\ }\textbf {\bibinfo {volume} {80}},\ \bibinfo {pages} {125118} (\bibinfo {year} {2009})}\BibitemShut {NoStop}%
\bibitem [{\citenamefont {Brenes}\ \emph {et~al.}(2018)\citenamefont {Brenes}, \citenamefont {Mascarenhas}, \citenamefont {Rigol},\ and\ \citenamefont {Goold}}]{2018.PhysRevB.98.235128}%
  \BibitemOpen
  \bibfield  {author} {\bibinfo {author} {\bibfnamefont {Marlon}\ \bibnamefont {Brenes}}, \bibinfo {author} {\bibfnamefont {Eduardo}\ \bibnamefont {Mascarenhas}}, \bibinfo {author} {\bibfnamefont {Marcos}\ \bibnamefont {Rigol}}, \ and\ \bibinfo {author} {\bibfnamefont {John}\ \bibnamefont {Goold}},\ }\bibfield  {title} {\enquote {\bibinfo {title} {High-temperature coherent transport in the xxz chain in the presence of an impurity},}\ }\href {\doibase 10.1103/PhysRevB.98.235128} {\bibfield  {journal} {\bibinfo  {journal} {Phys. Rev. B}\ }\textbf {\bibinfo {volume} {98}},\ \bibinfo {pages} {235128} (\bibinfo {year} {2018})}\BibitemShut {NoStop}%
\bibitem [{\citenamefont {LeBlond}\ \emph {et~al.}(2019)\citenamefont {LeBlond}, \citenamefont {Mallayya}, \citenamefont {Vidmar},\ and\ \citenamefont {Rigol}}]{2019.PhysRevE.100.062134}%
  \BibitemOpen
  \bibfield  {author} {\bibinfo {author} {\bibfnamefont {Tyler}\ \bibnamefont {LeBlond}}, \bibinfo {author} {\bibfnamefont {Krishnanand}\ \bibnamefont {Mallayya}}, \bibinfo {author} {\bibfnamefont {Lev}\ \bibnamefont {Vidmar}}, \ and\ \bibinfo {author} {\bibfnamefont {Marcos}\ \bibnamefont {Rigol}},\ }\bibfield  {title} {\enquote {\bibinfo {title} {Entanglement and matrix elements of observables in interacting integrable systems},}\ }\href {\doibase 10.1103/PhysRevE.100.062134} {\bibfield  {journal} {\bibinfo  {journal} {Phys. Rev. E}\ }\textbf {\bibinfo {volume} {100}},\ \bibinfo {pages} {062134} (\bibinfo {year} {2019})}\BibitemShut {NoStop}%
\bibitem [{\citenamefont {LeBlond}\ and\ \citenamefont {Rigol}(2020)}]{2020.PhysRevE.102.062113}%
  \BibitemOpen
  \bibfield  {author} {\bibinfo {author} {\bibfnamefont {Tyler}\ \bibnamefont {LeBlond}}\ and\ \bibinfo {author} {\bibfnamefont {Marcos}\ \bibnamefont {Rigol}},\ }\bibfield  {title} {\enquote {\bibinfo {title} {Eigenstate thermalization for observables that break hamiltonian symmetries and its counterpart in interacting integrable systems},}\ }\href {\doibase 10.1103/PhysRevE.102.062113} {\bibfield  {journal} {\bibinfo  {journal} {Phys. Rev. E}\ }\textbf {\bibinfo {volume} {102}},\ \bibinfo {pages} {062113} (\bibinfo {year} {2020})}\BibitemShut {NoStop}%
\bibitem [{\citenamefont {Richter}\ \emph {et~al.}(2020)\citenamefont {Richter}, \citenamefont {Dymarsky}, \citenamefont {Steinigeweg},\ and\ \citenamefont {Gemmer}}]{2020.PhysRevE.102.042127}%
  \BibitemOpen
  \bibfield  {author} {\bibinfo {author} {\bibfnamefont {Jonas}\ \bibnamefont {Richter}}, \bibinfo {author} {\bibfnamefont {Anatoly}\ \bibnamefont {Dymarsky}}, \bibinfo {author} {\bibfnamefont {Robin}\ \bibnamefont {Steinigeweg}}, \ and\ \bibinfo {author} {\bibfnamefont {Jochen}\ \bibnamefont {Gemmer}},\ }\bibfield  {title} {\enquote {\bibinfo {title} {Eigenstate thermalization hypothesis beyond standard indicators: Emergence of random-matrix behavior at small frequencies},}\ }\href {\doibase 10.1103/PhysRevE.102.042127} {\bibfield  {journal} {\bibinfo  {journal} {Phys. Rev. E}\ }\textbf {\bibinfo {volume} {102}},\ \bibinfo {pages} {042127} (\bibinfo {year} {2020})}\BibitemShut {NoStop}%
\bibitem [{\citenamefont {Brenes}\ \emph {et~al.}(2020{\natexlab{a}})\citenamefont {Brenes}, \citenamefont {Goold},\ and\ \citenamefont {Rigol}}]{2020.PhysRevB.102.075127}%
  \BibitemOpen
  \bibfield  {author} {\bibinfo {author} {\bibfnamefont {Marlon}\ \bibnamefont {Brenes}}, \bibinfo {author} {\bibfnamefont {John}\ \bibnamefont {Goold}}, \ and\ \bibinfo {author} {\bibfnamefont {Marcos}\ \bibnamefont {Rigol}},\ }\bibfield  {title} {\enquote {\bibinfo {title} {Low-frequency behavior of off-diagonal matrix elements in the integrable xxz chain and in a locally perturbed quantum-chaotic xxz chain},}\ }\href {\doibase 10.1103/PhysRevB.102.075127} {\bibfield  {journal} {\bibinfo  {journal} {Phys. Rev. B}\ }\textbf {\bibinfo {volume} {102}},\ \bibinfo {pages} {075127} (\bibinfo {year} {2020}{\natexlab{a}})}\BibitemShut {NoStop}%
\bibitem [{\citenamefont {Brenes}\ \emph {et~al.}(2020{\natexlab{b}})\citenamefont {Brenes}, \citenamefont {LeBlond}, \citenamefont {Goold},\ and\ \citenamefont {Rigol}}]{2020.PhysRevLett.125.070605}%
  \BibitemOpen
  \bibfield  {author} {\bibinfo {author} {\bibfnamefont {Marlon}\ \bibnamefont {Brenes}}, \bibinfo {author} {\bibfnamefont {Tyler}\ \bibnamefont {LeBlond}}, \bibinfo {author} {\bibfnamefont {John}\ \bibnamefont {Goold}}, \ and\ \bibinfo {author} {\bibfnamefont {Marcos}\ \bibnamefont {Rigol}},\ }\bibfield  {title} {\enquote {\bibinfo {title} {Eigenstate thermalization in a locally perturbed integrable system},}\ }\href {\doibase 10.1103/PhysRevLett.125.070605} {\bibfield  {journal} {\bibinfo  {journal} {Phys. Rev. Lett.}\ }\textbf {\bibinfo {volume} {125}},\ \bibinfo {pages} {070605} (\bibinfo {year} {2020}{\natexlab{b}})}\BibitemShut {NoStop}%
\bibitem [{\citenamefont {Vasseur}\ \emph {et~al.}(2014)\citenamefont {Vasseur}, \citenamefont {Dahlhaus},\ and\ \citenamefont {Moore}}]{2014.PhysRevX.4.041007}%
  \BibitemOpen
  \bibfield  {author} {\bibinfo {author} {\bibfnamefont {R.}~\bibnamefont {Vasseur}}, \bibinfo {author} {\bibfnamefont {J.~P.}\ \bibnamefont {Dahlhaus}}, \ and\ \bibinfo {author} {\bibfnamefont {J.~E.}\ \bibnamefont {Moore}},\ }\bibfield  {title} {\enquote {\bibinfo {title} {Universal nonequilibrium signatures of majorana zero modes in quench dynamics},}\ }\href {\doibase 10.1103/PhysRevX.4.041007} {\bibfield  {journal} {\bibinfo  {journal} {Phys. Rev. X}\ }\textbf {\bibinfo {volume} {4}},\ \bibinfo {pages} {041007} (\bibinfo {year} {2014})}\BibitemShut {NoStop}%
\bibitem [{\citenamefont {{Viti}}\ \emph {et~al.}(2016)\citenamefont {{Viti}}, \citenamefont {{St{\'e}phan}}, \citenamefont {{Dubail}},\ and\ \citenamefont {{Haque}}}]{2016.EL.11540011V}%
  \BibitemOpen
  \bibfield  {author} {\bibinfo {author} {\bibfnamefont {Jacopo}\ \bibnamefont {{Viti}}}, \bibinfo {author} {\bibfnamefont {Jean-Marie}\ \bibnamefont {{St{\'e}phan}}}, \bibinfo {author} {\bibfnamefont {J{\'e}r{\^o}me}\ \bibnamefont {{Dubail}}}, \ and\ \bibinfo {author} {\bibfnamefont {Masudul}\ \bibnamefont {{Haque}}},\ }\bibfield  {title} {\enquote {\bibinfo {title} {{Inhomogeneous quenches in a free fermionic chain: Exact results}},}\ }\href {\doibase 10.1209/0295-5075/115/40011} {\bibfield  {journal} {\bibinfo  {journal} {EPL (Europhysics Letters)}\ }\textbf {\bibinfo {volume} {115}},\ \bibinfo {pages} {40011} (\bibinfo {year} {2016})}\BibitemShut {NoStop}%
\bibitem [{\citenamefont {{Allegra}}\ \emph {et~al.}(2016)\citenamefont {{Allegra}}, \citenamefont {{Dubail}}, \citenamefont {{St{\'e}phan}},\ and\ \citenamefont {{Viti}}}]{2016.JSMTE.05.3108A}%
  \BibitemOpen
  \bibfield  {author} {\bibinfo {author} {\bibfnamefont {Nicolas}\ \bibnamefont {{Allegra}}}, \bibinfo {author} {\bibfnamefont {J{\'e}r{\^o}me}\ \bibnamefont {{Dubail}}}, \bibinfo {author} {\bibfnamefont {Jean-Marie}\ \bibnamefont {{St{\'e}phan}}}, \ and\ \bibinfo {author} {\bibfnamefont {Jacopo}\ \bibnamefont {{Viti}}},\ }\bibfield  {title} {\enquote {\bibinfo {title} {{Inhomogeneous field theory inside the arctic circle}},}\ }\href {\doibase 10.1088/1742-5468/2016/05/053108} {\bibfield  {journal} {\bibinfo  {journal} {Journal of Statistical Mechanics: Theory and Experiment}\ }\textbf {\bibinfo {volume} {5}},\ \bibinfo {pages} {053108} (\bibinfo {year} {2016})}\BibitemShut {NoStop}%
\bibitem [{\citenamefont {{Dubail}}\ \emph {et~al.}(2017)\citenamefont {{Dubail}}, \citenamefont {{St{\'e}phan}}, \citenamefont {{Viti}},\ and\ \citenamefont {{Calabrese}}}]{2017.ScPP.2.2D}%
  \BibitemOpen
  \bibfield  {author} {\bibinfo {author} {\bibfnamefont {Jerome}\ \bibnamefont {{Dubail}}}, \bibinfo {author} {\bibfnamefont {Jean-Marie}\ \bibnamefont {{St{\'e}phan}}}, \bibinfo {author} {\bibfnamefont {Jacopo}\ \bibnamefont {{Viti}}}, \ and\ \bibinfo {author} {\bibfnamefont {Pasquale}\ \bibnamefont {{Calabrese}}},\ }\bibfield  {title} {\enquote {\bibinfo {title} {{Conformal field theory for inhomogeneous one-dimensional quantum systems: the example of non-interacting Fermi gases}},}\ }\href {\doibase 10.21468/SciPostPhys.2.1.002} {\bibfield  {journal} {\bibinfo  {journal} {SciPost Physics}\ }\textbf {\bibinfo {volume} {2}},\ \bibinfo {eid} {002} (\bibinfo {year} {2017})}\BibitemShut {NoStop}%
\bibitem [{\citenamefont {Bondyopadhaya}\ and\ \citenamefont {Roy}(2019)}]{2019.PhysRevB.99.214514}%
  \BibitemOpen
  \bibfield  {author} {\bibinfo {author} {\bibfnamefont {Nilanjan}\ \bibnamefont {Bondyopadhaya}}\ and\ \bibinfo {author} {\bibfnamefont {Dibyendu}\ \bibnamefont {Roy}},\ }\bibfield  {title} {\enquote {\bibinfo {title} {Dynamics of hybrid junctions of majorana wires},}\ }\href {\doibase 10.1103/PhysRevB.99.214514} {\bibfield  {journal} {\bibinfo  {journal} {Phys. Rev. B}\ }\textbf {\bibinfo {volume} {99}},\ \bibinfo {pages} {214514} (\bibinfo {year} {2019})}\BibitemShut {NoStop}%
\bibitem [{\citenamefont {{Biella}}\ \emph {et~al.}(2019)\citenamefont {{Biella}}, \citenamefont {{Collura}}, \citenamefont {{Rossini}}, \citenamefont {{De Luca}},\ and\ \citenamefont {{Mazza}}}]{2019.NatCo.10.4820B}%
  \BibitemOpen
  \bibfield  {author} {\bibinfo {author} {\bibfnamefont {Alberto}\ \bibnamefont {{Biella}}}, \bibinfo {author} {\bibfnamefont {Mario}\ \bibnamefont {{Collura}}}, \bibinfo {author} {\bibfnamefont {Davide}\ \bibnamefont {{Rossini}}}, \bibinfo {author} {\bibfnamefont {Andrea}\ \bibnamefont {{De Luca}}}, \ and\ \bibinfo {author} {\bibfnamefont {Leonardo}\ \bibnamefont {{Mazza}}},\ }\bibfield  {title} {\enquote {\bibinfo {title} {{Ballistic transport and boundary resistances in inhomogeneous quantum spin chains}},}\ }\href {\doibase 10.1038/s41467-019-12784-4} {\bibfield  {journal} {\bibinfo  {journal} {Nature Communications}\ }\textbf {\bibinfo {volume} {10}},\ \bibinfo {eid} {4820} (\bibinfo {year} {2019})}\BibitemShut {NoStop}%
\bibitem [{\citenamefont {{Gaw{\`E}{\textcopyright}dzki}}\ and\ \citenamefont {{Koz{\l}owski}}(2020)}]{2020.CMaPh.377.1227G}%
  \BibitemOpen
  \bibfield  {author} {\bibinfo {author} {\bibfnamefont {Krzysztof}\ \bibnamefont {{Gaw{\`E}{\textcopyright}dzki}}}\ and\ \bibinfo {author} {\bibfnamefont {Karol~K.}\ \bibnamefont {{Koz{\l}owski}}},\ }\bibfield  {title} {\enquote {\bibinfo {title} {{Full Counting Statistics of Energy Transfers in Inhomogeneous Nonequilibrium States of (1 +1 )D CFT}},}\ }\href {\doibase 10.1007/s00220-020-03774-5} {\bibfield  {journal} {\bibinfo  {journal} {Communications in Mathematical Physics}\ }\textbf {\bibinfo {volume} {377}},\ \bibinfo {pages} {1227--1309} (\bibinfo {year} {2020})}\BibitemShut {NoStop}%
\bibitem [{\citenamefont {{del Vecchio del Vecchio}}\ \emph {et~al.}(2022)\citenamefont {{del Vecchio del Vecchio}}, \citenamefont {{De Luca}},\ and\ \citenamefont {{Bastianello}}}]{2022.ScPP.12.60D}%
  \BibitemOpen
  \bibfield  {author} {\bibinfo {author} {\bibfnamefont {Giuseppe}\ \bibnamefont {{del Vecchio del Vecchio}}}, \bibinfo {author} {\bibfnamefont {Andrea}\ \bibnamefont {{De Luca}}}, \ and\ \bibinfo {author} {\bibfnamefont {Alvise}\ \bibnamefont {{Bastianello}}},\ }\bibfield  {title} {\enquote {\bibinfo {title} {{Transport through interacting defects and lack of thermalisation}},}\ }\href {\doibase 10.21468/SciPostPhys.12.2.060} {\bibfield  {journal} {\bibinfo  {journal} {SciPost Physics}\ }\textbf {\bibinfo {volume} {12}},\ \bibinfo {eid} {060} (\bibinfo {year} {2022})}\BibitemShut {NoStop}%
\bibitem [{\citenamefont {Ljubotina}\ \emph {et~al.}(2022)\citenamefont {Ljubotina}, \citenamefont {Roy},\ and\ \citenamefont {Prosen}}]{2022.PhysRevB.106.054314}%
  \BibitemOpen
  \bibfield  {author} {\bibinfo {author} {\bibfnamefont {Marko}\ \bibnamefont {Ljubotina}}, \bibinfo {author} {\bibfnamefont {Dibyendu}\ \bibnamefont {Roy}}, \ and\ \bibinfo {author} {\bibfnamefont {Toma\ifmmode \check{z}\else~\v{z}\fi{}}\ \bibnamefont {Prosen}},\ }\bibfield  {title} {\enquote {\bibinfo {title} {Absence of thermalization of free systems coupled to gapped interacting reservoirs},}\ }\href {\doibase 10.1103/PhysRevB.106.054314} {\bibfield  {journal} {\bibinfo  {journal} {Phys. Rev. B}\ }\textbf {\bibinfo {volume} {106}},\ \bibinfo {pages} {054314} (\bibinfo {year} {2022})}\BibitemShut {NoStop}%
\bibitem [{\citenamefont {{Wang}}\ \emph {et~al.}(2024)\citenamefont {{Wang}}, \citenamefont {{Zhang}}, \citenamefont {{Lewenstein}},\ and\ \citenamefont {{Ran}}}]{2024.QST.9a5008W}%
  \BibitemOpen
  \bibfield  {author} {\bibinfo {author} {\bibfnamefont {Ding-Zu}\ \bibnamefont {{Wang}}}, \bibinfo {author} {\bibfnamefont {Guo-Feng}\ \bibnamefont {{Zhang}}}, \bibinfo {author} {\bibfnamefont {Maciej}\ \bibnamefont {{Lewenstein}}}, \ and\ \bibinfo {author} {\bibfnamefont {Shi-Ju}\ \bibnamefont {{Ran}}},\ }\bibfield  {title} {\enquote {\bibinfo {title} {{Boundary-induced singularity in strongly-correlated quantum systems at finite temperature}},}\ }\href {\doibase 10.1088/2058-9565/ad038a} {\bibfield  {journal} {\bibinfo  {journal} {Quantum Science and Technology}\ }\textbf {\bibinfo {volume} {9}},\ \bibinfo {eid} {015008} (\bibinfo {year} {2024})}\BibitemShut {NoStop}%
\bibitem [{\citenamefont {Bastianello}\ \emph {et~al.}(2019)\citenamefont {Bastianello}, \citenamefont {Alba},\ and\ \citenamefont {Caux}}]{2019.PhysRevLett.123.130602}%
  \BibitemOpen
  \bibfield  {author} {\bibinfo {author} {\bibfnamefont {Alvise}\ \bibnamefont {Bastianello}}, \bibinfo {author} {\bibfnamefont {Vincenzo}\ \bibnamefont {Alba}}, \ and\ \bibinfo {author} {\bibfnamefont {Jean-S\'ebastien}\ \bibnamefont {Caux}},\ }\bibfield  {title} {\enquote {\bibinfo {title} {Generalized hydrodynamics with space-time inhomogeneous interactions},}\ }\href {\doibase 10.1103/PhysRevLett.123.130602} {\bibfield  {journal} {\bibinfo  {journal} {Phys. Rev. Lett.}\ }\textbf {\bibinfo {volume} {123}},\ \bibinfo {pages} {130602} (\bibinfo {year} {2019})}\BibitemShut {NoStop}%
\bibitem [{\citenamefont {Collura}\ \emph {et~al.}(2020)\citenamefont {Collura}, \citenamefont {De~Luca}, \citenamefont {Calabrese},\ and\ \citenamefont {Dubail}}]{2020.PhysRevB.102.180409}%
  \BibitemOpen
  \bibfield  {author} {\bibinfo {author} {\bibfnamefont {Mario}\ \bibnamefont {Collura}}, \bibinfo {author} {\bibfnamefont {Andrea}\ \bibnamefont {De~Luca}}, \bibinfo {author} {\bibfnamefont {Pasquale}\ \bibnamefont {Calabrese}}, \ and\ \bibinfo {author} {\bibfnamefont {J\'er\^ome}\ \bibnamefont {Dubail}},\ }\bibfield  {title} {\enquote {\bibinfo {title} {Domain wall melting in the spin-$\frac{1}{2}$ xxz spin chain: Emergent luttinger liquid with a fractal quasiparticle charge},}\ }\href {\doibase 10.1103/PhysRevB.102.180409} {\bibfield  {journal} {\bibinfo  {journal} {Phys. Rev. B}\ }\textbf {\bibinfo {volume} {102}},\ \bibinfo {pages} {180409(R)} (\bibinfo {year} {2020})}\BibitemShut {NoStop}%
\bibitem [{\citenamefont {Koch}\ \emph {et~al.}(2021)\citenamefont {Koch}, \citenamefont {Bastianello},\ and\ \citenamefont {Caux}}]{2021.PhysRevB.103.165121}%
  \BibitemOpen
  \bibfield  {author} {\bibinfo {author} {\bibfnamefont {Rebekka}\ \bibnamefont {Koch}}, \bibinfo {author} {\bibfnamefont {Alvise}\ \bibnamefont {Bastianello}}, \ and\ \bibinfo {author} {\bibfnamefont {Jean-S\'ebastien}\ \bibnamefont {Caux}},\ }\bibfield  {title} {\enquote {\bibinfo {title} {Adiabatic formation of bound states in the one-dimensional bose gas},}\ }\href {\doibase 10.1103/PhysRevB.103.165121} {\bibfield  {journal} {\bibinfo  {journal} {Phys. Rev. B}\ }\textbf {\bibinfo {volume} {103}},\ \bibinfo {pages} {165121} (\bibinfo {year} {2021})}\BibitemShut {NoStop}%
\bibitem [{\citenamefont {{Durnin}}\ \emph {et~al.}(2021)\citenamefont {{Durnin}}, \citenamefont {{De Luca}}, \citenamefont {{De Nardis}},\ and\ \citenamefont {{Doyon}}}]{2021.JPhA.54W4001D}%
  \BibitemOpen
  \bibfield  {author} {\bibinfo {author} {\bibfnamefont {Joseph}\ \bibnamefont {{Durnin}}}, \bibinfo {author} {\bibfnamefont {Andrea}\ \bibnamefont {{De Luca}}}, \bibinfo {author} {\bibfnamefont {Jacopo}\ \bibnamefont {{De Nardis}}}, \ and\ \bibinfo {author} {\bibfnamefont {Benjamin}\ \bibnamefont {{Doyon}}},\ }\bibfield  {title} {\enquote {\bibinfo {title} {{Diffusive hydrodynamics of inhomogenous Hamiltonians}},}\ }\href {\doibase 10.1088/1751-8121/ac2c57} {\bibfield  {journal} {\bibinfo  {journal} {Journal of Physics A Mathematical General}\ }\textbf {\bibinfo {volume} {54}},\ \bibinfo {eid} {494001} (\bibinfo {year} {2021})}\BibitemShut {NoStop}%
\bibitem [{\citenamefont {{Alba}}\ \emph {et~al.}(2021)\citenamefont {{Alba}}, \citenamefont {{Bertini}}, \citenamefont {{Fagotti}}, \citenamefont {{Piroli}},\ and\ \citenamefont {{Ruggiero}}}]{2021.JSMTE2021k4004A}%
  \BibitemOpen
  \bibfield  {author} {\bibinfo {author} {\bibfnamefont {Vincenzo}\ \bibnamefont {{Alba}}}, \bibinfo {author} {\bibfnamefont {Bruno}\ \bibnamefont {{Bertini}}}, \bibinfo {author} {\bibfnamefont {Maurizio}\ \bibnamefont {{Fagotti}}}, \bibinfo {author} {\bibfnamefont {Lorenzo}\ \bibnamefont {{Piroli}}}, \ and\ \bibinfo {author} {\bibfnamefont {Paola}\ \bibnamefont {{Ruggiero}}},\ }\bibfield  {title} {\enquote {\bibinfo {title} {{Generalized-hydrodynamic approach to inhomogeneous quenches: correlations, entanglement and quantum effects}},}\ }\href {\doibase 10.1088/1742-5468/ac257d} {\bibfield  {journal} {\bibinfo  {journal} {Journal of Statistical Mechanics: Theory and Experiment}\ }\textbf {\bibinfo {volume} {2021}},\ \bibinfo {eid} {114004} (\bibinfo {year} {2021})}\BibitemShut {NoStop}%
\bibitem [{\citenamefont {{Bastianello}}\ \emph {et~al.}(2021)\citenamefont {{Bastianello}}, \citenamefont {{De Luca}},\ and\ \citenamefont {{Vasseur}}}]{2021.JSMTE2021k4003B}%
  \BibitemOpen
  \bibfield  {author} {\bibinfo {author} {\bibfnamefont {Alvise}\ \bibnamefont {{Bastianello}}}, \bibinfo {author} {\bibfnamefont {Andrea}\ \bibnamefont {{De Luca}}}, \ and\ \bibinfo {author} {\bibfnamefont {Romain}\ \bibnamefont {{Vasseur}}},\ }\bibfield  {title} {\enquote {\bibinfo {title} {{Hydrodynamics of weak integrability breaking}},}\ }\href {\doibase 10.1088/1742-5468/ac26b2} {\bibfield  {journal} {\bibinfo  {journal} {Journal of Statistical Mechanics: Theory and Experiment}\ }\textbf {\bibinfo {volume} {2021}},\ \bibinfo {eid} {114003} (\bibinfo {year} {2021})}\BibitemShut {NoStop}%
\bibitem [{\citenamefont {Shastry}\ and\ \citenamefont {Sutherland}(1990)}]{1990.PhysRevLett.65.243}%
  \BibitemOpen
  \bibfield  {author} {\bibinfo {author} {\bibfnamefont {B.~S.}\ \bibnamefont {Shastry}}\ and\ \bibinfo {author} {\bibfnamefont {Bill}\ \bibnamefont {Sutherland}},\ }\bibfield  {title} {\enquote {\bibinfo {title} {Twisted boundary conditions and effective mass in heisenberg-ising and hubbard rings},}\ }\href {\doibase 10.1103/PhysRevLett.65.243} {\bibfield  {journal} {\bibinfo  {journal} {Phys. Rev. Lett.}\ }\textbf {\bibinfo {volume} {65}},\ \bibinfo {pages} {243--246} (\bibinfo {year} {1990})}\BibitemShut {NoStop}%
\bibitem [{\citenamefont {Cazalilla}\ \emph {et~al.}(2011)\citenamefont {Cazalilla}, \citenamefont {Citro}, \citenamefont {Giamarchi}, \citenamefont {Orignac},\ and\ \citenamefont {Rigol}}]{2011.RevModPhys.83.1405}%
  \BibitemOpen
  \bibfield  {author} {\bibinfo {author} {\bibfnamefont {M.~A.}\ \bibnamefont {Cazalilla}}, \bibinfo {author} {\bibfnamefont {R.}~\bibnamefont {Citro}}, \bibinfo {author} {\bibfnamefont {T.}~\bibnamefont {Giamarchi}}, \bibinfo {author} {\bibfnamefont {E.}~\bibnamefont {Orignac}}, \ and\ \bibinfo {author} {\bibfnamefont {M.}~\bibnamefont {Rigol}},\ }\bibfield  {title} {\enquote {\bibinfo {title} {One dimensional bosons: From condensed matter systems to ultracold gases},}\ }\href {\doibase 10.1103/RevModPhys.83.1405} {\bibfield  {journal} {\bibinfo  {journal} {Rev. Mod. Phys.}\ }\textbf {\bibinfo {volume} {83}},\ \bibinfo {pages} {1405--1466} (\bibinfo {year} {2011})}\BibitemShut {NoStop}%
\bibitem [{\citenamefont {Brody}\ \emph {et~al.}(1981)\citenamefont {Brody}, \citenamefont {Flores}, \citenamefont {French}, \citenamefont {Mello}, \citenamefont {Pandey},\ and\ \citenamefont {Wong}}]{1981.RevModPhys.53.385}%
  \BibitemOpen
  \bibfield  {author} {\bibinfo {author} {\bibfnamefont {T.~A.}\ \bibnamefont {Brody}}, \bibinfo {author} {\bibfnamefont {J.}~\bibnamefont {Flores}}, \bibinfo {author} {\bibfnamefont {J.~B.}\ \bibnamefont {French}}, \bibinfo {author} {\bibfnamefont {P.~A.}\ \bibnamefont {Mello}}, \bibinfo {author} {\bibfnamefont {A.}~\bibnamefont {Pandey}}, \ and\ \bibinfo {author} {\bibfnamefont {S.~S.~M.}\ \bibnamefont {Wong}},\ }\bibfield  {title} {\enquote {\bibinfo {title} {Random-matrix physics: spectrum and strength fluctuations},}\ }\href {\doibase 10.1103/RevModPhys.53.385} {\bibfield  {journal} {\bibinfo  {journal} {Rev. Mod. Phys.}\ }\textbf {\bibinfo {volume} {53}},\ \bibinfo {pages} {385--479} (\bibinfo {year} {1981})}\BibitemShut {NoStop}%
\bibitem [{\citenamefont {Santos}\ and\ \citenamefont {Rigol}(2010)}]{2010.PhysRevE.81.036206}%
  \BibitemOpen
  \bibfield  {author} {\bibinfo {author} {\bibfnamefont {Lea~F.}\ \bibnamefont {Santos}}\ and\ \bibinfo {author} {\bibfnamefont {Marcos}\ \bibnamefont {Rigol}},\ }\bibfield  {title} {\enquote {\bibinfo {title} {Onset of quantum chaos in one-dimensional bosonic and fermionic systems and its relation to thermalization},}\ }\href {\doibase 10.1103/PhysRevE.81.036206} {\bibfield  {journal} {\bibinfo  {journal} {Phys. Rev. E}\ }\textbf {\bibinfo {volume} {81}},\ \bibinfo {pages} {036206} (\bibinfo {year} {2010})}\BibitemShut {NoStop}%
\bibitem [{\citenamefont {Sierant}\ and\ \citenamefont {Zakrzewski}(2019)}]{PhysRevB.99.104205}%
  \BibitemOpen
  \bibfield  {author} {\bibinfo {author} {\bibfnamefont {Piotr}\ \bibnamefont {Sierant}}\ and\ \bibinfo {author} {\bibfnamefont {Jakub}\ \bibnamefont {Zakrzewski}},\ }\bibfield  {title} {\enquote {\bibinfo {title} {Level statistics across the many-body localization transition},}\ }\href {\doibase 10.1103/PhysRevB.99.104205} {\bibfield  {journal} {\bibinfo  {journal} {Phys. Rev. B}\ }\textbf {\bibinfo {volume} {99}},\ \bibinfo {pages} {104205} (\bibinfo {year} {2019})}\BibitemShut {NoStop}%
\bibitem [{\citenamefont {Pandey}\ and\ \citenamefont {Ramaswamy}(1991)}]{1991.PhysRevA.43.4237}%
  \BibitemOpen
  \bibfield  {author} {\bibinfo {author} {\bibfnamefont {Akhilesh}\ \bibnamefont {Pandey}}\ and\ \bibinfo {author} {\bibfnamefont {Ramakrishna}\ \bibnamefont {Ramaswamy}},\ }\bibfield  {title} {\enquote {\bibinfo {title} {Level spacings for harmonic-oscillator systems},}\ }\href {\doibase 10.1103/PhysRevA.43.4237} {\bibfield  {journal} {\bibinfo  {journal} {Phys. Rev. A}\ }\textbf {\bibinfo {volume} {43}},\ \bibinfo {pages} {4237--4243} (\bibinfo {year} {1991})}\BibitemShut {NoStop}%
\bibitem [{\citenamefont {Zangara}\ \emph {et~al.}(2013)\citenamefont {Zangara}, \citenamefont {Dente}, \citenamefont {Torres-Herrera}, \citenamefont {Pastawski}, \citenamefont {Iucci},\ and\ \citenamefont {Santos}}]{2013.PhysRevE.88.032913}%
  \BibitemOpen
  \bibfield  {author} {\bibinfo {author} {\bibfnamefont {Pablo~R.}\ \bibnamefont {Zangara}}, \bibinfo {author} {\bibfnamefont {Axel~D.}\ \bibnamefont {Dente}}, \bibinfo {author} {\bibfnamefont {E.~J.}\ \bibnamefont {Torres-Herrera}}, \bibinfo {author} {\bibfnamefont {Horacio~M.}\ \bibnamefont {Pastawski}}, \bibinfo {author} {\bibfnamefont {An\'{\i}bal}\ \bibnamefont {Iucci}}, \ and\ \bibinfo {author} {\bibfnamefont {Lea~F.}\ \bibnamefont {Santos}},\ }\bibfield  {title} {\enquote {\bibinfo {title} {Time fluctuations in isolated quantum systems of interacting particles},}\ }\href {\doibase 10.1103/PhysRevE.88.032913} {\bibfield  {journal} {\bibinfo  {journal} {Phys. Rev. E}\ }\textbf {\bibinfo {volume} {88}},\ \bibinfo {pages} {032913} (\bibinfo {year} {2013})}\BibitemShut {NoStop}%
\bibitem [{\citenamefont {{Garcia-March}}\ \emph {et~al.}(2018)\citenamefont {{Garcia-March}}, \citenamefont {{van Frank}}, \citenamefont {{Bonneau}}, \citenamefont {{Schmiedmayer}}, \citenamefont {{Lewenstein}},\ and\ \citenamefont {{Santos}}}]{2018.NJPh.20k3039G}%
  \BibitemOpen
  \bibfield  {author} {\bibinfo {author} {\bibfnamefont {M.~A.}\ \bibnamefont {{Garcia-March}}}, \bibinfo {author} {\bibfnamefont {S.}~\bibnamefont {{van Frank}}}, \bibinfo {author} {\bibfnamefont {M.}~\bibnamefont {{Bonneau}}}, \bibinfo {author} {\bibfnamefont {J.}~\bibnamefont {{Schmiedmayer}}}, \bibinfo {author} {\bibfnamefont {M.}~\bibnamefont {{Lewenstein}}}, \ and\ \bibinfo {author} {\bibfnamefont {Lea~F.}\ \bibnamefont {{Santos}}},\ }\bibfield  {title} {\enquote {\bibinfo {title} {{Relaxation, chaos, and thermalization in a three-mode model of a Bose-Einstein condensate}},}\ }\href {\doibase 10.1088/1367-2630/aaed68} {\bibfield  {journal} {\bibinfo  {journal} {New Journal of Physics}\ }\textbf {\bibinfo {volume} {20}},\ \bibinfo {eid} {113039} (\bibinfo {year} {2018})}\BibitemShut {NoStop}%
\bibitem [{\citenamefont {{Guhr}}\ \emph {et~al.}(1998)\citenamefont {{Guhr}}, \citenamefont {{M{\"u}ller-Groeling}},\ and\ \citenamefont {{Weidenm{\"u}ller}}}]{1998.PhR.299.189G}%
  \BibitemOpen
  \bibfield  {author} {\bibinfo {author} {\bibfnamefont {Thomas}\ \bibnamefont {{Guhr}}}, \bibinfo {author} {\bibfnamefont {Axel}\ \bibnamefont {{M{\"u}ller-Groeling}}}, \ and\ \bibinfo {author} {\bibfnamefont {Hans~A.}\ \bibnamefont {{Weidenm{\"u}ller}}},\ }\bibfield  {title} {\enquote {\bibinfo {title} {{Random-matrix theories in quantum physics: common concepts}},}\ }\href {\doibase 10.1016/S0370-1573(97)00088-4} {\bibfield  {journal} {\bibinfo  {journal} {Physics Reports}\ }\textbf {\bibinfo {volume} {299}},\ \bibinfo {pages} {189--425} (\bibinfo {year} {1998})}\BibitemShut {NoStop}%
\bibitem [{\citenamefont {Oganesyan}\ and\ \citenamefont {Huse}(2007)}]{Oganesyan07}%
  \BibitemOpen
  \bibfield  {author} {\bibinfo {author} {\bibfnamefont {Vadim}\ \bibnamefont {Oganesyan}}\ and\ \bibinfo {author} {\bibfnamefont {David~A.}\ \bibnamefont {Huse}},\ }\bibfield  {title} {\enquote {\bibinfo {title} {Localization of interacting fermions at high temperature},}\ }\href {\doibase 10.1103/PhysRevB.75.155111} {\bibfield  {journal} {\bibinfo  {journal} {Phys. Rev. B}\ }\textbf {\bibinfo {volume} {75}},\ \bibinfo {pages} {155111} (\bibinfo {year} {2007})}\BibitemShut {NoStop}%
\bibitem [{\citenamefont {Atas}\ \emph {et~al.}(2013)\citenamefont {Atas}, \citenamefont {Bogomolny}, \citenamefont {Giraud},\ and\ \citenamefont {Roux}}]{2013.PhysRevLett.110.084101}%
  \BibitemOpen
  \bibfield  {author} {\bibinfo {author} {\bibfnamefont {Y.~Y.}\ \bibnamefont {Atas}}, \bibinfo {author} {\bibfnamefont {E.}~\bibnamefont {Bogomolny}}, \bibinfo {author} {\bibfnamefont {O.}~\bibnamefont {Giraud}}, \ and\ \bibinfo {author} {\bibfnamefont {G.}~\bibnamefont {Roux}},\ }\bibfield  {title} {\enquote {\bibinfo {title} {Distribution of the ratio of consecutive level spacings in random matrix ensembles},}\ }\href {\doibase 10.1103/PhysRevLett.110.084101} {\bibfield  {journal} {\bibinfo  {journal} {Phys. Rev. Lett.}\ }\textbf {\bibinfo {volume} {110}},\ \bibinfo {pages} {084101} (\bibinfo {year} {2013})}\BibitemShut {NoStop}%
\bibitem [{\citenamefont {Luitz}\ \emph {et~al.}(2015)\citenamefont {Luitz}, \citenamefont {Laflorencie},\ and\ \citenamefont {Alet}}]{Luitz15}%
  \BibitemOpen
  \bibfield  {author} {\bibinfo {author} {\bibfnamefont {David~J.}\ \bibnamefont {Luitz}}, \bibinfo {author} {\bibfnamefont {Nicolas}\ \bibnamefont {Laflorencie}}, \ and\ \bibinfo {author} {\bibfnamefont {Fabien}\ \bibnamefont {Alet}},\ }\bibfield  {title} {\enquote {\bibinfo {title} {Many-body localization edge in the random-field heisenberg chain},}\ }\href {\doibase 10.1103/PhysRevB.91.081103} {\bibfield  {journal} {\bibinfo  {journal} {Phys. Rev. B}\ }\textbf {\bibinfo {volume} {91}},\ \bibinfo {pages} {081103(R)} (\bibinfo {year} {2015})}\BibitemShut {NoStop}%
\bibitem [{\citenamefont {Sierant}\ \emph {et~al.}(2020{\natexlab{a}})\citenamefont {Sierant}, \citenamefont {Lewenstein},\ and\ \citenamefont {Zakrzewski}}]{Sierant20poly}%
  \BibitemOpen
  \bibfield  {author} {\bibinfo {author} {\bibfnamefont {Piotr}\ \bibnamefont {Sierant}}, \bibinfo {author} {\bibfnamefont {Maciej}\ \bibnamefont {Lewenstein}}, \ and\ \bibinfo {author} {\bibfnamefont {Jakub}\ \bibnamefont {Zakrzewski}},\ }\bibfield  {title} {\enquote {\bibinfo {title} {Polynomially filtered exact diagonalization approach to many-body localization},}\ }\href {\doibase 10.1103/PhysRevLett.125.156601} {\bibfield  {journal} {\bibinfo  {journal} {Phys. Rev. Lett.}\ }\textbf {\bibinfo {volume} {125}},\ \bibinfo {pages} {156601} (\bibinfo {year} {2020}{\natexlab{a}})}\BibitemShut {NoStop}%
\bibitem [{\citenamefont {\v{S}untajs}\ \emph {et~al.}(2020)\citenamefont {\v{S}untajs}, \citenamefont {Bon\v{c}a}, \citenamefont {Prosen},\ and\ \citenamefont {Vidmar}}]{Suntajs20}%
  \BibitemOpen
  \bibfield  {author} {\bibinfo {author} {\bibfnamefont {Jan}\ \bibnamefont {\v{S}untajs}}, \bibinfo {author} {\bibfnamefont {Janez}\ \bibnamefont {Bon\v{c}a}}, \bibinfo {author} {\bibfnamefont {Toma\v{z}}\ \bibnamefont {Prosen}}, \ and\ \bibinfo {author} {\bibfnamefont {Lev}\ \bibnamefont {Vidmar}},\ }\bibfield  {title} {\enquote {\bibinfo {title} {Quantum chaos challenges many-body localization},}\ }\href {\doibase 10.1103/PhysRevE.102.062144} {\bibfield  {journal} {\bibinfo  {journal} {Phys. Rev. E}\ }\textbf {\bibinfo {volume} {102}},\ \bibinfo {pages} {062144} (\bibinfo {year} {2020})}\BibitemShut {NoStop}%
\bibitem [{\citenamefont {Sierant}\ \emph {et~al.}(2020{\natexlab{b}})\citenamefont {Sierant}, \citenamefont {Delande},\ and\ \citenamefont {Zakrzewski}}]{Sierant20thou}%
  \BibitemOpen
  \bibfield  {author} {\bibinfo {author} {\bibfnamefont {Piotr}\ \bibnamefont {Sierant}}, \bibinfo {author} {\bibfnamefont {Dominique}\ \bibnamefont {Delande}}, \ and\ \bibinfo {author} {\bibfnamefont {Jakub}\ \bibnamefont {Zakrzewski}},\ }\bibfield  {title} {\enquote {\bibinfo {title} {Thouless time analysis of anderson and many-body localization transitions},}\ }\href {\doibase 10.1103/PhysRevLett.124.186601} {\bibfield  {journal} {\bibinfo  {journal} {Phys. Rev. Lett.}\ }\textbf {\bibinfo {volume} {124}},\ \bibinfo {pages} {186601} (\bibinfo {year} {2020}{\natexlab{b}})}\BibitemShut {NoStop}%
\bibitem [{\citenamefont {Abanin}\ \emph {et~al.}(2021)\citenamefont {Abanin}, \citenamefont {Bardarson}, \citenamefont {{De Tomasi}}, \citenamefont {Gopalakrishnan}, \citenamefont {Khemani}, \citenamefont {Parameswaran}, \citenamefont {Pollmann}, \citenamefont {Potter}, \citenamefont {Serbyn},\ and\ \citenamefont {Vasseur}}]{Abanin21}%
  \BibitemOpen
  \bibfield  {author} {\bibinfo {author} {\bibfnamefont {D.A.}\ \bibnamefont {Abanin}}, \bibinfo {author} {\bibfnamefont {J.H.}\ \bibnamefont {Bardarson}}, \bibinfo {author} {\bibfnamefont {G.}~\bibnamefont {{De Tomasi}}}, \bibinfo {author} {\bibfnamefont {S.}~\bibnamefont {Gopalakrishnan}}, \bibinfo {author} {\bibfnamefont {V.}~\bibnamefont {Khemani}}, \bibinfo {author} {\bibfnamefont {S.A.}\ \bibnamefont {Parameswaran}}, \bibinfo {author} {\bibfnamefont {F.}~\bibnamefont {Pollmann}}, \bibinfo {author} {\bibfnamefont {A.C.}\ \bibnamefont {Potter}}, \bibinfo {author} {\bibfnamefont {M.}~\bibnamefont {Serbyn}}, \ and\ \bibinfo {author} {\bibfnamefont {R.}~\bibnamefont {Vasseur}},\ }\bibfield  {title} {\enquote {\bibinfo {title} {Distinguishing localization from chaos: Challenges in finite-size systems},}\ }\href {\doibase https://doi.org/10.1016/j.aop.2021.168415} {\bibfield  {journal} {\bibinfo  {journal} {Annals of Physics}\ }\textbf {\bibinfo {volume} {427}},\ \bibinfo {pages} {168415} (\bibinfo {year} {2021})}\BibitemShut {NoStop}%
\bibitem [{\citenamefont {{Kiefer-Emmanouilidis, M.}}\ \emph {et~al.}(2020)\citenamefont {{Kiefer-Emmanouilidis, M.}}, \citenamefont {Unanyan}, \citenamefont {Fleischhauer},\ and\ \citenamefont {Sirker}}]{Kiefer20}%
  \BibitemOpen
  \bibfield  {author} {\bibinfo {author} {\bibnamefont {{Kiefer-Emmanouilidis, M.}}}, \bibinfo {author} {\bibfnamefont {Razmik}\ \bibnamefont {Unanyan}}, \bibinfo {author} {\bibfnamefont {Michael}\ \bibnamefont {Fleischhauer}}, \ and\ \bibinfo {author} {\bibfnamefont {Jesko}\ \bibnamefont {Sirker}},\ }\bibfield  {title} {\enquote {\bibinfo {title} {Evidence for unbounded growth of the number entropy in many-body localized phases},}\ }\href {\doibase 10.1103/PhysRevLett.124.243601} {\bibfield  {journal} {\bibinfo  {journal} {Phys. Rev. Lett.}\ }\textbf {\bibinfo {volume} {124}},\ \bibinfo {pages} {243601} (\bibinfo {year} {2020})}\BibitemShut {NoStop}%
\bibitem [{\citenamefont {Panda}\ \emph {et~al.}(2020)\citenamefont {Panda}, \citenamefont {Scardicchio}, \citenamefont {Schulz}, \citenamefont {Taylor},\ and\ \citenamefont {Žnidarič}}]{Panda19}%
  \BibitemOpen
  \bibfield  {author} {\bibinfo {author} {\bibfnamefont {R.~K.}\ \bibnamefont {Panda}}, \bibinfo {author} {\bibfnamefont {A.}~\bibnamefont {Scardicchio}}, \bibinfo {author} {\bibfnamefont {M.}~\bibnamefont {Schulz}}, \bibinfo {author} {\bibfnamefont {S.~R.}\ \bibnamefont {Taylor}}, \ and\ \bibinfo {author} {\bibfnamefont {M.}~\bibnamefont {Žnidarič}},\ }\bibfield  {title} {\enquote {\bibinfo {title} {Can we study the many-body localisation transition?}}\ }\href {\doibase 10.1209/0295-5075/128/67003} {\bibfield  {journal} {\bibinfo  {journal} {Europhysics Letters}\ }\textbf {\bibinfo {volume} {128}},\ \bibinfo {pages} {67003} (\bibinfo {year} {2020})}\BibitemShut {NoStop}%
\bibitem [{\citenamefont {Sierant}\ \emph {et~al.}(2021)\citenamefont {Sierant}, \citenamefont {Lazo}, \citenamefont {Dalmonte}, \citenamefont {Scardicchio},\ and\ \citenamefont {Zakrzewski}}]{Sierant21constr}%
  \BibitemOpen
  \bibfield  {author} {\bibinfo {author} {\bibfnamefont {Piotr}\ \bibnamefont {Sierant}}, \bibinfo {author} {\bibfnamefont {Eduardo~Gonzalez}\ \bibnamefont {Lazo}}, \bibinfo {author} {\bibfnamefont {Marcello}\ \bibnamefont {Dalmonte}}, \bibinfo {author} {\bibfnamefont {Antonello}\ \bibnamefont {Scardicchio}}, \ and\ \bibinfo {author} {\bibfnamefont {Jakub}\ \bibnamefont {Zakrzewski}},\ }\bibfield  {title} {\enquote {\bibinfo {title} {Constraint-induced delocalization},}\ }\href {\doibase 10.1103/PhysRevLett.127.126603} {\bibfield  {journal} {\bibinfo  {journal} {Phys. Rev. Lett.}\ }\textbf {\bibinfo {volume} {127}},\ \bibinfo {pages} {126603} (\bibinfo {year} {2021})}\BibitemShut {NoStop}%
\bibitem [{\citenamefont {{van Nieuwenburg}}\ \emph {et~al.}(2019)\citenamefont {{van Nieuwenburg}}, \citenamefont {{Baum}},\ and\ \citenamefont {{Refael}}}]{2019PNAS.116.9269V}%
  \BibitemOpen
  \bibfield  {author} {\bibinfo {author} {\bibfnamefont {Evert}\ \bibnamefont {{van Nieuwenburg}}}, \bibinfo {author} {\bibfnamefont {Yuval}\ \bibnamefont {{Baum}}}, \ and\ \bibinfo {author} {\bibfnamefont {Gil}\ \bibnamefont {{Refael}}},\ }\bibfield  {title} {\enquote {\bibinfo {title} {{From Bloch oscillations to many-body localization in clean interacting systems}},}\ }\href {\doibase 10.1073/pnas.1819316116} {\bibfield  {journal} {\bibinfo  {journal} {Proceedings of the National Academy of Science}\ }\textbf {\bibinfo {volume} {116}},\ \bibinfo {pages} {9269--9274} (\bibinfo {year} {2019})}\BibitemShut {NoStop}%
\bibitem [{\citenamefont {Schulz}\ \emph {et~al.}(2019)\citenamefont {Schulz}, \citenamefont {Hooley}, \citenamefont {Moessner},\ and\ \citenamefont {Pollmann}}]{2019.PhysRevLett.122.040606}%
  \BibitemOpen
  \bibfield  {author} {\bibinfo {author} {\bibfnamefont {M.}~\bibnamefont {Schulz}}, \bibinfo {author} {\bibfnamefont {C.~A.}\ \bibnamefont {Hooley}}, \bibinfo {author} {\bibfnamefont {R.}~\bibnamefont {Moessner}}, \ and\ \bibinfo {author} {\bibfnamefont {F.}~\bibnamefont {Pollmann}},\ }\bibfield  {title} {\enquote {\bibinfo {title} {Stark many-body localization},}\ }\href {\doibase 10.1103/PhysRevLett.122.040606} {\bibfield  {journal} {\bibinfo  {journal} {Phys. Rev. Lett.}\ }\textbf {\bibinfo {volume} {122}},\ \bibinfo {pages} {040606} (\bibinfo {year} {2019})}\BibitemShut {NoStop}%
\bibitem [{\citenamefont {Chanda}\ \emph {et~al.}(2020)\citenamefont {Chanda}, \citenamefont {Yao},\ and\ \citenamefont {Zakrzewski}}]{Chanda20}%
  \BibitemOpen
  \bibfield  {author} {\bibinfo {author} {\bibfnamefont {Titas}\ \bibnamefont {Chanda}}, \bibinfo {author} {\bibfnamefont {Ruixiao}\ \bibnamefont {Yao}}, \ and\ \bibinfo {author} {\bibfnamefont {Jakub}\ \bibnamefont {Zakrzewski}},\ }\bibfield  {title} {\enquote {\bibinfo {title} {Coexistence of localized and extended phases: Many-body localization in a harmonic trap},}\ }\href {\doibase 10.1103/PhysRevResearch.2.032039} {\bibfield  {journal} {\bibinfo  {journal} {Phys. Rev. Res.}\ }\textbf {\bibinfo {volume} {2}},\ \bibinfo {pages} {032039(R)} (\bibinfo {year} {2020})}\BibitemShut {NoStop}%
\bibitem [{\citenamefont {Yao}\ \emph {et~al.}(2021)\citenamefont {Yao}, \citenamefont {Chanda},\ and\ \citenamefont {Zakrzewski}}]{Yao21}%
  \BibitemOpen
  \bibfield  {author} {\bibinfo {author} {\bibfnamefont {Ruixiao}\ \bibnamefont {Yao}}, \bibinfo {author} {\bibfnamefont {Titas}\ \bibnamefont {Chanda}}, \ and\ \bibinfo {author} {\bibfnamefont {Jakub}\ \bibnamefont {Zakrzewski}},\ }\bibfield  {title} {\enquote {\bibinfo {title} {Many-body localization in tilted and harmonic potentials},}\ }\href {\doibase 10.1103/PhysRevB.104.014201} {\bibfield  {journal} {\bibinfo  {journal} {Phys. Rev. B}\ }\textbf {\bibinfo {volume} {104}},\ \bibinfo {pages} {014201} (\bibinfo {year} {2021})}\BibitemShut {NoStop}%
\bibitem [{\citenamefont {Bahovadinov}\ \emph {et~al.}(2022)\citenamefont {Bahovadinov}, \citenamefont {Kurlov}, \citenamefont {Matveenko}, \citenamefont {Altshuler},\ and\ \citenamefont {Shlyapnikov}}]{2022.PhysRevB.106.075107}%
  \BibitemOpen
  \bibfield  {author} {\bibinfo {author} {\bibfnamefont {M.~S.}\ \bibnamefont {Bahovadinov}}, \bibinfo {author} {\bibfnamefont {D.~V.}\ \bibnamefont {Kurlov}}, \bibinfo {author} {\bibfnamefont {S.~I.}\ \bibnamefont {Matveenko}}, \bibinfo {author} {\bibfnamefont {B.~L.}\ \bibnamefont {Altshuler}}, \ and\ \bibinfo {author} {\bibfnamefont {G.~V.}\ \bibnamefont {Shlyapnikov}},\ }\bibfield  {title} {\enquote {\bibinfo {title} {Many-body localization transition in a frustrated xy chain},}\ }\href {\doibase 10.1103/PhysRevB.106.075107} {\bibfield  {journal} {\bibinfo  {journal} {Phys. Rev. B}\ }\textbf {\bibinfo {volume} {106}},\ \bibinfo {pages} {075107} (\bibinfo {year} {2022})}\BibitemShut {NoStop}%
\bibitem [{\citenamefont {Khemani}\ \emph {et~al.}(2020)\citenamefont {Khemani}, \citenamefont {Hermele},\ and\ \citenamefont {Nandkishore}}]{Khemani20}%
  \BibitemOpen
  \bibfield  {author} {\bibinfo {author} {\bibfnamefont {Vedika}\ \bibnamefont {Khemani}}, \bibinfo {author} {\bibfnamefont {Michael}\ \bibnamefont {Hermele}}, \ and\ \bibinfo {author} {\bibfnamefont {Rahul}\ \bibnamefont {Nandkishore}},\ }\bibfield  {title} {\enquote {\bibinfo {title} {Localization from hilbert space shattering: From theory to physical realizations},}\ }\href {\doibase 10.1103/PhysRevB.101.174204} {\bibfield  {journal} {\bibinfo  {journal} {Phys. Rev. B}\ }\textbf {\bibinfo {volume} {101}},\ \bibinfo {pages} {174204} (\bibinfo {year} {2020})}\BibitemShut {NoStop}%
\bibitem [{\citenamefont {Sala}\ \emph {et~al.}(2020)\citenamefont {Sala}, \citenamefont {Rakovszky}, \citenamefont {Verresen}, \citenamefont {Knap},\ and\ \citenamefont {Pollmann}}]{Sala20}%
  \BibitemOpen
  \bibfield  {author} {\bibinfo {author} {\bibfnamefont {Pablo}\ \bibnamefont {Sala}}, \bibinfo {author} {\bibfnamefont {Tibor}\ \bibnamefont {Rakovszky}}, \bibinfo {author} {\bibfnamefont {Ruben}\ \bibnamefont {Verresen}}, \bibinfo {author} {\bibfnamefont {Michael}\ \bibnamefont {Knap}}, \ and\ \bibinfo {author} {\bibfnamefont {Frank}\ \bibnamefont {Pollmann}},\ }\bibfield  {title} {\enquote {\bibinfo {title} {Ergodicity breaking arising from hilbert space fragmentation in dipole-conserving hamiltonians},}\ }\href {\doibase 10.1103/PhysRevX.10.011047} {\bibfield  {journal} {\bibinfo  {journal} {Phys. Rev. X}\ }\textbf {\bibinfo {volume} {10}},\ \bibinfo {pages} {011047} (\bibinfo {year} {2020})}\BibitemShut {NoStop}%
\bibitem [{\citenamefont {Lan}\ \emph {et~al.}(2018)\citenamefont {Lan}, \citenamefont {van Horssen}, \citenamefont {Powell},\ and\ \citenamefont {Garrahan}}]{Lan18}%
  \BibitemOpen
  \bibfield  {author} {\bibinfo {author} {\bibfnamefont {Zhihao}\ \bibnamefont {Lan}}, \bibinfo {author} {\bibfnamefont {Merlijn}\ \bibnamefont {van Horssen}}, \bibinfo {author} {\bibfnamefont {Stephen}\ \bibnamefont {Powell}}, \ and\ \bibinfo {author} {\bibfnamefont {Juan~P.}\ \bibnamefont {Garrahan}},\ }\bibfield  {title} {\enquote {\bibinfo {title} {Quantum slow relaxation and metastability due to dynamical constraints},}\ }\href {\doibase 10.1103/PhysRevLett.121.040603} {\bibfield  {journal} {\bibinfo  {journal} {Phys. Rev. Lett.}\ }\textbf {\bibinfo {volume} {121}},\ \bibinfo {pages} {040603} (\bibinfo {year} {2018})}\BibitemShut {NoStop}%
\bibitem [{\citenamefont {Li}\ \emph {et~al.}(2021)\citenamefont {Li}, \citenamefont {Deng},\ and\ \citenamefont {Santos}}]{Li21}%
  \BibitemOpen
  \bibfield  {author} {\bibinfo {author} {\bibfnamefont {Wei-Han}\ \bibnamefont {Li}}, \bibinfo {author} {\bibfnamefont {Xiaolong}\ \bibnamefont {Deng}}, \ and\ \bibinfo {author} {\bibfnamefont {Luis}\ \bibnamefont {Santos}},\ }\bibfield  {title} {\enquote {\bibinfo {title} {Hilbert space shattering and disorder-free localization in polar lattice gases},}\ }\href {\doibase 10.1103/PhysRevLett.127.260601} {\bibfield  {journal} {\bibinfo  {journal} {Phys. Rev. Lett.}\ }\textbf {\bibinfo {volume} {127}},\ \bibinfo {pages} {260601} (\bibinfo {year} {2021})}\BibitemShut {NoStop}%
\bibitem [{\citenamefont {Korbmacher}\ \emph {et~al.}(2023)\citenamefont {Korbmacher}, \citenamefont {Sierant}, \citenamefont {Li}, \citenamefont {Deng}, \citenamefont {Zakrzewski},\ and\ \citenamefont {Santos}}]{Korbmacher23}%
  \BibitemOpen
  \bibfield  {author} {\bibinfo {author} {\bibfnamefont {H.}~\bibnamefont {Korbmacher}}, \bibinfo {author} {\bibfnamefont {P.}~\bibnamefont {Sierant}}, \bibinfo {author} {\bibfnamefont {W.}~\bibnamefont {Li}}, \bibinfo {author} {\bibfnamefont {X.}~\bibnamefont {Deng}}, \bibinfo {author} {\bibfnamefont {J.}~\bibnamefont {Zakrzewski}}, \ and\ \bibinfo {author} {\bibfnamefont {L.}~\bibnamefont {Santos}},\ }\bibfield  {title} {\enquote {\bibinfo {title} {Lattice control of nonergodicity in a polar lattice gas},}\ }\href {\doibase 10.1103/PhysRevA.107.013301} {\bibfield  {journal} {\bibinfo  {journal} {Phys. Rev. A}\ }\textbf {\bibinfo {volume} {107}},\ \bibinfo {pages} {013301} (\bibinfo {year} {2023})}\BibitemShut {NoStop}%
\bibitem [{\citenamefont {Scherg}\ \emph {et~al.}(2021)\citenamefont {Scherg}, \citenamefont {Kohlert}, \citenamefont {Sala}, \citenamefont {Pollmann}, \citenamefont {Hebbe~Madhusudhana}, \citenamefont {Bloch},\ and\ \citenamefont {Aidelsburger}}]{Scherg21}%
  \BibitemOpen
  \bibfield  {author} {\bibinfo {author} {\bibfnamefont {Sebastian}\ \bibnamefont {Scherg}}, \bibinfo {author} {\bibfnamefont {Thomas}\ \bibnamefont {Kohlert}}, \bibinfo {author} {\bibfnamefont {Pablo}\ \bibnamefont {Sala}}, \bibinfo {author} {\bibfnamefont {Frank}\ \bibnamefont {Pollmann}}, \bibinfo {author} {\bibfnamefont {Bharath}\ \bibnamefont {Hebbe~Madhusudhana}}, \bibinfo {author} {\bibfnamefont {Immanuel}\ \bibnamefont {Bloch}}, \ and\ \bibinfo {author} {\bibfnamefont {Monika}\ \bibnamefont {Aidelsburger}},\ }\bibfield  {title} {\enquote {\bibinfo {title} {Observing non-ergodicity due to kinetic constraints in tilted fermi-hubbard chains},}\ }\href {\doibase 10.1038/s41467-021-24726-0} {\bibfield  {journal} {\bibinfo  {journal} {Nature Communications}\ }\textbf {\bibinfo {volume} {12}},\ \bibinfo {pages} {4490} (\bibinfo {year} {2021})}\BibitemShut {NoStop}%
\bibitem [{\citenamefont {Kohlert}\ \emph {et~al.}(2023)\citenamefont {Kohlert}, \citenamefont {Scherg}, \citenamefont {Sala}, \citenamefont {Pollmann}, \citenamefont {{HebbeMadhusudhana,B.}}, \citenamefont {Bloch},\ and\ \citenamefont {Aidelsburger}}]{Kohlert23}%
  \BibitemOpen
  \bibfield  {author} {\bibinfo {author} {\bibfnamefont {Thomas}\ \bibnamefont {Kohlert}}, \bibinfo {author} {\bibfnamefont {Sebastian}\ \bibnamefont {Scherg}}, \bibinfo {author} {\bibfnamefont {Pablo}\ \bibnamefont {Sala}}, \bibinfo {author} {\bibfnamefont {Frank}\ \bibnamefont {Pollmann}}, \bibinfo {author} {\bibnamefont {{HebbeMadhusudhana,B.}}}, \bibinfo {author} {\bibfnamefont {Immanuel}\ \bibnamefont {Bloch}}, \ and\ \bibinfo {author} {\bibfnamefont {Monika}\ \bibnamefont {Aidelsburger}},\ }\bibfield  {title} {\enquote {\bibinfo {title} {Exploring the regime of fragmentation in strongly tilted fermi-hubbard chains},}\ }\href {\doibase 10.1103/PhysRevLett.130.010201} {\bibfield  {journal} {\bibinfo  {journal} {Phys. Rev. Lett.}\ }\textbf {\bibinfo {volume} {130}},\ \bibinfo {pages} {010201} (\bibinfo {year} {2023})}\BibitemShut {NoStop}%
\bibitem [{\citenamefont {Mallayya}\ and\ \citenamefont {Rigol}(2019)}]{2019.PhysRevLett.123.240603}%
  \BibitemOpen
  \bibfield  {author} {\bibinfo {author} {\bibfnamefont {Krishnanand}\ \bibnamefont {Mallayya}}\ and\ \bibinfo {author} {\bibfnamefont {Marcos}\ \bibnamefont {Rigol}},\ }\bibfield  {title} {\enquote {\bibinfo {title} {Heating rates in periodically driven strongly interacting quantum many-body systems},}\ }\href {\doibase 10.1103/PhysRevLett.123.240603} {\bibfield  {journal} {\bibinfo  {journal} {Phys. Rev. Lett.}\ }\textbf {\bibinfo {volume} {123}},\ \bibinfo {pages} {240603} (\bibinfo {year} {2019})}\BibitemShut {NoStop}%
\bibitem [{\citenamefont {Burke}\ and\ \citenamefont {Haque}(2023)}]{Burke23}%
  \BibitemOpen
  \bibfield  {author} {\bibinfo {author} {\bibfnamefont {Phillip~C.}\ \bibnamefont {Burke}}\ and\ \bibinfo {author} {\bibfnamefont {Masudul}\ \bibnamefont {Haque}},\ }\bibfield  {title} {\enquote {\bibinfo {title} {Entropy and temperature in finite isolated quantum systems},}\ }\href {\doibase 10.1103/PhysRevE.107.034125} {\bibfield  {journal} {\bibinfo  {journal} {Phys. Rev. E}\ }\textbf {\bibinfo {volume} {107}},\ \bibinfo {pages} {034125} (\bibinfo {year} {2023})}\BibitemShut {NoStop}%
\bibitem [{\citenamefont {Mondaini}\ and\ \citenamefont {Rigol}(2017)}]{2017.PhysRevE.96.012157}%
  \BibitemOpen
  \bibfield  {author} {\bibinfo {author} {\bibfnamefont {Rubem}\ \bibnamefont {Mondaini}}\ and\ \bibinfo {author} {\bibfnamefont {Marcos}\ \bibnamefont {Rigol}},\ }\bibfield  {title} {\enquote {\bibinfo {title} {Eigenstate thermalization in the two-dimensional transverse field ising model. ii. off-diagonal matrix elements of observables},}\ }\href {\doibase 10.1103/PhysRevE.96.012157} {\bibfield  {journal} {\bibinfo  {journal} {Phys. Rev. E}\ }\textbf {\bibinfo {volume} {96}},\ \bibinfo {pages} {012157} (\bibinfo {year} {2017})}\BibitemShut {NoStop}%
\bibitem [{\citenamefont {Essler}\ \emph {et~al.}(2012)\citenamefont {Essler}, \citenamefont {Evangelisti},\ and\ \citenamefont {Fagotti}}]{2012.PhysRevLett.109.247206}%
  \BibitemOpen
  \bibfield  {author} {\bibinfo {author} {\bibfnamefont {Fabian H.~L.}\ \bibnamefont {Essler}}, \bibinfo {author} {\bibfnamefont {Stefano}\ \bibnamefont {Evangelisti}}, \ and\ \bibinfo {author} {\bibfnamefont {Maurizio}\ \bibnamefont {Fagotti}},\ }\bibfield  {title} {\enquote {\bibinfo {title} {Dynamical correlations after a quantum quench},}\ }\href {\doibase 10.1103/PhysRevLett.109.247206} {\bibfield  {journal} {\bibinfo  {journal} {Phys. Rev. Lett.}\ }\textbf {\bibinfo {volume} {109}},\ \bibinfo {pages} {247206} (\bibinfo {year} {2012})}\BibitemShut {NoStop}%
\bibitem [{\citenamefont {{Srednicki}}(1999)}]{1999.JPhA.32.1163S}%
  \BibitemOpen
  \bibfield  {author} {\bibinfo {author} {\bibfnamefont {Mark}\ \bibnamefont {{Srednicki}}},\ }\bibfield  {title} {\enquote {\bibinfo {title} {{The approach to thermal equilibrium in quantized chaotic systems}},}\ }\href {\doibase 10.1088/0305-4470/32/7/007} {\bibfield  {journal} {\bibinfo  {journal} {Journal of Physics A Mathematical General}\ }\textbf {\bibinfo {volume} {32}},\ \bibinfo {pages} {1163--1175} (\bibinfo {year} {1999})}\BibitemShut {NoStop}%
\bibitem [{\citenamefont {Nation}\ and\ \citenamefont {Porras}(2019)}]{2019.PhysRevE.99.052139}%
  \BibitemOpen
  \bibfield  {author} {\bibinfo {author} {\bibfnamefont {Charlie}\ \bibnamefont {Nation}}\ and\ \bibinfo {author} {\bibfnamefont {Diego}\ \bibnamefont {Porras}},\ }\bibfield  {title} {\enquote {\bibinfo {title} {Quantum chaotic fluctuation-dissipation theorem: Effective brownian motion in closed quantum systems},}\ }\href {\doibase 10.1103/PhysRevE.99.052139} {\bibfield  {journal} {\bibinfo  {journal} {Phys. Rev. E}\ }\textbf {\bibinfo {volume} {99}},\ \bibinfo {pages} {052139} (\bibinfo {year} {2019})}\BibitemShut {NoStop}%
\bibitem [{\citenamefont {Noh}\ \emph {et~al.}(2020)\citenamefont {Noh}, \citenamefont {Sagawa},\ and\ \citenamefont {Yeo}}]{2020.PhysRevLett.125.050603}%
  \BibitemOpen
  \bibfield  {author} {\bibinfo {author} {\bibfnamefont {Jae~Dong}\ \bibnamefont {Noh}}, \bibinfo {author} {\bibfnamefont {Takahiro}\ \bibnamefont {Sagawa}}, \ and\ \bibinfo {author} {\bibfnamefont {Joonhyun}\ \bibnamefont {Yeo}},\ }\bibfield  {title} {\enquote {\bibinfo {title} {Numerical verification of the fluctuation-dissipation theorem for isolated quantum systems},}\ }\href {\doibase 10.1103/PhysRevLett.125.050603} {\bibfield  {journal} {\bibinfo  {journal} {Phys. Rev. Lett.}\ }\textbf {\bibinfo {volume} {125}},\ \bibinfo {pages} {050603} (\bibinfo {year} {2020})}\BibitemShut {NoStop}%
\bibitem [{\citenamefont {Serbyn}\ \emph {et~al.}(2017)\citenamefont {Serbyn}, \citenamefont {Papi\ifmmode~\acute{c}\else \'{c}\fi{}},\ and\ \citenamefont {Abanin}}]{Serbyn17}%
  \BibitemOpen
  \bibfield  {author} {\bibinfo {author} {\bibfnamefont {Maksym}\ \bibnamefont {Serbyn}}, \bibinfo {author} {\bibfnamefont {Z.}~\bibnamefont {Papi\ifmmode~\acute{c}\else \'{c}\fi{}}}, \ and\ \bibinfo {author} {\bibfnamefont {Dmitry~A.}\ \bibnamefont {Abanin}},\ }\bibfield  {title} {\enquote {\bibinfo {title} {Thouless energy and multifractality across the many-body localization transition},}\ }\href {\doibase 10.1103/PhysRevB.96.104201} {\bibfield  {journal} {\bibinfo  {journal} {Phys. Rev. B}\ }\textbf {\bibinfo {volume} {96}},\ \bibinfo {pages} {104201} (\bibinfo {year} {2017})}\BibitemShut {NoStop}%
\bibitem [{\citenamefont {Foini}\ and\ \citenamefont {Kurchan}(2019)}]{2019.PhysRevE.99.042139}%
  \BibitemOpen
  \bibfield  {author} {\bibinfo {author} {\bibfnamefont {Laura}\ \bibnamefont {Foini}}\ and\ \bibinfo {author} {\bibfnamefont {Jorge}\ \bibnamefont {Kurchan}},\ }\bibfield  {title} {\enquote {\bibinfo {title} {Eigenstate thermalization hypothesis and out of time order correlators},}\ }\href {\doibase 10.1103/PhysRevE.99.042139} {\bibfield  {journal} {\bibinfo  {journal} {Phys. Rev. E}\ }\textbf {\bibinfo {volume} {99}},\ \bibinfo {pages} {042139} (\bibinfo {year} {2019})}\BibitemShut {NoStop}%
\bibitem [{\citenamefont {Chan}\ \emph {et~al.}(2019)\citenamefont {Chan}, \citenamefont {De~Luca},\ and\ \citenamefont {Chalker}}]{2019.PhysRevLett.122.220601}%
  \BibitemOpen
  \bibfield  {author} {\bibinfo {author} {\bibfnamefont {Amos}\ \bibnamefont {Chan}}, \bibinfo {author} {\bibfnamefont {Andrea}\ \bibnamefont {De~Luca}}, \ and\ \bibinfo {author} {\bibfnamefont {J.~T.}\ \bibnamefont {Chalker}},\ }\bibfield  {title} {\enquote {\bibinfo {title} {Eigenstate correlations, thermalization, and the butterfly effect},}\ }\href {\doibase 10.1103/PhysRevLett.122.220601} {\bibfield  {journal} {\bibinfo  {journal} {Phys. Rev. Lett.}\ }\textbf {\bibinfo {volume} {122}},\ \bibinfo {pages} {220601} (\bibinfo {year} {2019})}\BibitemShut {NoStop}%
\bibitem [{\citenamefont {Brenes}\ \emph {et~al.}(2021)\citenamefont {Brenes}, \citenamefont {Pappalardi}, \citenamefont {Mitchison}, \citenamefont {Goold},\ and\ \citenamefont {Silva}}]{2021.PhysRevE.104.034120}%
  \BibitemOpen
  \bibfield  {author} {\bibinfo {author} {\bibfnamefont {Marlon}\ \bibnamefont {Brenes}}, \bibinfo {author} {\bibfnamefont {Silvia}\ \bibnamefont {Pappalardi}}, \bibinfo {author} {\bibfnamefont {Mark~T.}\ \bibnamefont {Mitchison}}, \bibinfo {author} {\bibfnamefont {John}\ \bibnamefont {Goold}}, \ and\ \bibinfo {author} {\bibfnamefont {Alessandro}\ \bibnamefont {Silva}},\ }\bibfield  {title} {\enquote {\bibinfo {title} {Out-of-time-order correlations and the fine structure of eigenstate thermalization},}\ }\href {\doibase 10.1103/PhysRevE.104.034120} {\bibfield  {journal} {\bibinfo  {journal} {Phys. Rev. E}\ }\textbf {\bibinfo {volume} {104}},\ \bibinfo {pages} {034120} (\bibinfo {year} {2021})}\BibitemShut {NoStop}%
\bibitem [{\citenamefont {Wang}\ \emph {et~al.}(2022)\citenamefont {Wang}, \citenamefont {Lamann}, \citenamefont {Richter}, \citenamefont {Steinigeweg}, \citenamefont {Dymarsky},\ and\ \citenamefont {Gemmer}}]{2022.PhysRevLett.128.180601}%
  \BibitemOpen
  \bibfield  {author} {\bibinfo {author} {\bibfnamefont {Jiaozi}\ \bibnamefont {Wang}}, \bibinfo {author} {\bibfnamefont {Mats~H.}\ \bibnamefont {Lamann}}, \bibinfo {author} {\bibfnamefont {Jonas}\ \bibnamefont {Richter}}, \bibinfo {author} {\bibfnamefont {Robin}\ \bibnamefont {Steinigeweg}}, \bibinfo {author} {\bibfnamefont {Anatoly}\ \bibnamefont {Dymarsky}}, \ and\ \bibinfo {author} {\bibfnamefont {Jochen}\ \bibnamefont {Gemmer}},\ }\bibfield  {title} {\enquote {\bibinfo {title} {Eigenstate thermalization hypothesis and its deviations from random-matrix theory beyond the thermalization time},}\ }\href {\doibase 10.1103/PhysRevLett.128.180601} {\bibfield  {journal} {\bibinfo  {journal} {Phys. Rev. Lett.}\ }\textbf {\bibinfo {volume} {128}},\ \bibinfo {pages} {180601} (\bibinfo {year} {2022})}\BibitemShut {NoStop}%
\bibitem [{\citenamefont {Kim}\ \emph {et~al.}(2014)\citenamefont {Kim}, \citenamefont {Ikeda},\ and\ \citenamefont {Huse}}]{2014.PhysRevE.90.052105}%
  \BibitemOpen
  \bibfield  {author} {\bibinfo {author} {\bibfnamefont {Hyungwon}\ \bibnamefont {Kim}}, \bibinfo {author} {\bibfnamefont {Tatsuhiko~N.}\ \bibnamefont {Ikeda}}, \ and\ \bibinfo {author} {\bibfnamefont {David~A.}\ \bibnamefont {Huse}},\ }\bibfield  {title} {\enquote {\bibinfo {title} {Testing whether all eigenstates obey the eigenstate thermalization hypothesis},}\ }\href {\doibase 10.1103/PhysRevE.90.052105} {\bibfield  {journal} {\bibinfo  {journal} {Phys. Rev. E}\ }\textbf {\bibinfo {volume} {90}},\ \bibinfo {pages} {052105} (\bibinfo {year} {2014})}\BibitemShut {NoStop}%
\bibitem [{\citenamefont {Luitz}(2016)}]{Luitz16long}%
  \BibitemOpen
  \bibfield  {author} {\bibinfo {author} {\bibfnamefont {David~J.}\ \bibnamefont {Luitz}},\ }\bibfield  {title} {\enquote {\bibinfo {title} {Long tail distributions near the many-body localization transition},}\ }\href {\doibase 10.1103/PhysRevB.93.134201} {\bibfield  {journal} {\bibinfo  {journal} {Phys. Rev. B}\ }\textbf {\bibinfo {volume} {93}},\ \bibinfo {pages} {134201} (\bibinfo {year} {2016})}\BibitemShut {NoStop}%
\bibitem [{\citenamefont {Mierzejewski}\ and\ \citenamefont {Vidmar}(2020)}]{2020.PhysRevLett.124.040603}%
  \BibitemOpen
  \bibfield  {author} {\bibinfo {author} {\bibfnamefont {Marcin}\ \bibnamefont {Mierzejewski}}\ and\ \bibinfo {author} {\bibfnamefont {Lev}\ \bibnamefont {Vidmar}},\ }\bibfield  {title} {\enquote {\bibinfo {title} {Quantitative impact of integrals of motion on the eigenstate thermalization hypothesis},}\ }\href {\doibase 10.1103/PhysRevLett.124.040603} {\bibfield  {journal} {\bibinfo  {journal} {Phys. Rev. Lett.}\ }\textbf {\bibinfo {volume} {124}},\ \bibinfo {pages} {040603} (\bibinfo {year} {2020})}\BibitemShut {NoStop}%
\bibitem [{\citenamefont {Kim}\ and\ \citenamefont {Huse}(2013)}]{2013.PhysRevLett.111.127205}%
  \BibitemOpen
  \bibfield  {author} {\bibinfo {author} {\bibfnamefont {Hyungwon}\ \bibnamefont {Kim}}\ and\ \bibinfo {author} {\bibfnamefont {David~A.}\ \bibnamefont {Huse}},\ }\bibfield  {title} {\enquote {\bibinfo {title} {Ballistic spreading of entanglement in a diffusive nonintegrable system},}\ }\href {\doibase 10.1103/PhysRevLett.111.127205} {\bibfield  {journal} {\bibinfo  {journal} {Phys. Rev. Lett.}\ }\textbf {\bibinfo {volume} {111}},\ \bibinfo {pages} {127205} (\bibinfo {year} {2013})}\BibitemShut {NoStop}%
\bibitem [{\citenamefont {Page}(1993)}]{1993.PhysRevLett.71.1291}%
  \BibitemOpen
  \bibfield  {author} {\bibinfo {author} {\bibfnamefont {Don~N.}\ \bibnamefont {Page}},\ }\bibfield  {title} {\enquote {\bibinfo {title} {Average entropy of a subsystem},}\ }\href {\doibase 10.1103/PhysRevLett.71.1291} {\bibfield  {journal} {\bibinfo  {journal} {Phys. Rev. Lett.}\ }\textbf {\bibinfo {volume} {71}},\ \bibinfo {pages} {1291--1294} (\bibinfo {year} {1993})}\BibitemShut {NoStop}%
\bibitem [{\citenamefont {Vidmar}\ and\ \citenamefont {Rigol}(2017)}]{Vidmar17}%
  \BibitemOpen
  \bibfield  {author} {\bibinfo {author} {\bibfnamefont {Lev}\ \bibnamefont {Vidmar}}\ and\ \bibinfo {author} {\bibfnamefont {Marcos}\ \bibnamefont {Rigol}},\ }\bibfield  {title} {\enquote {\bibinfo {title} {Entanglement entropy of eigenstates of quantum chaotic hamiltonians},}\ }\href {\doibase 10.1103/PhysRevLett.119.220603} {\bibfield  {journal} {\bibinfo  {journal} {Phys. Rev. Lett.}\ }\textbf {\bibinfo {volume} {119}},\ \bibinfo {pages} {220603} (\bibinfo {year} {2017})}\BibitemShut {NoStop}%
\bibitem [{\citenamefont {{Huang}}(2019)}]{2019.NuPhB.938.594H}%
  \BibitemOpen
  \bibfield  {author} {\bibinfo {author} {\bibfnamefont {Yichen}\ \bibnamefont {{Huang}}},\ }\bibfield  {title} {\enquote {\bibinfo {title} {{Universal eigenstate entanglement of chaotic local Hamiltonians}},}\ }\href {\doibase 10.1016/j.nuclphysb.2018.09.013} {\bibfield  {journal} {\bibinfo  {journal} {Nuclear Physics B}\ }\textbf {\bibinfo {volume} {938}},\ \bibinfo {pages} {594--604} (\bibinfo {year} {2019})}\BibitemShut {NoStop}%
\bibitem [{\citenamefont {Chiara}\ \emph {et~al.}(2006)\citenamefont {Chiara}, \citenamefont {Montangero}, \citenamefont {Calabrese},\ and\ \citenamefont {Fazio}}]{Chiara06}%
  \BibitemOpen
  \bibfield  {author} {\bibinfo {author} {\bibfnamefont {Gabriele~De}\ \bibnamefont {Chiara}}, \bibinfo {author} {\bibfnamefont {Simone}\ \bibnamefont {Montangero}}, \bibinfo {author} {\bibfnamefont {Pasquale}\ \bibnamefont {Calabrese}}, \ and\ \bibinfo {author} {\bibfnamefont {Rosario}\ \bibnamefont {Fazio}},\ }\bibfield  {title} {\enquote {\bibinfo {title} {Entanglement entropy dynamics of heisenberg chains},}\ }\href {\doibase 10.1088/1742-5468/2006/03/P03001} {\bibfield  {journal} {\bibinfo  {journal} {Journal of Statistical Mechanics: Theory and Experiment}\ }\textbf {\bibinfo {volume} {2006}},\ \bibinfo {pages} {P03001} (\bibinfo {year} {2006})}\BibitemShut {NoStop}%
\bibitem [{\citenamefont {\v{Z}nidari\v{c}}\ \emph {et~al.}(2008)\citenamefont {\v{Z}nidari\v{c}}, \citenamefont {Prosen},\ and\ \citenamefont {Prelov\v{s}ek}}]{Znidaric08}%
  \BibitemOpen
  \bibfield  {author} {\bibinfo {author} {\bibfnamefont {Marko}\ \bibnamefont {\v{Z}nidari\v{c}}}, \bibinfo {author} {\bibfnamefont {Toma\v{z}}\ \bibnamefont {Prosen}}, \ and\ \bibinfo {author} {\bibfnamefont {Peter}\ \bibnamefont {Prelov\v{s}ek}},\ }\bibfield  {title} {\enquote {\bibinfo {title} {Many-body localization in the {H}eisenberg {X}{X}{Z} magnet in a random field},}\ }\href {\doibase 10.1103/PhysRevB.77.064426} {\bibfield  {journal} {\bibinfo  {journal} {Phys. Rev. B}\ }\textbf {\bibinfo {volume} {77}},\ \bibinfo {pages} {064426} (\bibinfo {year} {2008})}\BibitemShut {NoStop}%
\bibitem [{\citenamefont {Bardarson}\ \emph {et~al.}(2012)\citenamefont {Bardarson}, \citenamefont {Pollmann},\ and\ \citenamefont {Moore}}]{Bardarson12}%
  \BibitemOpen
  \bibfield  {author} {\bibinfo {author} {\bibfnamefont {Jens~H.}\ \bibnamefont {Bardarson}}, \bibinfo {author} {\bibfnamefont {Frank}\ \bibnamefont {Pollmann}}, \ and\ \bibinfo {author} {\bibfnamefont {Joel~E.}\ \bibnamefont {Moore}},\ }\bibfield  {title} {\enquote {\bibinfo {title} {Unbounded growth of entanglement in models of many-body localization},}\ }\href {\doibase 10.1103/PhysRevLett.109.017202} {\bibfield  {journal} {\bibinfo  {journal} {Phys. Rev. Lett.}\ }\textbf {\bibinfo {volume} {109}},\ \bibinfo {pages} {017202} (\bibinfo {year} {2012})}\BibitemShut {NoStop}%
\bibitem [{\citenamefont {Serbyn}\ \emph {et~al.}(2013{\natexlab{a}})\citenamefont {Serbyn}, \citenamefont {Papi\'{c}},\ and\ \citenamefont {Abanin}}]{Serbyn13a}%
  \BibitemOpen
  \bibfield  {author} {\bibinfo {author} {\bibfnamefont {Maksym}\ \bibnamefont {Serbyn}}, \bibinfo {author} {\bibfnamefont {Z.}~\bibnamefont {Papi\'{c}}}, \ and\ \bibinfo {author} {\bibfnamefont {Dmitry~A.}\ \bibnamefont {Abanin}},\ }\bibfield  {title} {\enquote {\bibinfo {title} {Universal slow growth of entanglement in interacting strongly disordered systems},}\ }\href {\doibase 10.1103/PhysRevLett.110.260601} {\bibfield  {journal} {\bibinfo  {journal} {Phys. Rev. Lett.}\ }\textbf {\bibinfo {volume} {110}},\ \bibinfo {pages} {260601} (\bibinfo {year} {2013}{\natexlab{a}})}\BibitemShut {NoStop}%
\bibitem [{\citenamefont {Iemini}\ \emph {et~al.}(2016)\citenamefont {Iemini}, \citenamefont {Russomanno}, \citenamefont {Rossini}, \citenamefont {Scardicchio},\ and\ \citenamefont {Fazio}}]{iemini2016signatures}%
  \BibitemOpen
  \bibfield  {author} {\bibinfo {author} {\bibfnamefont {Fernando}\ \bibnamefont {Iemini}}, \bibinfo {author} {\bibfnamefont {Angelo}\ \bibnamefont {Russomanno}}, \bibinfo {author} {\bibfnamefont {Davide}\ \bibnamefont {Rossini}}, \bibinfo {author} {\bibfnamefont {Antonello}\ \bibnamefont {Scardicchio}}, \ and\ \bibinfo {author} {\bibfnamefont {Rosario}\ \bibnamefont {Fazio}},\ }\bibfield  {title} {\enquote {\bibinfo {title} {Signatures of many-body localization in the dynamics of two-site entanglement},}\ }\href@noop {} {\bibfield  {journal} {\bibinfo  {journal} {Physical Review B}\ }\textbf {\bibinfo {volume} {94}},\ \bibinfo {pages} {214206} (\bibinfo {year} {2016})}\BibitemShut {NoStop}%
\bibitem [{\citenamefont {Serbyn}\ \emph {et~al.}(2013{\natexlab{b}})\citenamefont {Serbyn}, \citenamefont {Papi\'{c}},\ and\ \citenamefont {Abanin}}]{Serbyn13b}%
  \BibitemOpen
  \bibfield  {author} {\bibinfo {author} {\bibfnamefont {Maksym}\ \bibnamefont {Serbyn}}, \bibinfo {author} {\bibfnamefont {Z.}~\bibnamefont {Papi\'{c}}}, \ and\ \bibinfo {author} {\bibfnamefont {Dmitry~A.}\ \bibnamefont {Abanin}},\ }\bibfield  {title} {\enquote {\bibinfo {title} {Local conservation laws and the structure of the many-body localized states},}\ }\href {\doibase 10.1103/PhysRevLett.111.127201} {\bibfield  {journal} {\bibinfo  {journal} {Phys. Rev. Lett.}\ }\textbf {\bibinfo {volume} {111}},\ \bibinfo {pages} {127201} (\bibinfo {year} {2013}{\natexlab{b}})}\BibitemShut {NoStop}%
\bibitem [{\citenamefont {Huse}\ \emph {et~al.}(2014)\citenamefont {Huse}, \citenamefont {Nandkishore},\ and\ \citenamefont {Oganesyan}}]{Huse14}%
  \BibitemOpen
  \bibfield  {author} {\bibinfo {author} {\bibfnamefont {David~A.}\ \bibnamefont {Huse}}, \bibinfo {author} {\bibfnamefont {Rahul}\ \bibnamefont {Nandkishore}}, \ and\ \bibinfo {author} {\bibfnamefont {Vadim}\ \bibnamefont {Oganesyan}},\ }\bibfield  {title} {\enquote {\bibinfo {title} {Phenomenology of fully many-body-localized systems},}\ }\href {\doibase 10.1103/PhysRevB.90.174202} {\bibfield  {journal} {\bibinfo  {journal} {Phys. Rev. B}\ }\textbf {\bibinfo {volume} {90}},\ \bibinfo {pages} {174202} (\bibinfo {year} {2014})}\BibitemShut {NoStop}%
\bibitem [{\citenamefont {Ros}\ \emph {et~al.}(2015)\citenamefont {Ros}, \citenamefont {Mueller},\ and\ \citenamefont {Scardicchio}}]{Ros15}%
  \BibitemOpen
  \bibfield  {author} {\bibinfo {author} {\bibfnamefont {V.}~\bibnamefont {Ros}}, \bibinfo {author} {\bibfnamefont {M.}~\bibnamefont {Mueller}}, \ and\ \bibinfo {author} {\bibfnamefont {A.}~\bibnamefont {Scardicchio}},\ }\bibfield  {title} {\enquote {\bibinfo {title} {Integrals of motion in the many-body localized phase},}\ }\href {\doibase https://doi.org/10.1016/j.nuclphysb.2014.12.014} {\bibfield  {journal} {\bibinfo  {journal} {Nuclear Physics B}\ }\textbf {\bibinfo {volume} {891}},\ \bibinfo {pages} {420 -- 465} (\bibinfo {year} {2015})}\BibitemShut {NoStop}%
\bibitem [{\citenamefont {Mierzejewski}\ \emph {et~al.}(2018)\citenamefont {Mierzejewski}, \citenamefont {Kozarzewski},\ and\ \citenamefont {Prelov\v{s}ek}}]{Mierzejewski18}%
  \BibitemOpen
  \bibfield  {author} {\bibinfo {author} {\bibfnamefont {Marcin}\ \bibnamefont {Mierzejewski}}, \bibinfo {author} {\bibfnamefont {Maciej}\ \bibnamefont {Kozarzewski}}, \ and\ \bibinfo {author} {\bibfnamefont {Peter}\ \bibnamefont {Prelov\v{s}ek}},\ }\bibfield  {title} {\enquote {\bibinfo {title} {Counting local integrals of motion in disordered spinless-fermion and {H}ubbard chains},}\ }\href {\doibase 10.1103/PhysRevB.97.064204} {\bibfield  {journal} {\bibinfo  {journal} {Phys. Rev. B}\ }\textbf {\bibinfo {volume} {97}},\ \bibinfo {pages} {064204} (\bibinfo {year} {2018})}\BibitemShut {NoStop}%
\bibitem [{\citenamefont {Flambaum}\ and\ \citenamefont {Izrailev}(2001)}]{2001.PhysRevE.64.026124}%
  \BibitemOpen
  \bibfield  {author} {\bibinfo {author} {\bibfnamefont {V.~V.}\ \bibnamefont {Flambaum}}\ and\ \bibinfo {author} {\bibfnamefont {F.~M.}\ \bibnamefont {Izrailev}},\ }\bibfield  {title} {\enquote {\bibinfo {title} {Unconventional decay law for excited states in closed many-body systems},}\ }\href {\doibase 10.1103/PhysRevE.64.026124} {\bibfield  {journal} {\bibinfo  {journal} {Phys. Rev. E}\ }\textbf {\bibinfo {volume} {64}},\ \bibinfo {pages} {026124} (\bibinfo {year} {2001})}\BibitemShut {NoStop}%
\bibitem [{\citenamefont {{Izrailev}}\ and\ \citenamefont {{Casta{\~n}eda-Mendoza}}(2006)}]{2006.PhLA.350.355I}%
  \BibitemOpen
  \bibfield  {author} {\bibinfo {author} {\bibfnamefont {F.~M.}\ \bibnamefont {{Izrailev}}}\ and\ \bibinfo {author} {\bibfnamefont {A.}~\bibnamefont {{Casta{\~n}eda-Mendoza}}},\ }\bibfield  {title} {\enquote {\bibinfo {title} {{Return probability: Exponential versus Gaussian decay}},}\ }\href {\doibase 10.1016/j.physleta.2005.10.077} {\bibfield  {journal} {\bibinfo  {journal} {Physics Letters A}\ }\textbf {\bibinfo {volume} {350}},\ \bibinfo {pages} {355--362} (\bibinfo {year} {2006})}\BibitemShut {NoStop}%
\bibitem [{\citenamefont {Torres-Herrera}\ and\ \citenamefont {Santos}(2014)}]{2014.PhysRevA.89.043620}%
  \BibitemOpen
  \bibfield  {author} {\bibinfo {author} {\bibfnamefont {E.~J.}\ \bibnamefont {Torres-Herrera}}\ and\ \bibinfo {author} {\bibfnamefont {Lea~F.}\ \bibnamefont {Santos}},\ }\bibfield  {title} {\enquote {\bibinfo {title} {Quench dynamics of isolated many-body quantum systems},}\ }\href {\doibase 10.1103/PhysRevA.89.043620} {\bibfield  {journal} {\bibinfo  {journal} {Phys. Rev. A}\ }\textbf {\bibinfo {volume} {89}},\ \bibinfo {pages} {043620} (\bibinfo {year} {2014})}\BibitemShut {NoStop}%
\bibitem [{\citenamefont {T\'avora}\ \emph {et~al.}(2016)\citenamefont {T\'avora}, \citenamefont {Torres-Herrera},\ and\ \citenamefont {Santos}}]{2016.PhysRevA.94.041603}%
  \BibitemOpen
  \bibfield  {author} {\bibinfo {author} {\bibfnamefont {Marco}\ \bibnamefont {T\'avora}}, \bibinfo {author} {\bibfnamefont {E.~J.}\ \bibnamefont {Torres-Herrera}}, \ and\ \bibinfo {author} {\bibfnamefont {Lea~F.}\ \bibnamefont {Santos}},\ }\bibfield  {title} {\enquote {\bibinfo {title} {Inevitable power-law behavior of isolated many-body quantum systems and how it anticipates thermalization},}\ }\href {\doibase 10.1103/PhysRevA.94.041603} {\bibfield  {journal} {\bibinfo  {journal} {Phys. Rev. A}\ }\textbf {\bibinfo {volume} {94}},\ \bibinfo {pages} {041603(R)} (\bibinfo {year} {2016})}\BibitemShut {NoStop}%
\bibitem [{\citenamefont {Schiulaz}\ \emph {et~al.}(2019)\citenamefont {Schiulaz}, \citenamefont {Torres-Herrera},\ and\ \citenamefont {Santos}}]{Schiulaz19}%
  \BibitemOpen
  \bibfield  {author} {\bibinfo {author} {\bibfnamefont {Mauro}\ \bibnamefont {Schiulaz}}, \bibinfo {author} {\bibfnamefont {E.~J.}\ \bibnamefont {Torres-Herrera}}, \ and\ \bibinfo {author} {\bibfnamefont {Lea~F.}\ \bibnamefont {Santos}},\ }\bibfield  {title} {\enquote {\bibinfo {title} {Thouless and relaxation time scales in many-body quantum systems},}\ }\href {\doibase 10.1103/PhysRevB.99.174313} {\bibfield  {journal} {\bibinfo  {journal} {Phys. Rev. B}\ }\textbf {\bibinfo {volume} {99}},\ \bibinfo {pages} {174313} (\bibinfo {year} {2019})}\BibitemShut {NoStop}%
\bibitem [{\citenamefont {Lezama}\ \emph {et~al.}(2021)\citenamefont {Lezama}, \citenamefont {Torres-Herrera}, \citenamefont {Perez-Bernal}, \citenamefont {{BarLev, Y.}},\ and\ \citenamefont {Santos}}]{Lezama21}%
  \BibitemOpen
  \bibfield  {author} {\bibinfo {author} {\bibfnamefont {T.~L.~M.}\ \bibnamefont {Lezama}}, \bibinfo {author} {\bibfnamefont {E.~J.}\ \bibnamefont {Torres-Herrera}}, \bibinfo {author} {\bibfnamefont {Francisco}\ \bibnamefont {Perez-Bernal}}, \bibinfo {author} {\bibnamefont {{BarLev, Y.}}}, \ and\ \bibinfo {author} {\bibfnamefont {Lea~F.}\ \bibnamefont {Santos}},\ }\bibfield  {title} {\enquote {\bibinfo {title} {Equilibration time in many-body quantum systems},}\ }\href {\doibase 10.1103/PhysRevB.104.085117} {\bibfield  {journal} {\bibinfo  {journal} {Phys. Rev. B}\ }\textbf {\bibinfo {volume} {104}},\ \bibinfo {pages} {085117} (\bibinfo {year} {2021})}\BibitemShut {NoStop}%
\bibitem [{\citenamefont {Luitz}\ \emph {et~al.}(2016)\citenamefont {Luitz}, \citenamefont {Laflorencie},\ and\ \citenamefont {Alet}}]{Luitz16slow}%
  \BibitemOpen
  \bibfield  {author} {\bibinfo {author} {\bibfnamefont {David~J.}\ \bibnamefont {Luitz}}, \bibinfo {author} {\bibfnamefont {Nicolas}\ \bibnamefont {Laflorencie}}, \ and\ \bibinfo {author} {\bibfnamefont {Fabien}\ \bibnamefont {Alet}},\ }\bibfield  {title} {\enquote {\bibinfo {title} {Extended slow dynamical regime close to the many-body localization transition},}\ }\href {\doibase 10.1103/PhysRevB.93.060201} {\bibfield  {journal} {\bibinfo  {journal} {Phys. Rev. B}\ }\textbf {\bibinfo {volume} {93}},\ \bibinfo {pages} {060201(R)} (\bibinfo {year} {2016})}\BibitemShut {NoStop}%
\bibitem [{\citenamefont {Bera}\ \emph {et~al.}(2017)\citenamefont {Bera}, \citenamefont {De~Tomasi}, \citenamefont {Weiner},\ and\ \citenamefont {Evers}}]{Bera17}%
  \BibitemOpen
  \bibfield  {author} {\bibinfo {author} {\bibfnamefont {Soumya}\ \bibnamefont {Bera}}, \bibinfo {author} {\bibfnamefont {Giuseppe}\ \bibnamefont {De~Tomasi}}, \bibinfo {author} {\bibfnamefont {Felix}\ \bibnamefont {Weiner}}, \ and\ \bibinfo {author} {\bibfnamefont {Ferdinand}\ \bibnamefont {Evers}},\ }\bibfield  {title} {\enquote {\bibinfo {title} {Density propagator for many-body localization: Finite-size effects, transient subdiffusion, and exponential decay},}\ }\href {\doibase 10.1103/PhysRevLett.118.196801} {\bibfield  {journal} {\bibinfo  {journal} {Phys. Rev. Lett.}\ }\textbf {\bibinfo {volume} {118}},\ \bibinfo {pages} {196801} (\bibinfo {year} {2017})}\BibitemShut {NoStop}%
\bibitem [{\citenamefont {Sierant}\ and\ \citenamefont {Zakrzewski}(2022)}]{Sierant22obs}%
  \BibitemOpen
  \bibfield  {author} {\bibinfo {author} {\bibfnamefont {Piotr}\ \bibnamefont {Sierant}}\ and\ \bibinfo {author} {\bibfnamefont {Jakub}\ \bibnamefont {Zakrzewski}},\ }\bibfield  {title} {\enquote {\bibinfo {title} {Challenges to observation of many-body localization},}\ }\href {\doibase 10.1103/PhysRevB.105.224203} {\bibfield  {journal} {\bibinfo  {journal} {Phys. Rev. B}\ }\textbf {\bibinfo {volume} {105}},\ \bibinfo {pages} {224203} (\bibinfo {year} {2022})}\BibitemShut {NoStop}%
\bibitem [{\citenamefont {Scholl}\ \emph {et~al.}(2022{\natexlab{a}})\citenamefont {Scholl}, \citenamefont {Williams}, \citenamefont {Bornet}, \citenamefont {Wallner}, \citenamefont {Barredo}, \citenamefont {Henriet}, \citenamefont {Signoles}, \citenamefont {Hainaut}, \citenamefont {Franz}, \citenamefont {Geier}, \citenamefont {Tebben}, \citenamefont {Salzinger}, \citenamefont {Z{\"u}rn}, \citenamefont {Lahaye}, \citenamefont {Weidem{\"u}ller},\ and\ \citenamefont {Browaeys}}]{schollMicrowave2022}%
  \BibitemOpen
  \bibfield  {author} {\bibinfo {author} {\bibfnamefont {P.}~\bibnamefont {Scholl}}, \bibinfo {author} {\bibfnamefont {H.~J.}\ \bibnamefont {Williams}}, \bibinfo {author} {\bibfnamefont {G.}~\bibnamefont {Bornet}}, \bibinfo {author} {\bibfnamefont {F.}~\bibnamefont {Wallner}}, \bibinfo {author} {\bibfnamefont {D.}~\bibnamefont {Barredo}}, \bibinfo {author} {\bibfnamefont {L.}~\bibnamefont {Henriet}}, \bibinfo {author} {\bibfnamefont {A.}~\bibnamefont {Signoles}}, \bibinfo {author} {\bibfnamefont {C.}~\bibnamefont {Hainaut}}, \bibinfo {author} {\bibfnamefont {T.}~\bibnamefont {Franz}}, \bibinfo {author} {\bibfnamefont {S.}~\bibnamefont {Geier}}, \bibinfo {author} {\bibfnamefont {A.}~\bibnamefont {Tebben}}, \bibinfo {author} {\bibfnamefont {A.}~\bibnamefont {Salzinger}}, \bibinfo {author} {\bibfnamefont {G.}~\bibnamefont {Z{\"u}rn}}, \bibinfo {author} {\bibfnamefont {T.}~\bibnamefont {Lahaye}}, \bibinfo {author} {\bibfnamefont {M.}~\bibnamefont {Weidem{\"u}ller}}, \ and\ \bibinfo {author} {\bibfnamefont {A.}~\bibnamefont {Browaeys}},\ }\bibfield  {title} {\enquote {\bibinfo {title} {Microwave {{Engineering}} of {{Programmable}} \${{XXZ}}\$ {{Hamiltonians}} in {{Arrays}} of {{Rydberg Atoms}}},}\ }\href {\doibase 10.1103/PRXQuantum.3.020303} {\bibfield  {journal} {\bibinfo  {journal} {PRX Quantum}\ }\textbf {\bibinfo {volume} {3}},\ \bibinfo {pages} {020303} (\bibinfo {year} {2022}{\natexlab{a}})}\BibitemShut {NoStop}%
\bibitem [{\citenamefont {Hung}\ \emph {et~al.}(2016)\citenamefont {Hung}, \citenamefont {{Gonz{\'a}lez-Tudela}}, \citenamefont {Cirac},\ and\ \citenamefont {Kimble}}]{hungQuantum2016}%
  \BibitemOpen
  \bibfield  {author} {\bibinfo {author} {\bibfnamefont {C.-L.}\ \bibnamefont {Hung}}, \bibinfo {author} {\bibfnamefont {Alejandro}\ \bibnamefont {{Gonz{\'a}lez-Tudela}}}, \bibinfo {author} {\bibfnamefont {J.~Ignacio}\ \bibnamefont {Cirac}}, \ and\ \bibinfo {author} {\bibfnamefont {H.~J.}\ \bibnamefont {Kimble}},\ }\bibfield  {title} {\enquote {\bibinfo {title} {Quantum spin dynamics with pairwise-tunable, long-range interactions},}\ }\href@noop {} {\bibfield  {journal} {\bibinfo  {journal} {Proceedings of the National Academy of Sciences}\ }\textbf {\bibinfo {volume} {113}},\ \bibinfo {pages} {E4946--E4955} (\bibinfo {year} {2016})}\BibitemShut {NoStop}%
\bibitem [{\citenamefont {Tabares}\ \emph {et~al.}(2023)\citenamefont {Tabares}, \citenamefont {{MunozdelasHeras, A.}}, \citenamefont {Tagliacozzo}, \citenamefont {Porras},\ and\ \citenamefont {{Gonz{\'a}lez-Tudela}}}]{tabaresVariational2023}%
  \BibitemOpen
  \bibfield  {author} {\bibinfo {author} {\bibfnamefont {C.}~\bibnamefont {Tabares}}, \bibinfo {author} {\bibnamefont {{MunozdelasHeras, A.}}}, \bibinfo {author} {\bibfnamefont {L.}~\bibnamefont {Tagliacozzo}}, \bibinfo {author} {\bibfnamefont {D.}~\bibnamefont {Porras}}, \ and\ \bibinfo {author} {\bibfnamefont {A.}~\bibnamefont {{Gonz{\'a}lez-Tudela}}},\ }\bibfield  {title} {\enquote {\bibinfo {title} {Variational {{Quantum Simulators Based}} on {{Waveguide QED}}},}\ }\href {\doibase 10.1103/PhysRevLett.131.073602} {\bibfield  {journal} {\bibinfo  {journal} {Physical Review Letters}\ }\textbf {\bibinfo {volume} {131}},\ \bibinfo {pages} {073602} (\bibinfo {year} {2023})}\BibitemShut {NoStop}%
\bibitem [{\citenamefont {Scholl}\ \emph {et~al.}(2022{\natexlab{b}})\citenamefont {Scholl}, \citenamefont {Williams}, \citenamefont {Bornet}, \citenamefont {Wallner}, \citenamefont {Barredo}, \citenamefont {Henriet}, \citenamefont {Signoles}, \citenamefont {Hainaut}, \citenamefont {Franz}, \citenamefont {Geier}, \citenamefont {Tebben}, \citenamefont {Salzinger}, \citenamefont {Z\"urn}, \citenamefont {Lahaye}, \citenamefont {Weidem\"uller},\ and\ \citenamefont {Browaeys}}]{PRXQuantum.3.020303}%
  \BibitemOpen
  \bibfield  {author} {\bibinfo {author} {\bibfnamefont {P.}~\bibnamefont {Scholl}}, \bibinfo {author} {\bibfnamefont {H.~J.}\ \bibnamefont {Williams}}, \bibinfo {author} {\bibfnamefont {G.}~\bibnamefont {Bornet}}, \bibinfo {author} {\bibfnamefont {F.}~\bibnamefont {Wallner}}, \bibinfo {author} {\bibfnamefont {D.}~\bibnamefont {Barredo}}, \bibinfo {author} {\bibfnamefont {L.}~\bibnamefont {Henriet}}, \bibinfo {author} {\bibfnamefont {A.}~\bibnamefont {Signoles}}, \bibinfo {author} {\bibfnamefont {C.}~\bibnamefont {Hainaut}}, \bibinfo {author} {\bibfnamefont {T.}~\bibnamefont {Franz}}, \bibinfo {author} {\bibfnamefont {S.}~\bibnamefont {Geier}}, \bibinfo {author} {\bibfnamefont {A.}~\bibnamefont {Tebben}}, \bibinfo {author} {\bibfnamefont {A.}~\bibnamefont {Salzinger}}, \bibinfo {author} {\bibfnamefont {G.}~\bibnamefont {Z\"urn}}, \bibinfo {author} {\bibfnamefont {T.}~\bibnamefont {Lahaye}}, \bibinfo {author} {\bibfnamefont {M.}~\bibnamefont {Weidem\"uller}}, \ and\ \bibinfo {author} {\bibfnamefont {A.}~\bibnamefont {Browaeys}},\ }\bibfield  {title} {\enquote {\bibinfo {title} {Microwave engineering of programmable $xxz$ hamiltonians in arrays of rydberg atoms},}\ }\href {\doibase 10.1103/PRXQuantum.3.020303} {\bibfield  {journal} {\bibinfo  {journal} {PRX Quantum}\ }\textbf {\bibinfo {volume} {3}},\ \bibinfo {pages} {020303} (\bibinfo {year} {2022}{\natexlab{b}})}\BibitemShut {NoStop}%
\bibitem [{\citenamefont {Notarnicola}\ \emph {et~al.}(2023)\citenamefont {Notarnicola}, \citenamefont {Elben}, \citenamefont {Lahaye}, \citenamefont {Browaeys}, \citenamefont {Montangero},\ and\ \citenamefont {Vermersch}}]{notarnicolaRandomized2023}%
  \BibitemOpen
  \bibfield  {author} {\bibinfo {author} {\bibfnamefont {Simone}\ \bibnamefont {Notarnicola}}, \bibinfo {author} {\bibfnamefont {A.}~\bibnamefont {Elben}}, \bibinfo {author} {\bibfnamefont {Thierry}\ \bibnamefont {Lahaye}}, \bibinfo {author} {\bibfnamefont {Antoine}\ \bibnamefont {Browaeys}}, \bibinfo {author} {\bibfnamefont {Simone}\ \bibnamefont {Montangero}}, \ and\ \bibinfo {author} {\bibfnamefont {B.}~\bibnamefont {Vermersch}},\ }\bibfield  {title} {\enquote {\bibinfo {title} {A randomized measurement toolbox for an interacting {{Rydberg-atom}} quantum simulator},}\ }\href {\doibase 10.1088/1367-2630/acfcd3} {\bibfield  {journal} {\bibinfo  {journal} {New Journal of Physics}\ } (\bibinfo {year} {2023}),\ 10.1088/1367-2630/acfcd3}\BibitemShut {NoStop}%
\bibitem [{\citenamefont {Bluvstein}\ \emph {et~al.}(2022)\citenamefont {Bluvstein}, \citenamefont {Levine}, \citenamefont {Semeghini}, \citenamefont {Wang}, \citenamefont {Ebadi}, \citenamefont {Kalinowski}, \citenamefont {Keesling}, \citenamefont {Maskara}, \citenamefont {Pichler}, \citenamefont {Greiner}, \citenamefont {Vuleti{\'c}},\ and\ \citenamefont {Lukin}}]{bluvsteinQuantum2022}%
  \BibitemOpen
  \bibfield  {author} {\bibinfo {author} {\bibfnamefont {Dolev}\ \bibnamefont {Bluvstein}}, \bibinfo {author} {\bibfnamefont {Harry}\ \bibnamefont {Levine}}, \bibinfo {author} {\bibfnamefont {Giulia}\ \bibnamefont {Semeghini}}, \bibinfo {author} {\bibfnamefont {Tout~T.}\ \bibnamefont {Wang}}, \bibinfo {author} {\bibfnamefont {Sepehr}\ \bibnamefont {Ebadi}}, \bibinfo {author} {\bibfnamefont {Marcin}\ \bibnamefont {Kalinowski}}, \bibinfo {author} {\bibfnamefont {Alexander}\ \bibnamefont {Keesling}}, \bibinfo {author} {\bibfnamefont {Nishad}\ \bibnamefont {Maskara}}, \bibinfo {author} {\bibfnamefont {Hannes}\ \bibnamefont {Pichler}}, \bibinfo {author} {\bibfnamefont {Markus}\ \bibnamefont {Greiner}}, \bibinfo {author} {\bibfnamefont {Vladan}\ \bibnamefont {Vuleti{\'c}}}, \ and\ \bibinfo {author} {\bibfnamefont {Mikhail~D.}\ \bibnamefont {Lukin}},\ }\bibfield  {title} {\enquote {\bibinfo {title} {A quantum processor based on coherent transport of entangled atom arrays},}\ }\href {\doibase 10.1038/s41586-022-04592-6} {\bibfield  {journal} {\bibinfo  {journal} {Nature}\ }\textbf {\bibinfo {volume} {604}},\ \bibinfo {pages} {451--456} (\bibinfo {year} {2022})}\BibitemShut {NoStop}%
\bibitem [{\citenamefont {Sierant}\ \emph {et~al.}(2017)\citenamefont {Sierant}, \citenamefont {Delande},\ and\ \citenamefont {Zakrzewski}}]{Sierant17ran}%
  \BibitemOpen
  \bibfield  {author} {\bibinfo {author} {\bibfnamefont {Piotr}\ \bibnamefont {Sierant}}, \bibinfo {author} {\bibfnamefont {Dominique}\ \bibnamefont {Delande}}, \ and\ \bibinfo {author} {\bibfnamefont {Jakub}\ \bibnamefont {Zakrzewski}},\ }\bibfield  {title} {\enquote {\bibinfo {title} {Many-body localization due to random interactions},}\ }\href {\doibase 10.1103/PhysRevA.95.021601} {\bibfield  {journal} {\bibinfo  {journal} {Phys. Rev. A}\ }\textbf {\bibinfo {volume} {95}},\ \bibinfo {pages} {021601(R)} (\bibinfo {year} {2017})}\BibitemShut {NoStop}%
\bibitem [{\citenamefont {{BarLev, Y.}}\ \emph {et~al.}(2016)\citenamefont {{BarLev, Y.}}, \citenamefont {Reichman},\ and\ \citenamefont {Sagi}}]{Bar16}%
  \BibitemOpen
  \bibfield  {author} {\bibinfo {author} {\bibnamefont {{BarLev, Y.}}}, \bibinfo {author} {\bibfnamefont {David~R.}\ \bibnamefont {Reichman}}, \ and\ \bibinfo {author} {\bibfnamefont {Yoav}\ \bibnamefont {Sagi}},\ }\bibfield  {title} {\enquote {\bibinfo {title} {Many-body localization in system with a completely delocalized single-particle spectrum},}\ }\href {\doibase 10.1103/PhysRevB.94.201116} {\bibfield  {journal} {\bibinfo  {journal} {Phys. Rev. B}\ }\textbf {\bibinfo {volume} {94}},\ \bibinfo {pages} {201116(R)} (\bibinfo {year} {2016})}\BibitemShut {NoStop}%
\bibitem [{\citenamefont {Li}\ \emph {et~al.}(2017)\citenamefont {Li}, \citenamefont {Deng}, \citenamefont {Wu},\ and\ \citenamefont {Das~Sarma}}]{Li17}%
  \BibitemOpen
  \bibfield  {author} {\bibinfo {author} {\bibfnamefont {Xiaopeng}\ \bibnamefont {Li}}, \bibinfo {author} {\bibfnamefont {Dong-Ling}\ \bibnamefont {Deng}}, \bibinfo {author} {\bibfnamefont {Yang-Le}\ \bibnamefont {Wu}}, \ and\ \bibinfo {author} {\bibfnamefont {S.}~\bibnamefont {Das~Sarma}},\ }\bibfield  {title} {\enquote {\bibinfo {title} {Statistical bubble localization with random interactions},}\ }\href {\doibase 10.1103/PhysRevB.95.020201} {\bibfield  {journal} {\bibinfo  {journal} {Phys. Rev. B}\ }\textbf {\bibinfo {volume} {95}},\ \bibinfo {pages} {020201(R)} (\bibinfo {year} {2017})}\BibitemShut {NoStop}%
\bibitem [{\citenamefont {Sierant}\ and\ \citenamefont {Zakrzewski}(2018)}]{Sierant18bos}%
  \BibitemOpen
  \bibfield  {author} {\bibinfo {author} {\bibfnamefont {Piotr}\ \bibnamefont {Sierant}}\ and\ \bibinfo {author} {\bibfnamefont {Jakub}\ \bibnamefont {Zakrzewski}},\ }\bibfield  {title} {\enquote {\bibinfo {title} {Many-body localization of bosons in optical lattices},}\ }\href {\doibase 10.1088/1367-2630/aabb17} {\bibfield  {journal} {\bibinfo  {journal} {New Journal of Physics}\ }\textbf {\bibinfo {volume} {20}},\ \bibinfo {pages} {043032} (\bibinfo {year} {2018})}\BibitemShut {NoStop}%
\end{thebibliography}%
\end{document}